\documentclass[
superscriptaddress,
twocolumn,
preprintnumbers,
nofootinbib,
 amsmath,amssymb,
 aps,prd,
]{revtex4-2}

\usepackage{graphicx}
\usepackage{dcolumn}
\usepackage{bm}
\usepackage[colorlinks,linkcolor=blue,citecolor=blue,urlcolor=blue ]{hyperref}
\usepackage{amsmath,amssymb,natbib,latexsym}
\usepackage[T1]{fontenc}
\usepackage[utf8]{inputenc}

\usepackage{bbm}
\usepackage{float}
\usepackage{url}
\usepackage[normalem]{ulem}
\usepackage{enumerate}
\usepackage[makeroom]{cancel}
\usepackage{array}
\usepackage{booktabs}
\usepackage{mathtools} 

\usepackage{xcolor}
\usepackage{comment}
\usepackage{xspace}
\usepackage{lineno}

\newcolumntype{Z}{>{\setbox0=\hbox\bgroup}c<{\egroup}@{}}

\newcommand{\appendixcite}[1]{\hyperref[#1]{\textcolor{blue}{Appendix \ref*{#1}}}}

\newcommand{\maglimplusplus}{\textsc{MagLim++}\xspace}
\newcommand{\maglim}{\textsc{MagLim}\xspace}
\newcommand{\Balrog}{\textsc{Balrog}\xspace}
\newcommand{\mdet}{\textsc{Metadetection}\xspace}
\newcommand{\Csample}{C_{\rm{sample}}}

\newcommand{\sqdeg}{{\rm deg}^2}

\usepackage{xcolor}

\usepackage[normalem]{ulem}

\usepackage{ulem}

\begin{document}



\preprint{DES-2025-0949}
\preprint{FERMILAB-PUB-25-1006-PPD}

\title[DES Y6 Magnification]{Dark Energy Survey Year 6 Results: Magnification modeling and its impact on galaxy clustering and galaxy-galaxy lensing cosmology}


\author{E.~Legnani}\email{elegnani@ifae.es}
\affiliation{Institut de F\'{\i}sica d'Altes Energies (IFAE), The Barcelona Institute of Science and Technology, Campus UAB, 08193 Bellaterra (Barcelona) Spain}

\author{J.~Elvin-Poole}
\affiliation{Department of Physics and Astronomy, University of Waterloo, 200 University Ave W, Waterloo, ON N2L 3G1, Canada}

\author{D.~Anbajagane}
\affiliation{Kavli Institute for Cosmological Physics, University of Chicago, Chicago, IL 60637, USA}

\author{D.~Sanchez Cid}
\affiliation{Centro de Investigaciones Energ\'eticas, Medioambientales y Tecnol\'ogicas (CIEMAT), Madrid, Spain}
\affiliation{Physik-Institut, University of Zürich, Winterthurerstrasse 190, CH-8057 Zürich, Switzerland}

\author{A.~Fert\'e}
\affiliation{SLAC National Accelerator Laboratory, Menlo Park, CA 94025, USA}

\author{N.~Weaverdyck}
\affiliation{Berkeley Center for Cosmological Physics, Department of Physics, University of California, Berkeley, CA 94720, US}
\affiliation{Lawrence Berkeley National Laboratory, 1 Cyclotron Road, Berkeley, CA 94720, USA}

\author{A.~Porredon}
\affiliation{Centro de Investigaciones Energ\'eticas, Medioambientales y Tecnol\'ogicas (CIEMAT), Madrid, Spain}
\affiliation{Ruhr University Bochum, Faculty of Physics and Astronomy, Astronomical Institute, German Centre for Cosmological Lensing, 44780 Bochum, Germany}

\author{S.~Avila}
\affiliation{Centro de Investigaciones Energ\'eticas, Medioambientales y Tecnol\'ogicas (CIEMAT), Madrid, Spain}

\author{R.~Miquel}
\affiliation{Instituci\'o Catalana de Recerca i Estudis Avan\c{c}ats, E-08010 Barcelona, Spain}
\affiliation{Institut de F\'{\i}sica d'Altes Energies (IFAE), The Barcelona Institute of Science and Technology, Campus UAB, 08193 Bellaterra (Barcelona) Spain}

\author{J.~De~Vicente}
\affiliation{Centro de Investigaciones Energ\'eticas, Medioambientales y Tecnol\'ogicas (CIEMAT), Madrid, Spain}

\author{J.~Coloma}
\affiliation{Institute of Space Sciences (ICE, CSIC),  Campus UAB, Carrer de Can Magrans, s/n,  08193 Barcelona, Spain}

\author{S.~Samuroff}
\affiliation{Department of Physics, Northeastern University, Boston, MA 02115, USA}
\affiliation{Institut de F\'{\i}sica d'Altes Energies (IFAE), The Barcelona Institute of Science and Technology, Campus UAB, 08193 Bellaterra (Barcelona) Spain}

\author{W.~d'Assignies}
\affiliation{Institut de F\'{\i}sica d'Altes Energies (IFAE), The Barcelona Institute of Science and Technology, Campus UAB, 08193 Bellaterra (Barcelona) Spain}

\author{A.~Alarcon}
\affiliation{Institute of Space Sciences (ICE, CSIC),  Campus UAB, Carrer de Can Magrans, s/n,  08193 Barcelona, Spain}

\author{C.~S{\'a}nchez}
\affiliation{Departament de F\'{\i}sica, Universitat Aut\`{o}noma de Barcelona (UAB), 08193 Bellaterra, Barcelona, Spain}
\affiliation{Institut de F\'{\i}sica d'Altes Energies (IFAE), The Barcelona Institute of Science and Technology, Campus UAB, 08193 Bellaterra (Barcelona) Spain}

\author{J.~Muir}
\affiliation{Department of Physics, University of Cincinnati, Cincinnati, Ohio 45221, USA}
\affiliation{Perimeter Institute for Theoretical Physics, 31 Caroline St. North, Waterloo, ON N2L 2Y5, Canada}

\author{J.~Prat}
\affiliation{Department of Astronomy and Astrophysics, University of Chicago, Chicago, IL 60637, USA}
\affiliation{Nordita, KTH Royal Institute of Technology and Stockholm University, Hannes Alfv\'ens v\"ag 12, SE-10691 Stockholm, Sweden}

\author{N.~MacCrann}
\affiliation{Department of Applied Mathematics and Theoretical Physics, University of Cambridge, Cambridge CB3 0WA, UK}

\author{D.~Bacon}
\affiliation{Institute of Cosmology and Gravitation, University of Portsmouth, Portsmouth, PO1 3FX, UK}

\author{M.~A.~Troxel}
\affiliation{Department of Physics, Duke University Durham, NC 27708, USA}

\author{C.~Chang}
\affiliation{Department of Astronomy and Astrophysics, University of Chicago, Chicago, IL 60637, USA}
\affiliation{Kavli Institute for Cosmological Physics, University of Chicago, Chicago, IL 60637, USA}

\author{M.~Crocce}
\affiliation{Institut d'Estudis Espacials de Catalunya (IEEC), 08034 Barcelona, Spain}
\affiliation{Institute of Space Sciences (ICE, CSIC),  Campus UAB, Carrer de Can Magrans, s/n,  08193 Barcelona, Spain}

\author{M.~R.~Becker}
\affiliation{Argonne National Laboratory, 9700 South Cass Avenue, Lemont, IL 60439, USA}

\author{J.~Blazek}
\affiliation{Department of Physics, Northeastern University, Boston, MA 02115, USA}

\author{M.~Yamamoto}
\affiliation{Department of Astrophysical Sciences, Princeton University, Peyton Hall, Princeton, NJ 08544, USA}
\affiliation{Department of Physics, Duke University Durham, NC 27708, USA}

\author{T.~Schutt}
\affiliation{Department of Physics, Stanford University, 382 Via Pueblo Mall, Stanford, CA 94305, USA}
\affiliation{Kavli Institute for Particle Astrophysics \& Cosmology, P. O. Box 2450, Stanford University, Stanford, CA 94305, USA}
\affiliation{SLAC National Accelerator Laboratory, Menlo Park, CA 94025, USA}

\author{M.~Rodriguez-Monroy}
\affiliation{Instituto de Física Teórica UAM/CSIC, Universidad Autónoma de Madrid, 28049 Madrid, Spain}
\affiliation{Laboratoire de physique des 2 infinis Irène Joliot-Curie}

\author{G.~Giannini}
\affiliation{Institute of Space Sciences (ICE, CSIC),  Campus UAB, Carrer de Can Magrans, s/n,  08193 Barcelona, Spain}
\affiliation{Kavli Institute for Cosmological Physics, University of Chicago, Chicago, IL 60637, USA}

\author{B.~Yin}
\affiliation{Department of Physics, Duke University Durham, NC 27708, USA}

\author{A.~Amon}
\affiliation{Department of Astrophysical Sciences, Princeton University, Peyton Hall, Princeton, NJ 08544, USA}

\author{K.~Bechtol}
\affiliation{Physics Department, 2320 Chamberlin Hall, University of Wisconsin-Madison, 1150 University Avenue Madison, WI  53706-1390}

\author{I.~Sevilla-Noarbe}
\affiliation{Centro de Investigaciones Energ\'eticas, Medioambientales y Tecnol\'ogicas (CIEMAT), Madrid, Spain}

\author{T.~M.~C.~Abbott}
\affiliation{Cerro Tololo Inter-American Observatory, NSF's National Optical-Infrared Astronomy Research Laboratory, Casilla 603, La Serena, Chile}
\author{M.~Aguena}
\affiliation{INAF-Osservatorio Astronomico di Trieste, via G. B. Tiepolo 11, I-34143 Trieste, Italy}
\affiliation{Laborat\'orio Interinstitucional de e-Astronomia - LIneA, Av. Pastor Martin Luther King Jr, 126 Del Castilho, Nova Am\'erica Offices, Torre 3000/sala 817 CEP: 20765-000, Brazil}
\author{S.~Allam}
\affiliation{Fermi National Accelerator Laboratory, P. O. Box 500, Batavia, IL 60510, USA}
\author{O.~Alves}
\affiliation{Department of Physics, University of Michigan, Ann Arbor, MI 48109, USA}
\author{F.~Andrade-Oliveira}
\affiliation{Physik-Institut, University of Zürich, Winterthurerstrasse 190, CH-8057 Zürich, Switzerland}
\author{G.~M.~Bernstein}
\affiliation{Department of Physics and Astronomy, University of Pennsylvania, Philadelphia, PA 19104, USA}
\author{S.~Bocquet}
\affiliation{University Observatory, LMU Faculty of Physics, Scheinerstr. 1, 81679 Munich, Germany}
\author{D.~Brooks}
\affiliation{Department of Physics \& Astronomy, University College London, Gower Street, London, WC1E 6BT, UK}
\author{R.~Camilleri}
\affiliation{School of Mathematics and Physics, University of Queensland,  Brisbane, QLD 4072, Australia}
\author{A.~Carnero~Rosell}
\affiliation{Instituto de Astrofisica de Canarias, E-38205 La Laguna, Tenerife, Spain}
\affiliation{Laborat\'orio Interinstitucional de e-Astronomia - LIneA, Av. Pastor Martin Luther King Jr, 126 Del Castilho, Nova Am\'erica Offices, Torre 3000/sala 817 CEP: 20765-000, Brazil}
\affiliation{Universidad de La Laguna, Dpto. Astrofísica, E-38206 La Laguna, Tenerife, Spain}
\author{J.~Carretero}
\affiliation{Institut de F\'{\i}sica d'Altes Energies (IFAE), The Barcelona Institute of Science and Technology, Campus UAB, 08193 Bellaterra (Barcelona) Spain}
\author{L.~N.~da Costa}
\affiliation{Laborat\'orio Interinstitucional de e-Astronomia - LIneA, Av. Pastor Martin Luther King Jr, 126 Del Castilho, Nova Am\'erica Offices, Torre 3000/sala 817 CEP: 20765-000, Brazil}
\author{M.~E.~da Silva Pereira}
\affiliation{Hamburger Sternwarte, Universit\"{a}t Hamburg, Gojenbergsweg 112, 21029 Hamburg, Germany}
\author{T.~M.~Davis}
\affiliation{School of Mathematics and Physics, University of Queensland,  Brisbane, QLD 4072, Australia}
\author{S.~Desai}
\affiliation{Department of Physics, IIT Hyderabad, Kandi, Telangana 502285, India}
\author{S.~Dodelson}
\affiliation{Department of Astronomy and Astrophysics, University of Chicago, Chicago, IL 60637, USA}
\affiliation{Fermi National Accelerator Laboratory, P. O. Box 500, Batavia, IL 60510, USA}
\affiliation{Kavli Institute for Cosmological Physics, University of Chicago, Chicago, IL 60637, USA}
\author{P.~Doel}
\affiliation{Department of Physics \& Astronomy, University College London, Gower Street, London, WC1E 6BT, UK}
\author{C.~Doux}
\affiliation{Department of Physics and Astronomy, University of Pennsylvania, Philadelphia, PA 19104, USA}
\affiliation{Universit\'e Grenoble Alpes, CNRS, LPSC-IN2P3, 38000 Grenoble, France}
\author{J.~Garc\'ia-Bellido}
\affiliation{Instituto de Fisica Teorica UAM/CSIC, Universidad Autonoma de Madrid, 28049 Madrid, Spain}
\author{D.~Gruen}
\affiliation{University Observatory, LMU Faculty of Physics, Scheinerstr. 1, 81679 Munich, Germany}
\author{G.~Gutierrez}
\affiliation{Fermi National Accelerator Laboratory, P. O. Box 500, Batavia, IL 60510, USA}
\author{S.~R.~Hinton}
\affiliation{School of Mathematics and Physics, University of Queensland,  Brisbane, QLD 4072, Australia}
\author{D.~L.~Hollowood}
\affiliation{Santa Cruz Institute for Particle Physics, Santa Cruz, CA 95064, USA}
\author{K.~Honscheid}
\affiliation{Center for Cosmology and Astro-Particle Physics, The Ohio State University, Columbus, OH 43210, USA}
\affiliation{Department of Physics, The Ohio State University, Columbus, OH 43210, USA}
\author{D.~Huterer}
\affiliation{Department of Physics, University of Michigan, Ann Arbor, MI 48109, USA}
\author{D.~J.~James}
\affiliation{Center for Astrophysics $\vert$ Harvard \& Smithsonian, 60 Garden Street, Cambridge, MA 02138, USA}
\author{K.~Kuehn}
\affiliation{Australian Astronomical Optics, Macquarie University, North Ryde, NSW 2113, Australia}
\affiliation{Lowell Observatory, 1400 Mars Hill Rd, Flagstaff, AZ 86001, USA}
\author{O.~Lahav}
\affiliation{Department of Physics \& Astronomy, University College London, Gower Street, London, WC1E 6BT, UK}
\author{S.~Lee}
\affiliation{Jet Propulsion Laboratory, California Institute of Technology, 4800 Oak Grove Dr., Pasadena, CA 91109, USA} 
\affiliation{Department of Physics and Astronomy, Ohio University, Clippinger Labs, Athens, OH 45701}
\author{J.~L.~Marshall}
\affiliation{George P. and Cynthia Woods Mitchell Institute for Fundamental Physics and Astronomy, and Department of Physics and Astronomy, Texas A\&M University, College Station, TX 77843,  USA}
\author{J. Mena-Fern{\'a}ndez}
\affiliation{Universit\'e Grenoble Alpes, CNRS, LPSC-IN2P3, 38000 Grenoble, France}
\author{F.~Menanteau}
\affiliation{Center for Astrophysical Surveys, National Center for Supercomputing Applications, 1205 West Clark St., Urbana, IL 61801, USA}
\affiliation{Department of Astronomy, University of Illinois at Urbana-Champaign, 1002 W. Green Street, Urbana, IL 61801, USA}
\author{J.~J.~Mohr}
\affiliation{University Observatory, LMU Faculty of Physics, Scheinerstr. 1, 81679 Munich, Germany}
\author{J.~Myles}
\affiliation{Department of Astrophysical Sciences, Princeton University, Peyton Hall, Princeton, NJ 08544, USA}
\author{R.~L.~C.~Ogando}
\affiliation{Centro de Tecnologia da Informa\c{c}\~ao Renato Archer, Campinas, SP, Brazil - 13069-901}
\affiliation{Observat\'orio Nacional, Rua Gal. Jos\'e Cristino 77, Rio de Janeiro, RJ - 20921-400, Brazil}
\author{M.~Paterno}
\affiliation{Fermi National Accelerator Laboratory, P. O. Box 500, Batavia, IL 60510, USA}
\author{A.~A.~Plazas~Malag\'on}
\affiliation{Kavli Institute for Particle Astrophysics \& Cosmology, P. O. Box 2450, Stanford University, Stanford, CA 94305, USA}
\affiliation{SLAC National Accelerator Laboratory, Menlo Park, CA 94025, USA}
\author{R.~Rosenfeld}
\affiliation{ICTP South American Institute for Fundamental Research\\ Instituto de F\'{\i}sica Te\'orica, Universidade Estadual Paulista, S\~ao Paulo, Brazil}
\affiliation{Laborat\'orio Interinstitucional de e-Astronomia - LIneA, Av. Pastor Martin Luther King Jr, 126 Del Castilho, Nova Am\'erica Offices, Torre 3000/sala 817 CEP: 20765-000, Brazil}
\author{E.~Sanchez}
\affiliation{Centro de Investigaciones Energ\'eticas, Medioambientales y Tecnol\'ogicas (CIEMAT), Madrid, Spain}
\author{M.~Smith}
\affiliation{Physics Department, Lancaster University, Lancaster, LA1 4YB, UK}
\author{M.~Soares-Santos}
\affiliation{Physik-Institut, University of Zürich, Winterthurerstrasse 190, CH-8057 Zürich, Switzerland}
\author{E.~Suchyta}
\affiliation{Computer Science and Mathematics Division, Oak Ridge National Laboratory, Oak Ridge, TN 37831}
\author{V.~Vikram}
\affiliation{Department of Physics and Astronomy, University of Pennsylvania,  Philadelphia, PA 19104, USA}

\collaboration{DES Collaboration}

\date{\today}


\begin{abstract}
Gravitational lensing magnification alters the observed spatial distribution of galaxies and must be accounted for to prevent biases in cosmological probes of the large-scale structure. We investigate its effects on the Dark Energy Survey Year 6 galaxy clustering and galaxy-galaxy lensing analyses using the fiducial lens (position tracer) sample \maglimplusplus. Magnification bias is parameterized by a coefficient that describes the response of the number of selected objects per unlensed area element to a change in the lensing convergence. We quantify this coefficient using the \Balrog synthetic source injection catalog to account for the complexity of the selection function, and compare these results with simplified estimates. The resulting values of the magnification coefficients for each redshift bin are $[3.16 \pm 0.08, 2.76 \pm 0.21, 4.09 \pm 0.15, 4.42 \pm 0.16, 4.90 \pm 0.29, 4.83 \pm 0.25]$. Relative to Year 3, this analysis provides more precise and accurate magnification bias estimates through a larger \Balrog area and reweighting to better match the data properties. The cosmological results are robust when tested against various magnification parameter prior choices and also when adding cross-clustering between lens redshift bins. Neglecting magnification, however, introduces significant systematic shifts: relative to the fiducial analysis with Gaussian priors centered on the \Balrog-derived estimates, we observe shifts of $1.37\sigma$ in $S_8$ and $-0.84\sigma$ in $\Omega_m$ (with cosmic shear included: $-0.61\sigma$ in $S_8$ and $-0.71\sigma$ in $\Omega_m$), in agreement with findings from simulated data, demonstrating that magnification must be modeled to avoid biases. Freeing the magnification bias in lens bin 2 leads to unphysical negative values, further justifying its exclusion from the fiducial Year 6 analysis.


\end{abstract}



\maketitle


\section{Introduction}
\label{sec:introduction}

Observations of the large-scale structure (LSS) of the Universe provide key insights into both the growth of cosmic structure and the expansion history of the Universe. Over the past decades, wide-field imaging surveys have enabled increasingly precise cosmological measurements, playing a central role in the advancement of LSS studies. By combining complementary probes---cosmic shear (the weak gravitational lensing of background galaxies), foreground galaxy clustering, and their cross-correlation, known as galaxy-galaxy lensing---these surveys can extract maximal cosmological information and break degeneracies among key parameters. The combination of these three two-point (2pt) correlation functions is commonly referred to as a ``3$\times$2pt'' analysis, while a ``2$\times$2pt'' analysis includes only galaxy clustering and galaxy-galaxy lensing.

Stage-III surveys such as the Dark Energy Survey (DES\footnote{\url{http://www.darkenergysurvey.org/}}, \citet{DECam}), the Kilo-Degree Survey (KiDS\footnote{\url{http://kids.strw.leidenuniv.nl/}}, \citet{de_jong_kilo-degree_2013}), and the Hyper Suprime-Cam Survey (HSC\footnote{\url{https://www.naoj.org/Projects/HSC/}}, \citet{aihara_hyper_2018}), have been particularly impactful, providing unprecedented depth and area for precision cosmology. The DES Year 6 (Y6) analysis---based on the full six years of observations, whereas the earlier Year 3 (Y3) analysis used only the data from the first three years---provides the final results of the survey and marks a significant advance over the previous Y3 analysis (\citet{y3-3x2ptkp}). Improvements in imaging depth, sample selection and calibration, and modeling strategy increased both the statistical power and the robustness of systematic error mitigation. Several key Y6 cosmology papers present these results, including analyses of the 2$\times$2pt (\citet{y6-2x2pt}) and 3$\times$2pt combinations (\citet{y6-3x2pt}), alongside detailed descriptions of the galaxy samples and their calibration. Of particular relevance for this work are the galaxy clustering measurement and sample selection described in \citet*{y6-maglim}, and the galaxy-galaxy lensing measurement presented in \citet{y6-gglens}).

An effect of growing importance in such analyses is weak lensing magnification. Magnification arises when foreground mass distributions alter the observed positions, fluxes and sizes of background galaxies. This modifies the number of galaxies that meet survey selection thresholds and, consequently, the observed clustering and lensing signals. While magnification effects are typically subdominant compared to intrinsic clustering or shear correlations, their impact grows with survey depth and precision, and becomes particularly relevant in the era of Stage-IV surveys such as Euclid (\citet{euclid_collaboration_euclid_2025}) and LSST (\citet{ivezic_lsst_2019}). Accounting for magnification, and in particular using informative priors on the magnification coefficients, can both reduce biases and enhance the constraining power of cosmological analyses.

Forecasts for next-generation surveys underscore this point. Studies for LSST (\citet{mahony_forecasting_2022}), Euclid (\citet{von_wietersheim-kramsta_magnification_2021, euclid_collaboration_euclid_2024}), and simulation-based analyses (\citet{duncan_cosmological_2022, thiele_disentangling_2020, lorenz_impact_2018, deshpande_post-limber_2020}) consistently show that magnification must be accurately modeled to obtain unbiased results and fully exploit the statistical power of these datasets.

Other works have estimated magnification bias for specific survey samples: for example, \citet{unruh_importance_2020} used the Millennium Simulation to estimate magnification for KiDS, with their results later applied in the KiDS-1000 analyses (\citet{dvornik_kids-1000_2023}). More recently, magnification has also been modeled in the context of spectroscopic galaxy clustering, including in BOSS (\citet{wenzl_magnification_2024, von_wietersheim-kramsta_magnification_2021}), WISE (\citet{wiselensing}) and forecast for Euclid (\citet{euclid_collaboration_euclid_2024}).

Within DES, lens magnification has been incorporated into the theoretical modeling of galaxy clustering and galaxy-galaxy lensing. This work follows the procedure established in the DES Y3 analyses (\citet*{y3-2x2ptmagnification}), extending and adapting it to the improved data and modeling framework of the Y6 analysis.

This paper is organized as follows: in Sec.~\ref{sec:theory}, we provide an overview of the theoretical framework for magnification and its role in LSS observables; Sec.~\ref{sec:data} provides details on the DES Y6 data and the simulations used in our analysis; in Sec.~\ref{sec:mag_coeff}, we estimate the magnification contributions to our theoretical predictions using multiple approaches; Sec.~\ref{sec:exp_mag_cosmology} extends these estimates to assess the expected magnification effects on DES Y6 cosmology; in Sec.~\ref{sec:mag_cosmology}, we discuss the impact of magnification on cosmological results using the DES Y6 data; we conclude in Sec.\ref{sec:conclusions} with a summary and outlook for future Stage-IV surveys.


\section{Theory}
\label{sec:theory}

\subsection{Galaxy Density Field}
\label{sec:density_field}

In large-scale structure surveys, the observed distribution of galaxies is used to probe the underlying matter density field through tomographic clustering and galaxy-galaxy lensing measurements. These measurements rely on accurate modeling of the observed number density of galaxies, which can be affected by various astrophysical and observational effects. In the absence of magnification and redshift space distortions (RSD), the line-of-sight projection of the three-dimensional galaxy density contrast in a tomographic bin $i$ can be expressed as:
\begin{equation}
    \delta_{g,\rm{int}}^i (\hat{\textbf{n}}) = \int {\rm{d}} \chi W_g^i(\chi) \delta_g^{\rm{3D}}(\hat{\textbf{n}}\chi, \chi),
\label{eq:density_field}
\end{equation}
where $\chi$ is the comoving distance, $\hat{\textbf{n}}$ represents a given direction in the sky, and $W_g^i(\chi)=n_g^i(z)\ \rm{d}z/\rm{d}\chi$ the normalized selection function of galaxies in redshift bin $i$. 

In this work, we explore the impact of magnification on the observed projected galaxy density contrast and its contribution to two-point statistics. For simplicity, we omit RSD from the following equations, although we note that RSD effects are fully incorporated in both the DES Y6 modeling and all analyses presented in this paper. Because RSD affect only the line-of-sight positions of galaxies, while magnification modifies the projected galaxy density through lensing, the two effects are independent at first order and uncorrelated (any second-order couplings from relativistic effects are negligible for the scales and precision of this analysis).

\subsection{Magnification bias}
\label{sec:magnification_bias}

Gravitational lensing magnification distorts the apparent distribution of galaxies by modifying their positions, sizes, and brightness while preserving surface brightness. The strength of this effect is quantified by the magnification factor $\mu$, which depends on the local weak lensing convergence $\kappa$ and shear $\gamma$ (\citet{bartelmann_weak_2001}):
\begin{equation}
    \mu 
    = \frac{1}{(1-\kappa)^2 - \vert \gamma \vert ^2 } \approx \frac{1}{1-2\kappa}\approx 1+ 2\kappa,
\label{eq:magnification_kappa}
\end{equation}
in the limit of weak lensing, where $\kappa \ll 1$ and $\gamma \ll 1$. Magnification affects the apparent distribution of galaxies by altering the solid angle subtended by a given patch of the sky. This change in the observed area impacts the observed number density of sources in two competing ways: by modifying the apparent separation between objects, and by altering the selection probability of individual galaxies through the change in the amount of light collected from each by a telescope.

First, because lensing stretches the sky by a factor of $\mu$, a given solid angle element $\Delta \Omega$ in the unlensed sky is mapped to an expanded area $\mu \Delta \Omega$, causing the observed number density of galaxies per unit solid angle to decrease by the same factor. This dilution effect reduces the apparent abundance of galaxies in magnified regions.  

Second, the same stretching of the sky also affects the observed flux and apparent size of individual sources. Since the solid angle captured by the telescope increases while surface brightness remains unchanged, the total observed flux of a galaxy is boosted by a factor of $\mu$, and its apparent size increases. These changes influence sample selection, particularly for flux or size limited surveys: sources that would otherwise fall below a survey’s detection threshold can become observable due to magnification. This leads to an excess of detected galaxies in magnified regions.

The net outcome of these competing effects, known as \textit{magnification bias}, depends on the properties of the galaxy population, selection criteria, and survey characteristics. In realistic galaxy surveys, selection effects are complex and depend on multiple factors beyond total flux. Accurately predicting the net impact of magnification thus requires detailed modeling and simulations to account for the full selection function of a given dataset.  

To quantify the impact of magnification on galaxy clustering and galaxy-galaxy lensing, we decompose the observed density contrast of the lens sample into two components: the intrinsic galaxy density contrast and an additional term induced by magnification,
\begin{equation}
     \delta_g^{\rm{obs}}(\hat{\textbf{n}})  =  \delta_g^{\rm{int}}(\hat{\textbf{n}}) 
 + \delta_g^{\rm{mag}}(\hat{\textbf{n}}, \kappa).
\label{eq:density}
\end{equation}
The magnification-induced overdensity, $\delta_g^{\rm{mag}}(\hat{\textbf{n}}, \kappa)$, can be expressed in terms of the observed galaxy number density, $n(\hat{\textbf{n}}, \kappa)$, and the same quantity in the absence of lensing ($\kappa = 0$):
\begin{equation}
    \delta_g^{\rm{mag}}(\hat{\textbf{n}}, \kappa) = \frac{n(\hat{\textbf{n}},\kappa)}{n(\hat{\textbf{n}},0)} - 1.
\label{eq:density_mag}
\end{equation}

The galaxy sample is selected based on thresholds applied to various observed (i.e., lensed) properties, represented by a vector $\mathbf{F}'$. The observed number density of galaxies at a given position $\hat{\textbf{n}}$ can be expressed as an integral over $N(\mathbf{F},\hat{\textbf{n}})$, which represents the absolute number of galaxies with intrinsic (unlensed) properties $\mathbf{F}$, divided by the lensed sky area element $\Delta\Omega(\kappa)$:
\begin{equation}
n(\hat{\textbf{n}},\kappa)
= \frac{1}{\Delta\Omega(\kappa)} \int {\rm{d}}{\mathbf{F}} S(\mathbf{F}') N(\mathbf{F},\hat{\textbf{n}}),
\end{equation}
where $S(\mathbf{F}')$ denotes the selection function applied on the lensed properties $\mathbf{F}'$. For small convergence $\kappa$, the area element transforms as $\Delta\Omega(\kappa) = \Delta\Omega(0)/(1-2\kappa)$, allowing us to rewrite the number density as
\begin{equation}
n(\hat{\textbf{n}},\kappa) = \frac{1-2\kappa}{\Delta\Omega(0)} \int {\rm{d}}{\mathbf{F}} S(\mathbf{F}') N(\mathbf{F},\hat{\textbf{n}}).
\end{equation}
Expanding the selection function $S(\mathbf{F}')$ in a Taylor expansion around $\kappa=0$, we obtain
\begin{equation}
n(\hat{\textbf{n}},\kappa)
\approx \frac{1-2\kappa}{\Delta\Omega(0)} \int {\rm{d}}{\mathbf{F}} \left[S(\mathbf{F})+\kappa \left. \frac{\partial{S}}{\partial{\kappa}}\right|_\kappa \right] N(\mathbf{F},\hat{\textbf{n}}).
\end{equation}
Neglecting terms of order $\kappa^2$, this leads to 
\begin{equation}
\begin{aligned}
n(\hat{\textbf{n}},\kappa)
&\approx \frac{1-2\kappa}{\Delta\Omega(0)} \int {\rm{d}}{\mathbf{F}} S(\mathbf{F})N(\mathbf{F},\hat{\textbf{n}}) \\
&\quad + \kappa \frac{1}{\Delta\Omega(0)}\int {\rm{d}}{\mathbf{F}} \frac{\partial{S}}{\partial{\kappa}} N(\mathbf{F},\hat{\textbf{n}}) \\[0.5em]
&\approx (1-2\kappa) n(\hat{\textbf{n}},0) +  \frac{\kappa}{\Delta\Omega(0)}\int {\rm{d}}{\mathbf{F}} \frac{\partial{S}}{\partial{\kappa}}N(\mathbf{F},\hat{\textbf{n}}).
\end{aligned}
\label{eq:n_sel}
\end{equation}
Substituting this into \eqref{eq:density_mag}, we obtain
\begin{equation}
\begin{aligned}
    \delta_g^{\rm{mag}}(\hat{\textbf{n}}) &= \kappa(\hat{\textbf{n}}) \left(-2 + \frac{1}{n(\hat{\textbf{n}},0)\Delta\Omega(0)}\int {\rm{d}}{\mathbf{F}} \frac{\partial{S}}{\partial{\kappa}}N(\mathbf{F},\hat{\textbf{n}})\right)\\
    &= \kappa(\hat{\textbf{n}}) \left[-2 + \frac{1}{N(\hat{\textbf{n}},0)}
    \frac{\partial N(\hat{\textbf{n}},0)}{\partial \kappa} \right],
\end{aligned}
\label{eq:density_mag2}
\end{equation}
where
\begin{equation}
    N(\hat{\textbf{n}},0) = \int {\rm{d}}{\mathbf{F}} S(\mathbf{F})N(\mathbf{F},\hat{\textbf{n}}).
\end{equation}
In Eq.~\eqref{eq:n_sel}, the first term corresponds to the number density in the absence of lensing, $n(\hat{\textbf{n}},0)$, modified by a factor of $(1-2\kappa)$ due to the change in area element. The second term accounts for the response of the number of selected galaxies per unit (unlensed) area to variations in $\kappa$. Consequently, the total magnification-induced contribution to the number density contrast can be expressed as
\begin{equation}
    \delta_g^{\rm{mag}}(\hat{\textbf{n}}) =  \kappa(\hat{\textbf{n}})\left[ C_{\rm{area}} + \Csample \right],
\label{eq:c_tot}
\end{equation}
where the area contribution is given by $C_{\rm{area}}=-2$, and the overall magnification effect is determined by a single coefficient $C=C_{\rm{area}} + \Csample$.
For cases where the galaxy sample is selected based solely on an observed magnitude threshold, $m_{\textrm{cut}}$, this expression simplifies to (\citet{joachimi_simultaneous_2010}, \citet{garcia-fernandez_weak_2018})\footnote{There is a wide variety of notations in the literature for the magnification bias, which can be ambiguous. For instance, other DES papers define the magnification coefficient as $2(\alpha-1)$ with $\alpha=2\Csample$; this should not be confused with the definition of $\alpha$ in Eq.~\eqref{eq:mag_alpha} for a magnitude-limited sample.}:
\begin{equation}
   \delta_g^{\rm{mag}}(\hat{\textbf{n}}) = 2[\alpha(m_{\rm{cut}})-1]\kappa (\hat{\textbf{n}}),
\label{eq:density_mag_alpha}
\end{equation}
where 
\begin{equation}
    \alpha(m_{\rm{cut}}) = 2.5\left. \frac{\rm{d}}{{\rm{d}}m} \log_{10}{N_{\mu}(m)} \right|_{m=m_{\rm{cut}}},
\label{eq:mag_alpha}
\end{equation}
and $N_{\mu}(m)$ represents the (lensed) cumulative number of galaxies as a function of limiting magnitude $m$. Whether magnification increases or decreases the observed galaxy density depends on the interplay between the boost in observed flux and the dilution caused by area distortion. In this context, the balance is governed by the intrinsic slope of the cumulative luminosity function. The higher the ratio of faint to bright objects in the sample, the more dominant the flux magnification effect becomes.

Since real galaxy selections involve a complex combination of flux, color, position, and shape criteria, the selection response constant, $\Csample$, cannot be easily computed analytically. Instead, in the DES Y6 analysis, this quantity is estimated using the \Balrog image simulation (\citet*{y6-balrog}), as detailed in Sec.~\ref{sec:mag_coeff_balrog}.

\subsection{Two-point statistics}
\label{sec:2pt}

As discussed in Sec.~\ref{sec:magnification_bias}, gravitational lensing magnification modifies the observed galaxy density contrast, $\delta_g^{\rm{obs}}$, relative to its intrinsic form, $\delta_g^{\rm{int}}$. This effect is parameterized by the coefficient $C^i$ and is proportional to the convergence $\kappa$, such that:
\begin{equation}
    \delta_{g\rm{,mag}}^i(\ell) =  C^i \ \kappa^i(\ell),
\label{eq:c_tot_short}
\end{equation}
where the index $i$ accounts for tomographic redshift measurements, denoting the $i$-th redshift bin, and the density contrast is now expressed in harmonic space as a function of the angular multipole $\ell$.

In this study, we focus on the magnification of the lens sample, as its impact on clustering and lensing statistics is significant. While galaxies in the source sample also experience magnification, its effect enters the two-point functions at higher order and can be safely neglected in our analysis. Appendix~\ref{app:source_mag} provides further details on source magnification in DES Y6, and additional justification for its exclusion in 2$\times$2pt analyses can be found in \citet{y3-gglensing} and \citet{duncan_cosmological_2022}.

Magnification is important because the induced change in the observed number density of lens galaxies is not random, but correlated with the lensing convergence, $\kappa$. This effect propagates into the two-point correlation functions, affecting both galaxy clustering (i.e., the galaxy auto-correlation, Eq.~\ref{eq:density_corr}) and galaxy-galaxy lensing (i.e., the galaxy-shear cross-correlation, Eq.~\ref{eq:density_shear_corr}), collectively referred to as the 2$\times$2pt analysis, and becomes more pronounced for high-redshift lens bins, where $\kappa$ is larger due to more intervening large-scale structure. Properly incorporating these magnification contributions is crucial to avoiding biases in cosmological analyses.

By averaging over all galaxy pairs in redshift bins $i$ and $j$, the observed galaxy density correlation function can be expressed as:
\begin{equation}
\begin{aligned}
    \langle \delta_{g\rm{,obs}}^i \delta_{g\rm{,obs}}^j \rangle	
    =& \langle \delta_{g\rm{,int}}^i \delta_{g\rm{,int}}^j \rangle + C^iC^j \langle \kappa^i \kappa^j \rangle \\ 
    &+ C^j  \langle \delta_{g\rm{,int}}^i \kappa^j \rangle + C^i  \langle \delta_{g\rm{,int}}^j \kappa^i \rangle,
\label{eq:density_corr}
\end{aligned}
\end{equation}
where we have dropped the explicit dependence on the angular multipole $\ell$ for brevity. Similarly, the correlation between the observed galaxy density and the shear of background source galaxies is given by:
\begin{equation}    
    \begin{aligned}
        \langle \delta_{g\rm{,obs}}^i \gamma_{\rm{obs}}^j \rangle	&= \langle \delta_{g\rm{,int}}^i (\gamma_{\rm{G}}^j+\gamma_{\rm{IA}}^j) \rangle + C^i \langle \kappa^i (\gamma_{\rm{G}}^j+\gamma_{\rm{IA}}^j) \rangle  \\
        &= \langle \delta_{g\rm{,int}}^i (\gamma_{\rm{G}}^j+\gamma_{\rm{IA}}^j) \rangle + C^i \langle \kappa^i \gamma_{\rm{G}}^j \rangle + C^i \langle \kappa^i \gamma_{\rm{IA}}^j \rangle, 
    \end{aligned}
\label{eq:density_shear_corr}
\end{equation}
where the observed shear $\gamma_{\rm{obs}}$ of the source galaxies consists of two contributions: the shear due to gravitational lensing, $\gamma_{\rm{G}}$, and their intrinsic alignment, $\gamma_{\rm{IA}}$, which arises from their intrinsic orientation within the large-scale structure.

The modeling of the two-point correlation functions is described in detail in \citet*{y6-methods}. Here, we summarize the key aspects of this computation.

The power spectra are related to the angular correlation functions via a full-sky projection (e.g., \citet{stebbins_weak_1996}, \citet{kamionkowski_statistics_1997}). Galaxy clustering, denoted as $\omega(\theta)$, represents the angular correlation of lens galaxy positions between tomographic redshift bins $i$ and $j$ and is given by:
\begin{equation} 
    \omega^{ij}(\theta) = \sum_\ell \frac{2\ell+1}{4\pi}P_\ell(\cos\theta) C^{ij}_{\delta_{\rm{g,obs}}\delta_{\rm{g,obs}}}(\ell)~.
\label{eq:2pt_w}
\end{equation}
The galaxy-galaxy lensing signal, $\gamma_t(\theta)$, which captures the cross-correlation between the galaxy overdensity field in tomographic bin $i$ and the E-mode component of the shear field in bin $j$, is given by:
\begin{equation} 
    \gamma_t^{ij}(\theta) = \sum_\ell \frac{2\ell+1}{4\pi}\frac{P^2_\ell(\cos\theta)}{\ell(\ell+1)} C^{ij}_{\delta_{\rm{g,obs}}\rm{E}}(\ell)~.
\label{eq:2pt_gammat}
\end{equation}
Here, $P_\ell$ and $P_\ell^2$ are Legendre polynomials.


To compute the galaxy-galaxy lensing power spectrum, we employ the Limber approximation, which simplifies the line-of-sight integral by assuming that only small-scale modes contribute significantly. For two general fields $A$ and $B$, the power spectrum is given by:
\begin{equation} 
     C_{AB}^{ij}(\ell) = \int \mathrm{d}\chi \frac{W_A^i(\chi)W_B^j(\chi)}{\chi^2}
     P_{AB}\left(k = \frac{\ell+0.5}{\chi},z(\chi)\right),
     \label{eq:cl}
\end{equation} 
where $W_A^i(\chi)$ and $W_B^i(\chi)$ are the tomographic lens efficiencies for the respective fields, and $P_{AB}(k,z)$ is the corresponding three-dimensional power spectrum evaluated at wave number $k$ and redshift $z(\chi)$. For details, see \citet*{y6-methods}.

However, when computing the galaxy angular clustering power spectrum, the Limber approximation is insufficient, particularly on large scales. Instead, we follow the approach outlined in \citet{fang_beyond_2020}. The exact expression for the galaxy angular power spectrum (neglecting magnification and RSD) is given by:
\begin{equation} 
    \begin{aligned}
        C_{gg}^{ij} (\ell) =& \frac{2}{\pi}\int \mathrm{d} \chi_1\,W^i_g(\chi_1)\int \mathrm{d}\chi_2\,W^j_g(\chi_2)\\
        &\int\frac{\mathrm{d}k}{k}k^3 P_{gg}(k,\chi_1,\chi_2)j_\ell(k\chi_1)j_\ell(k\chi_2)\,,
    \end{aligned}
\label{eq:Cl_gg}
\end{equation} 
where $j_\ell(x)$ are spherical Bessel functions, and $P_{gg}(k,\chi_1,\chi_2)$ is the galaxy power spectrum as a function of redshift. In practice, the computation of Eq.~\eqref{eq:Cl_gg} is simplified by applying the Limber approximation on small, nonlinear scales, where it remains valid, and by treating the large, linear scales exactly. On these linear scales, the power spectrum can be separated into scale- and time-dependent components, which allows for a faster and more accurate evaluation of the full integrals. The complete expression, including magnification and RSD, is given in \citet{fang_beyond_2020}.

\subsection{Parameter inference}
\label{sec:inference}

Constraints on the cosmological parameters are obtained using a Bayesian analysis. The \textit{posterior}, i.e., the a posteriori knowledge of the parameters $\textbf{p}$ of the model $M$ given the observed data $\hat{\textbf{D}}$, denoted by $\mathcal{P}(\textbf{p}|\hat{\textbf{D}}, M)$, is related via Bayes’ theorem to the \textit{likelihood} $\mathcal{L}(\hat{\textbf{D}}| \textbf{p}, M)$ and to the \textit{prior} $\Pi(\textbf{p}|M)$ as

\begin{equation}
    \mathcal{P}(\textbf{p}|\hat{\textbf{D}}, M) = \frac{\mathcal{L}(\hat{\textbf{D}}| \textbf{p}, M) \Pi(\textbf{p}|M)}{P(\hat{\textbf{D}}| M)},
\end{equation}
where $P(\hat{\textbf{D}}| M)$ is the \textit{evidence} of the data. The likelihood is assumed to follow a multivariate Gaussian distribution:

\begin{equation} 
     {\rm{ln}} \ \mathcal{L}(\hat{\textbf{D}}|\textbf{p}, M) = -\frac{1}{2} \ (\hat{\textbf{D}} - \textbf{T}_M(\textbf{p}))^{\rm{T}} \ \textbf{C}^{-1} \ (\hat{\textbf{D}} - \textbf{T}_M(\textbf{p})),
     \label{eq:likelihood}
\end{equation} 
where $\textbf{C}$ is the data covariance matrix and $\textbf{T}_{M}(\textbf{p})$ is the theoretical prediction for the data vector $\hat{\textbf{D}}$. The data vector is constructed by concatenating all tomographic 2-point correlation function measurements over $N_\theta$ angular bins and $N_z$ redshift bins pairs, after applying scale cuts---which remove small- and large-scale measurements where modeling uncertainties (e.g., baryonic effects, non-linear galaxy bias) are significant. The covariance is modeled analytically, as described and validated in \citet*{y6-methods}. 

The posterior is sampled using the importance nested sampling algorithm \textsc{Nautilus}\footnote{\url{https://nautilus-sampler.readthedocs.io}} (\citet{lange_nautilus_2023}), and the analysis pipeline is built upon \textsc{CosmoSIS}\footnote{\url{https://cosmosis.readthedocs.io}} (\citet{zuntz_cosmosis_2015}), a modular cosmological parameter estimation framework using \textsc{CAMB} (\citet{lewis_efficient_2000}, \citet{howlett_cmb_2012}) to calculate the linear matter power spectrum. We perform parameter inference on the 2-point correlation function measurements over a set of cosmological, astrophysical, and systematic parameters. The fiducial values and priors for these parameters are detailed in \citet*{y6-methods}. We conduct this analysis in two cosmological models, $\Lambda$CDM and $w$CDM, both assuming a flat universe and allowing for a free neutrino mass.

DES Y6 follows a three-stage blinding process to prevent confirmation bias: (1) catalog-level: shear values are rescaled by an unknown factor until the catalog passes validation; (2) data vector-level: data vectors are blinded using the method of \citet{y3-blinding}, which shifts them by a cosmology-dependent factor; (3) parameter-level: chains on unblinded data vectors are run but remain blinded through random parameter offsets until all analysis choices and validation steps are finalized.

For a comprehensive discussion of the DES Y6 analysis choices, as well as validation of the modeling and pipeline robustness, see \citet*{y6-methods}.


\section{Data}
\label{sec:data}

The Dark Energy Survey (DES) conducted imaging from 2013 to 2019 using the 570-megapixel Dark Energy Camera (DECam; \citet{DECam}) on the 4-meter Blanco Telescope at Cerro Tololo Inter-American Observatory in Chile. The survey covered approximately 5000 $\sqdeg$ of the southern sky in five broadband filters (\textit{grizY}), spanning visible to near-infrared wavelengths.

DES collected 76217 high-quality exposures, which were processed through detrending, calibration, coaddition and object detection, resulting in Data Release 2 (DR2; \citet{abbott_dark_2021}) with approximately 691 million detected objects. Data processing was handled by the DES Data Management system (DESDM; \citet{imageproc}), which generated object catalogs using \textsc{SExtractor} (\citet{bertin_sextractor_1996}). The Y6 Gold dataset (\citet{y6-gold}) was built upon DR2 with additional processing pipelines that enhance measurement accuracy, photometric calibration, object classification, and incorporate other essential metadata, while retaining the same objects. The Y6 dataset provides a deeper imaging depth than Y3, reaching a median depth of $i_{\text{AB}} \sim 23.4$ mag for extended objects at $S/N \sim 10$, compared to $i_{\text{AB}} \sim 23.0$ mag for Y3 (\citet{y3-gold}).

\subsection{\maglimplusplus lens sample}
\label{sec:maglim}

The DES Y6 fiducial lens sample, \maglimplusplus (\citet*{y6-maglim}), follows a selection similar to that of the Y3 \maglim catalog (\citet{y3-2x2ptaltlensresults}), with additional quality cuts applied to refine the sample. Objects are selected from the Y6 Gold catalog based on an $i$-band magnitude limit that varies with the mean photometric redshift, estimated using the Directional Neighborhood Fitting (DNF) algorithm (\citet{de_vicente_dnf_2016}):
\begin{equation}
    17.5 < m_i < 18 + 4 z.
    \label{eq:maglim}
\end{equation}

Objects with \texttt{FLAGS\_GOLD} = 0 are selected to exclude those with unusual features in the measurement process or those considered unphysical. A star-galaxy separation is performed using \texttt{EXT\_MASH} = 4, where \texttt{EXT\_MASH} assigns each object in the Y6 Gold catalog to a morphological class, with larger values indicating objects more confidently classified as extended sources (see \citet{y6-gold}).

The \maglimplusplus sample uses the DES Y6 \textsc{joint} mask (\citet*{y6-mask}, 4031.04 deg$^2$), which defines the survey footprint for both galaxy clustering and weak lensing analyses. This mask removes regions affected by imaging artifacts, poor observing conditions, and extreme values of spatially varying observational systematics. By limiting these systematics, the mask improves the performance of decontamination methods and ensures more accurate galaxy weights.

Weights are applied to each galaxy to account for variations in number densities that are correlated with observing properties, as detailed in \citet*{y6-maglim}.

For the \maglimplusplus sample, additional selections are implemented to improve sample quality beyond the Y3 \maglim catalog. These include: (1) a near-infrared star-galaxy separation technique using forced photometry magnitudes from the unWISE data release (\citet{unwise}), which applies redshift-dependent cuts in $m_r-m_z$ and $m_z-m_{W1}$ color space to reduce stellar contamination optimized for each redshift bin, and (2) a self-organizing map (SOM)-based method to exclude regions in color space affected by significant systematic contamination from interlopers. Further details on these additional cuts can be found in \citet*{y6-maglim}.

The sample is divided into six tomographic redshift bins, ensuring uniform number densities and robustness to systematics (\citet{y3-2x2ptaltlensresults}). Table~\ref{tab:maglimplusplus} lists the redshift bin edges, number of galaxies, and number densities for each bin.

\begin{table}
\setlength{\tabcolsep}{4pt}
\begin{tabular}{cccc}
\hline
\hline
\rule{0pt}{1.2em}
Bin & Redshift range & $N_{\rm gal}$ & $n_{\rm gal}$ [arcmin$^{-2}$] \rule[-0.8em]{0pt}{0pt} \\ 
\hline
 1 & (0.20, 0.40] & 1852541 & 0.128 \\ 
 2 & (0.40, 0.55] & 1335298 & 0.092 \\ 
 3 & (0.55, 0.70] & 1413743 & 0.097 \\ 
 4 & (0.70, 0.85] & 1783837 & 0.123 \\ 
 5 & (0.85, 0.95] & 1391524 & 0.096 \\ 
 6 & (0.95, 1.05] & 1409284 & 0.097 \\ 
\hline 
\end{tabular}

\caption{Redshift bin edges, number of galaxies $N_{\rm gal}$, and number densities $n_{\rm gal}$ for the six tomographic redshift bins of the \maglimplusplus lens sample.}
\label{tab:maglimplusplus}
\end{table}

\subsection{\mdet source sample}
\label{sec:mdet}

The \mdet sample (\citet*{y6-metadetect}) is a weak lensing galaxy shape catalog based on the full six years of DES imaging. Covering the entire DES footprint, it includes around 152 million source galaxies detected in the \textit{griz} bands, with an effective number density of $8.22$ galaxies per $\rm{arcmin}^2$ and a shape noise of $\sigma_e = 0.29$. \mdet incorporates advanced image-processing techniques, including a cell-based image coaddition method and a dedicated shear calibration algorithm, ensuring precise shape measurements for weak lensing studies. After applying star and foreground masks (\citet*{y6-gold}), flux and color outlier cuts (\citet*{y6-metadetect}; \citet{y6-sourcepz}), and the \textsc{joint} mask (\citet*{y6-mask}), the galaxies are grouped into four tomographic bins.

\subsection{\Balrog synthetic source injection}
\label{sec:balrog}

The \Balrog catalog (\citet*{y6-balrog}) is a synthetic source injection (SSI) dataset designed to characterize detection and measurement biases in the DES wide-field survey. It is created by injecting artificial galaxies---whose light profiles are derived from model fits to DES Y3 deep-field observations (\citet*{y3-deepfields})---into real DES on-sky images. These injected sources, with known photometry and morphology, are processed through the full DES photometric pipeline to produce object catalogs that can be matched to their original inputs. This provides a distribution of realistic, noise-affected measurements for each injected source, enabling a detailed study of the survey transfer function. The final catalog contains 146 million injected sources distributed across the full 5000 deg$^2$ DES footprint.

In the Y6 \Balrog simulations, alongside the fiducial run, a magnification run was performed in which the same sources were injected but with their fluxes and areas increased by 2\%, simulating the effect of magnification corresponding to a convergence of $\delta\kappa = 0.01$. This magnified variant of \Balrog was run on 2000 tiles, covering 1000 deg$^2$ (or 20\% of the full DES Y6 footprint), with each tile maintaining identical source injections, including the same sky positions, as in the fiducial run. 

\begin{figure}
    \centering
    \includegraphics[width=0.45\textwidth]{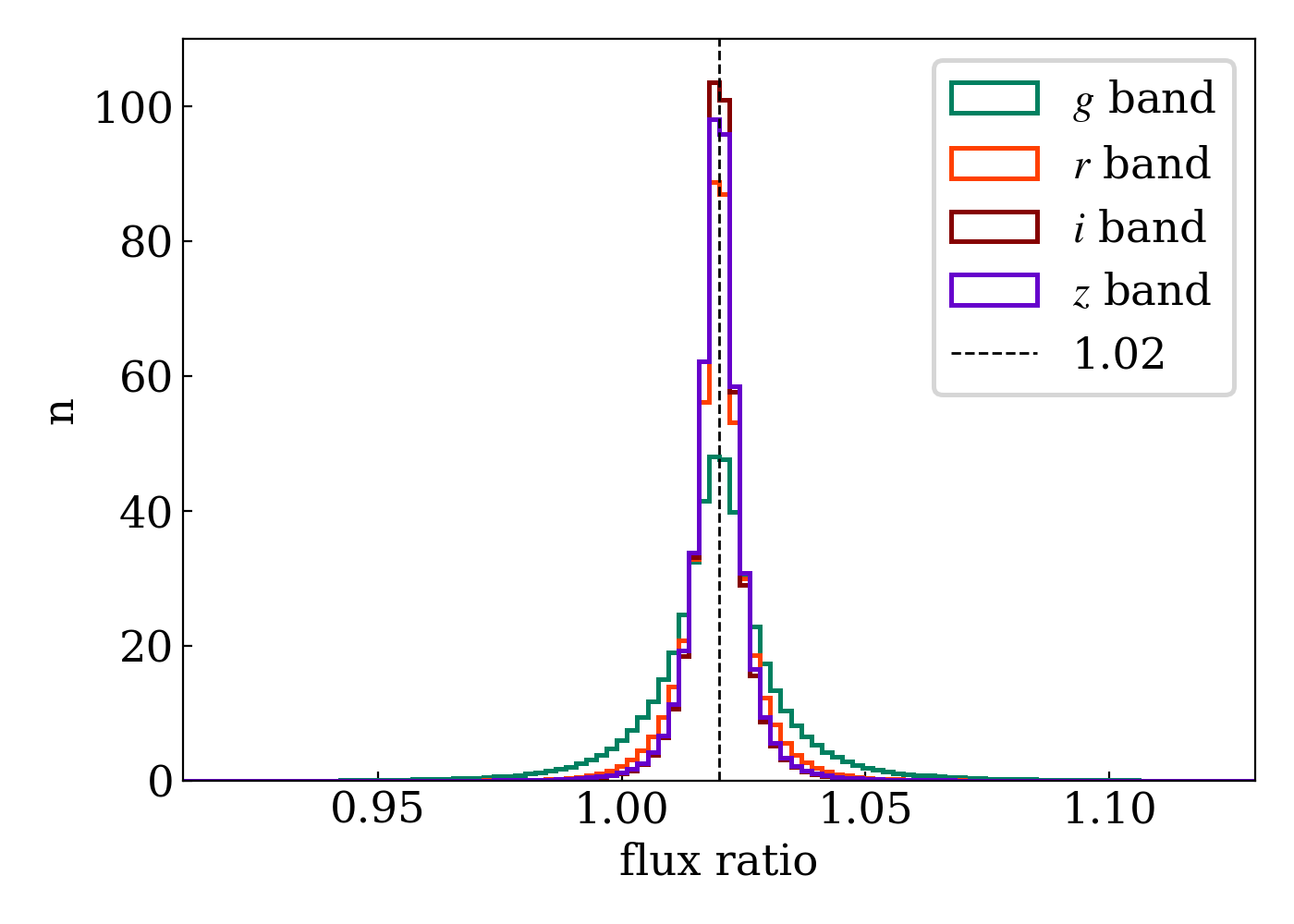}
    \caption{Distributions of measured flux ratio of the same \maglimplusplus objects in the magnified and regular \Balrog runs in the \textit{griz} bands. The distributions are centered at 1.02 with minimal tails, in agreement with the expected input 2\% magnification.} 
    \label{fig:hist_flux_ratio}
\end{figure}

Fig.~\ref{fig:hist_flux_ratio} presents the distributions of measured flux ratio of the same \maglimplusplus objects in the magnified and fiducial \Balrog runs in the \textit{griz} bands. The peak at 1.02 is consistent with the applied 2\% magnification. 


In the Y6 \Balrog simulations, source injections are divided into five categories to enhance statistics for key DES Y6 cosmology analyses. These include an unweighted ``Y3-scheme'' sample, as well as targeted selections for large-scale structure (LSS) and weak lensing (WL) samples, and additional subsets requiring high-quality redshift estimates. This refined weighting scheme significantly increases the number of detected sources in Y6 \Balrog, improving the statistical power of analyses while maintaining a representative distribution of objects across the full magnitude range relevant to Y6 cosmology studies. See Appendix~\ref{app:balrog_injweights} and \citet*{y6-balrog} for details.

A key systematic in our estimation of the magnification coefficients arises from discrepancies between the \Balrog catalog and the real data. To mitigate this, we apply a reweighting scheme that assigns deep-field galaxy weights, ensuring that the weighted property distribution of the \maglimplusplus sample in \Balrog matches that of the data. Appendix~\ref{app:balrog_reweighting} details this method and demonstrates the close agreement between the weighted \Balrog and real data \maglimplusplus in measured properties such as magnitudes, colors, photometric redshifts, and sizes.

To estimate magnification bias with \Balrog, we use the simulations that incorporate all injection weighting schemes and apply reweighting to match the distribution of data properties.


\section{Magnification coefficient estimation}
\label{sec:mag_coeff}

As discussed in Sec.~\ref{sec:theory}, the response of galaxy number density to convergence $\kappa$ is characterized by $C = \Csample + C_{\rm{area}}$, where $C_{\rm{area}} = -2$ accounts for the area change. The term $\Csample$ reflects variations in galaxy selection resulting from changes in flux and size.  

To estimate $\Csample$, we rely on simulations. Specifically, we examine how the number of selected galaxies changes when a small convergence perturbation $\delta\kappa$ is applied to the input galaxy properties (flux and size), while keeping their spatial distribution unchanged to exclude the area effect from the measurement. The fractional variation in the number of selected galaxies gives:  
\begin{equation}
    \Csample = \frac{N(\delta\kappa) - N(0)}{N(0) \ \delta\kappa},
\label{eq:csample}
\end{equation}
where $N(0)$ and $N(\delta\kappa)$ represent the total number of galaxies selected from the simulations without and with the applied convergence perturbation, respectively.

These estimates are affected by shot noise due to the limited number density of objects in the simulation. We quantify this uncertainty as:
\begin{equation}
    \resizebox{\columnwidth}{!}{$
    \frac{\sigma_{\Csample}}{\Csample} 
    = \sqrt{ \frac{N(0 \text{ only}) + N(\delta\kappa \text{ only})}{[N(\delta\kappa)-N(0)]^2} + \frac{1}{N(0)} + \frac{2N(0 \text{ only})}{N(0)[N(\delta\kappa) - N(0)]}},
    $}
    \label{eq:stat_error}
\end{equation}
where $N(0 \text{ only})$ represents the number of objects selected from the $\kappa =
0$ simulation but not from the $\kappa = \delta\kappa$ simulation, and $N(\delta\kappa \text{ only})$ represents those selected from the $\kappa = \delta\kappa$ simulation but not from the $\kappa = 0$ simulation. This is the statistical contribution to the error bars shown in Fig.~\ref{fig:mag_coeff}. The full derivation of this uncertainty is available in Appendix~\ref{app:stat_uncertainty}.

These equations serve as the basis for estimating $\Csample$ using different datasets, as detailed in the following sections.

\subsection{Estimation with \Balrog}
\label{sec:mag_coeff_balrog}

We compute the magnification coefficients $\Csample$ using the fiducial and magnification Balrog runs described in Sec.~\ref{sec:data}. The latter applies a constant magnification $\delta\mu = 1.02$ ($\delta\kappa \sim 0.01$) to the same galaxy models and positions as the fiducial run. The choice of $\delta\kappa \sim 0.01$ is deliberate---it is large enough to ensure a precise estimate of $\Csample$, as a sufficient number of galaxies are shifted above the detection threshold, yet small enough that higher-order $\kappa^2$ contributions to the change in number density remain negligible (on the order of $10^{-4}$). We apply the \maglimplusplus lens sample selection to the galaxy catalogs from both the $\kappa = 0$ and $\kappa = \delta\kappa$ runs. We then use Eq.~\eqref{eq:csample} to estimate $\Csample$, capturing the effects of magnification on the specific selection function, including cuts on color, magnitude, and size-dependent selection, such as star-galaxy separation.

\subsection{Estimation via flux perturbation}
\label{sec:mag_coeff_flux-only}

In this method, we create a magnified sample by adding a constant offset $\Delta m$ to the data magnitudes used for sample selection,
\begin{equation}
    \Delta m = -2.5 \log_{10}(1+2 \delta \kappa),
\end{equation}
where $\delta \kappa = 0.01$. We then re-select the \maglimplusplus sample based on these perturbed magnitudes, and we compute $\Csample$ using Eq.~\eqref{eq:csample}.

Notably, we do not re-run the DNF algorithm to compute the photometric redshifts, meaning they remain derived from the original, unperturbed magnitudes. However, to account for the impact of the flux perturbation, we introduce an additional redshift offset $\Delta z$, which is computed as follows:
\begin{equation}
    \Delta z = \sum_j c_j \ \Delta m_j,
\label{eq:deltaz}
\end{equation}
where $c_j$ are the DNF fit parameters, with $j$ indexing over the \textit{griz} bands. For each galaxy, DNF performs a linear fit between spectroscopic redshift and multi-band magnitudes of nearby spectroscopic galaxies in color space, estimating a parameter vector $\boldsymbol{c}$ that best describes this relationship. This vector is then used to compute a photometric redshift estimate $z$ from the observed magnitudes. Instead of retraining DNF, one can reuse these coefficients to predict a new $z$ for a different set of magnitudes, assuming minimal magnitude changes. This is an addition to the Y3 flux-only estimation method, and its impact is shown in Appendix \ref{deltaz}, where we compare to the results we would have obtained if magnification did not affect the photometric redshifts. Note that $\Delta m_j = \Delta m$ $\forall j$, since the same magnitude perturbation is applied across all bands.

This approach provides an estimate of how magnification influences fluxes, but does not account for the broader survey transfer function, excluding effects such as selection based on galaxy size, photometric noise, and observational systematics.

Additionally, we apply this flux-only approach to the fiducial \Balrog catalog. By comparing these results to those obtained from real data, we assess how well the \Balrog sample reflects actual observations.

\subsection{Magnification coefficient estimates}
\label{sec:mag_coeff_values}

\begin{figure}
    \centering
    \includegraphics[width=0.48\textwidth]{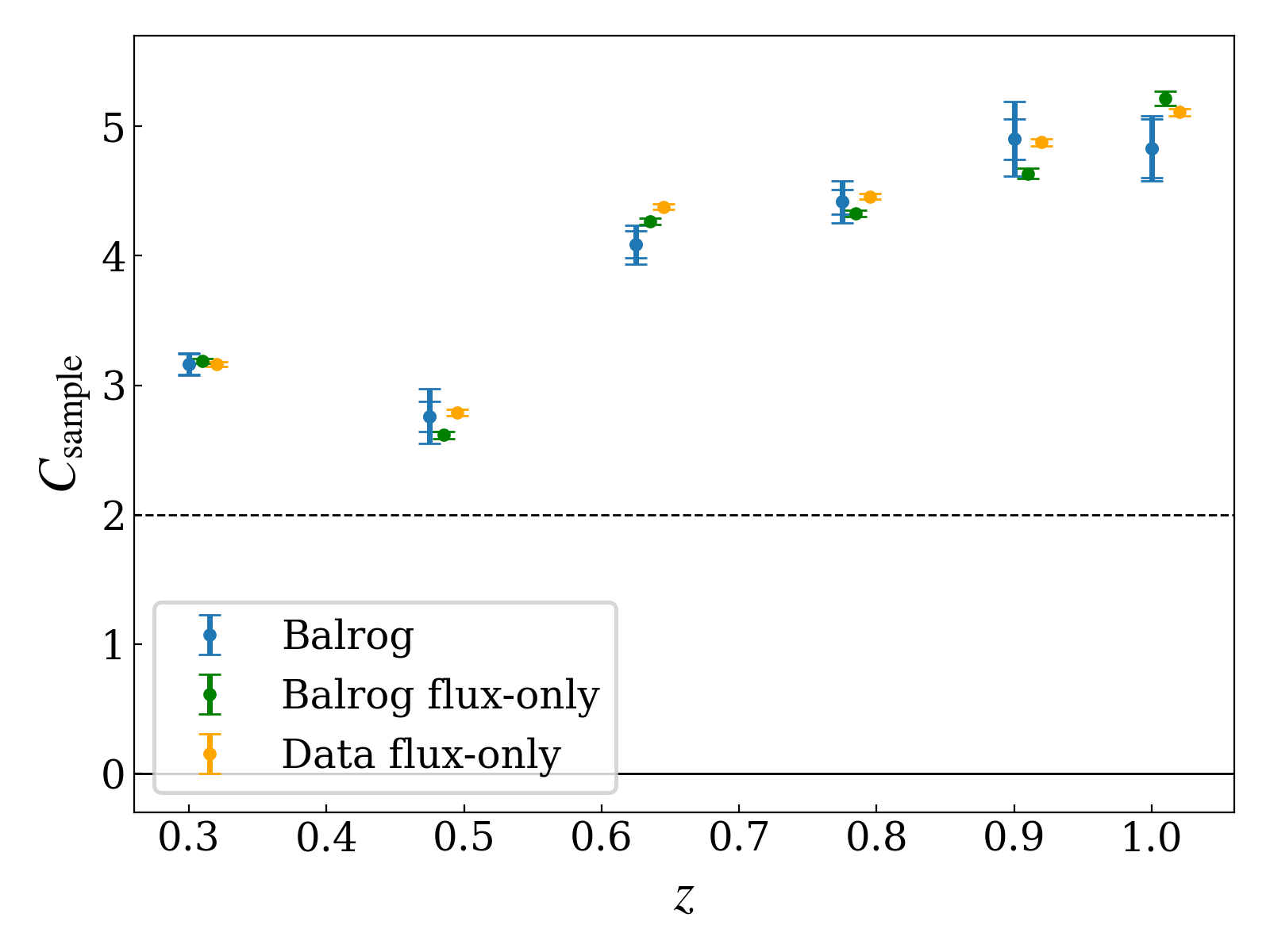}
    \caption{Magnification coefficient estimates for the six redshift bins of the \maglimplusplus lens sample, computed using the three methods described in Sec.~\ref{sec:mag_coeff}. The blue points represent our fiducial estimates from \Balrog simulations (Sec.~\ref{sec:mag_coeff_balrog}), while the green and yellow points correspond to approximate flux-only estimates derived from \Balrog and real data, respectively (Sec.~\ref{sec:mag_coeff_flux-only}). If the \Balrog sample were perfectly representative of real data, the flux-only estimates would be identical; therefore, their difference serves as a measure of systematic error in the blue \Balrog estimates. We present both statistical errors and the total uncertainty---obtained by adding statistical and systematic errors in quadrature---for the \Balrog estimates. The solid line represents zero magnification bias from sample selection alone, while the dashed line accounts for zero magnification bias when also considering changes in the area element. The blue \Balrog measurements will be used as the default choice in the DES Y6 3$\times$2pt analysis.} 
    \label{fig:mag_coeff}
\end{figure}

Table \ref{tab:mag_coeff} presents the magnification coefficient $\Csample$ estimates for each \maglimplusplus redshift bin, obtained using the three different methods. Fig.~\ref{fig:mag_coeff} provides a visual comparison of the estimates across methods.

\begin{table*}
\setlength{\tabcolsep}{4pt}
\begin{tabular}{ccccc}
\hline
\hline
\rule{0pt}{1.5em}
Bin & Redshift range & $\Csample^\text{\Balrog}$ & $\Csample^\text{\Balrog flux-only}$ &  $\Csample^\text{Data flux-only}$ \rule[-0.8em]{0pt}{0pt} \\ 
\hline
 1 & (0.20, 0.40] & $3.163 \pm 0.083$ & $3.190 \pm 0.019$ & $3.164 \pm 0.018$ \\ 
 2 & (0.40, 0.55] & $2.760 \pm 0.211$ & $2.615 \pm 0.029$ & $2.790 \pm 0.025$ \\ 
 3 & (0.55, 0.70] & $4.086 \pm 0.152$ & $4.267 \pm 0.025$ & $4.378 \pm 0.022$ \\ 
 4 & (0.70, 0.85] & $4.417 \pm 0.162$ & $4.327 \pm 0.025$ & $4.458 \pm 0.021$ \\ 
 5 & (0.85, 0.95] & $4.902 \pm 0.287$ & $4.636 \pm 0.039$ & $4.878 \pm 0.028$ \\ 
 6 & (0.95, 1.05] & $4.832 \pm 0.250$ & $5.214 \pm 0.054$ & $5.111 \pm 0.028$ \\
\hline 
\end{tabular}
\caption{Magnification coefficient estimates, $\Csample$, along with their associated errors for the six redshift bins of the \maglimplusplus lens sample, computed using the three different approaches outlined in Sec.~\ref{sec:mag_coeff}.}
\label{tab:mag_coeff}
\end{table*}

The $\Csample^{\rm{\Balrog}}$ estimates incorporate a systematic error to account for differences in color, magnitude, and size selection between \Balrog and the real data. While \Balrog properties closely match the real data, especially after re-weighting (see Appendix~\ref{app:balrog_reweighting}), minor deviations remain. If \Balrog perfectly represented the real data, flux-only estimates from both would be identical; thus, the systematic uncertainty in the \Balrog estimates is quantified by their difference, and added in quadrature to the statistical error. 

The $\Csample$ flux-only \Balrog and data estimates are consistent with each other across all redshift bins, with the largest deviations observed in bins 2 and 5, suggesting that the \Balrog \maglimplusplus sample may be less representative of the real data at these redshifts. Overall, these estimates also align well with our fiducial \Balrog measurements. This contrasts with the Y3 results (\citet*{y3-2x2ptmagnification}), where significant discrepancies between flux-only and \Balrog estimates were particularly evident in bin 2. As shown in Appendix \ref{deltaz}, these differences in Y3 were driven by the impact of magnification on photometric redshifts, which was not accounted for in the flux-only method at the time.

Compared to the Y3 \maglim estimates (\citet*{y3-2x2ptmagnification}), the statistical uncertainties are smaller in Y6 due to the larger size of the samples used for measurement (see Sec.~\ref{app:balrog_injweights}). Additionally, the flux-only estimates show better agreement in Y6, thanks to \Balrog reweighting, which reduces systematic uncertainties relative to Y3. Differences in $\Csample$ values may arise from variations in sample selection between \maglim and the Y6 \maglimplusplus.

We also tested different values of $\delta\kappa$ around 0.01 using the flux-only method and found no significant shifts in the magnification coefficients.

We consider \Balrog, which incorporates a broad range of observational effects and accounts for selection on quantities beyond just flux, to be the most accurate method for estimating magnification coefficients. Thus, we use the $\Csample^{\rm{\Balrog}}$ estimates as fiducial moving forward. However, for this specific sample, using flux-only estimates from the data---without the need to run an image simulation---can be considered safe. In Appendix~\ref{app:size_selection}, we demonstrate that this approximate approach is not always reliable, particularly when selection criteria involve size cuts.


\section{Expected impact on cosmology}
\label{sec:exp_mag_cosmology}

We evaluate the impact of magnification on the DES Y6 galaxy clustering and galaxy-galaxy lensing (2$\times$2pt) analysis, using a noiseless data vector generated from our theoretical model \citet*{y6-methods}. For these tests, we adopt the fiducial $\Csample = C_{\rm{sample}}^{\rm{Balrog}}$ values obtained from the Balrog simulations. Additionally, we assess the impact of including the cross-correlation clustering signal between different lens redshift bins, which---although not included in the fiducial Y6 analysis---helps self-calibrate the magnification signal, and we also extend this investigation to the full 3$\times$2pt analysis incorporating cosmic shear. We perform cosmological parameter inference following the DES Y6 analysis choices \citet*{y6-methods}, and test this within a $\Lambda$CDM model with linear galaxy bias. Note that all of this analysis is carried out before unblinding the DES Y6 cosmology results. 

\begin{figure*}
    \centering
    \includegraphics[width=0.9\textwidth]{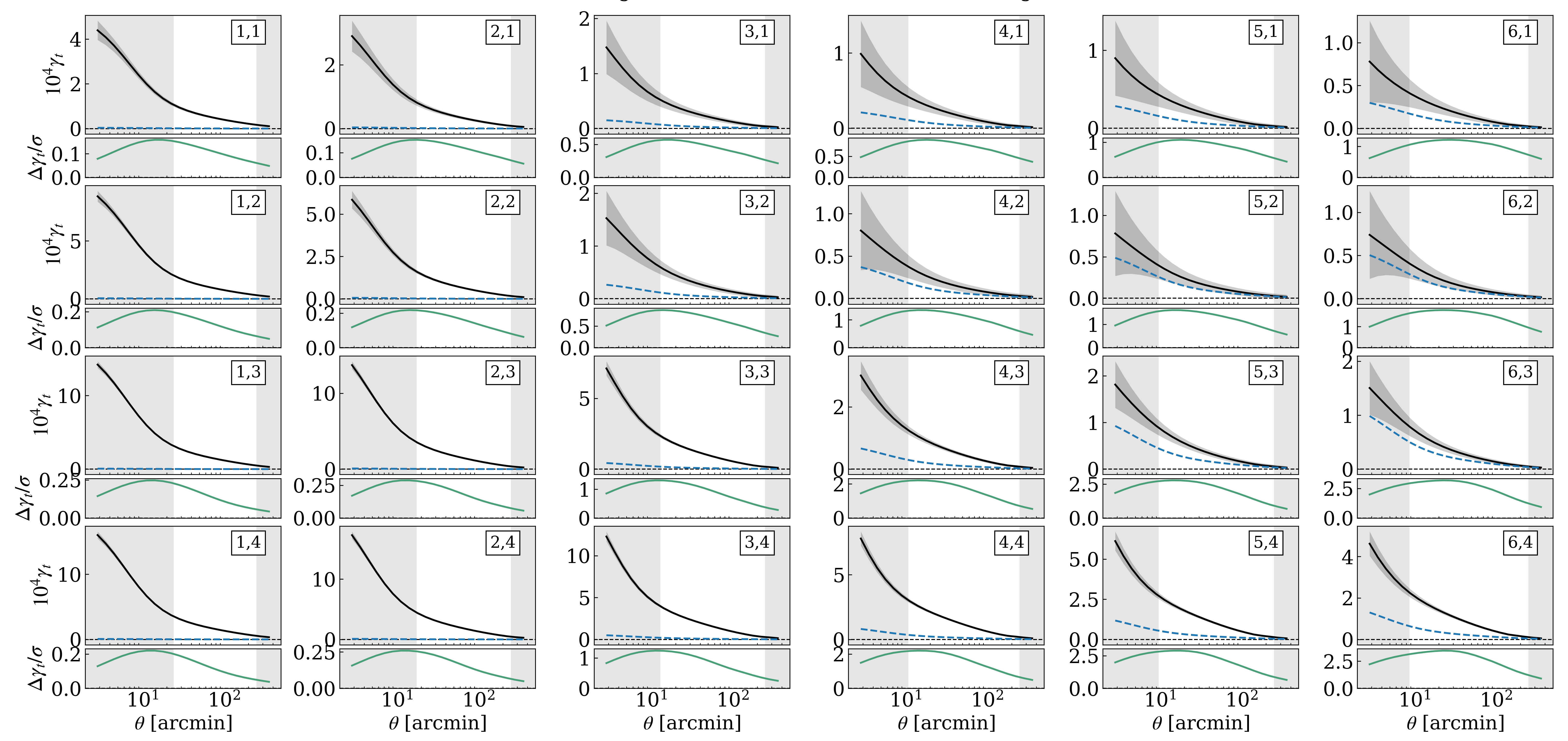}
    \includegraphics[width=0.9\textwidth]{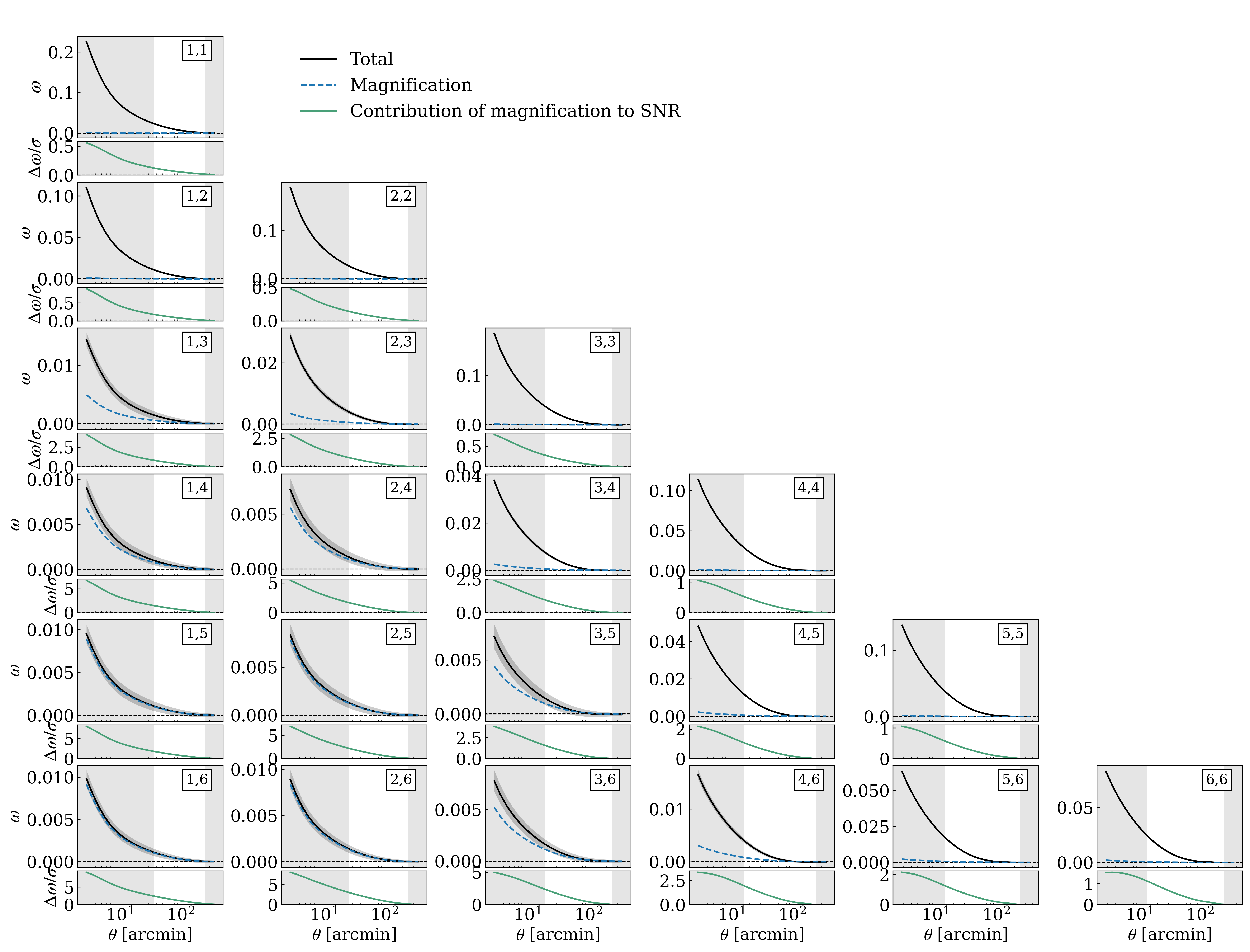}
    \caption{The impact of magnification on model predictions for galaxy-galaxy lensing ($\gamma_t(\theta)$; top panel) and galaxy clustering ($\omega(\theta)$; bottom panel) across all redshift bin combinations. In each panel, the black line represents the model prediction with lens magnification included, with the surrounding shaded region indicating the corresponding statistical uncertainty. The blue dotted line isolates the lens magnification contribution, derived using fiducial magnification coefficient values estimated from \Balrog. The lower sub-panels show the fractional change in the signal relative to its statistical uncertainty, highlighting the contribution of magnification to the overall signal-to-noise ratio. Gray shaded regions indicate the Y6 scale cuts. For galaxy-galaxy lensing, magnification primarily affects combinations involving high-redshift lenses, while most of the signal-to-noise arises from the three lowest-redshift lens bins where magnification biases are minimal. Its impact on clustering auto-correlations is small, and becomes relevant only for cross-correlations between widely separated redshift bins, which are excluded in the fiducial DES Y6 3$\times$2pt analysis.} 
    \label{fig:predictions}
\end{figure*}

In Fig.~\ref{fig:predictions}, we illustrate the effect of lens magnification on the various components of the 2$\times$2pt data vector. For galaxy-galaxy lensing, magnification has the most significant impact on source-lens pairs involving high-redshift lenses, which are more strongly magnified by the greater amount of foreground structure along the line of sight. The main lensing signal is also weaker when lens and source redshift distributions overlap, making the relative contribution from magnification naturally larger in those cases. Magnification also plays a crucial role in clustering cross-correlations, particularly for widely separated redshift bins, where it becomes the dominant contribution to the signal. However, its impact on clustering auto-correlations is relatively small. In the fiducial DES Y6 3$\times$2pt analysis, cross-correlation clustering between different lens redshift bins is not included, and most of the signal-to-noise ratio in galaxy-galaxy lensing comes from the three lowest-redshift lens bins, where magnification biases are small. As a result, lens magnification is expected to have a limited impact on cosmological parameter constraints in the fiducial DES Y6 cosmology analysis. However, magnification bias can still be significant even without cross-redshift bin clustering, as the galaxy bias introduced by magnification-induced cross-correlations acts opposite to galaxy-galaxy lensing, potentially shifting inferred cosmological parameters (see \citet{thiele_disentangling_2020}).


We explore five prior choices on the magnification coefficients:
\begin{itemize}
    \item  Gaussian prior: Gaussian priors centered on the true $\Csample$ values, with widths set to the total uncertainties in the \Balrog estimates.
    \item Fixed mag: $\Csample$ values fixed to their true values used to generate the datavector.
    \item No mag: $\Csample$ values fixed to 2, effectively removing the magnification contribution from the data vector.
    \item  Gaussian prior shifted: Gaussian priors centered on the true $\Csample+3\sigma_{\Csample,\rm{tot}}$, where $\sigma_{\Csample,\rm{tot}}$ is the total uncertainty in the \Balrog estimates.
    \item Flat prior: Wide uniform priors on $\Csample$ in the range $[-4, 12]$.
\end{itemize}

\begin{figure*}
    \centering
    \begin{minipage}{0.49\textwidth}
        \centering
        \includegraphics[width=\linewidth]{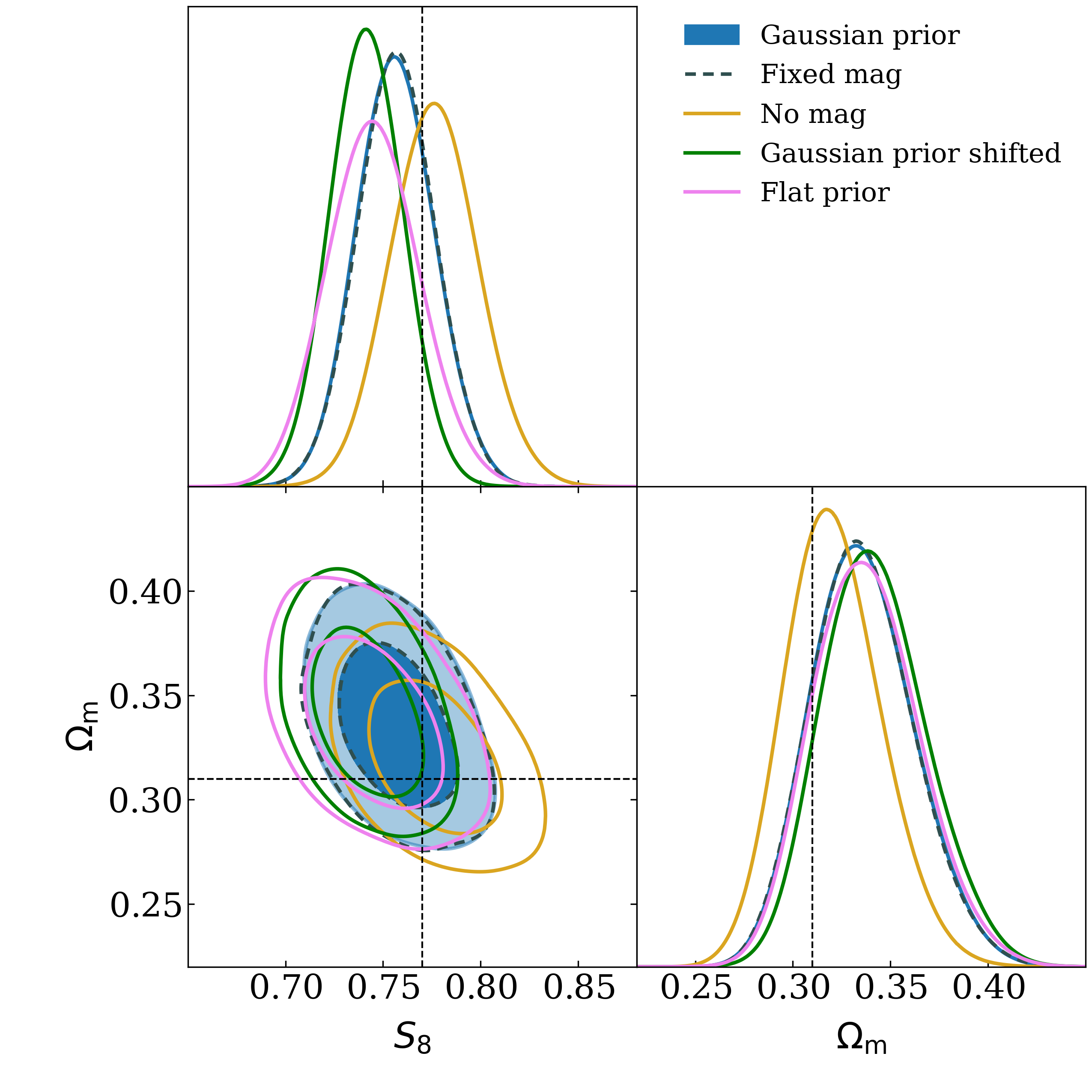}
        {(a) \textbf{Simulated 2$\times$2pt}.}
        \label{fig:contours_fid}
    \end{minipage}\hfill
    \begin{minipage}{0.49\textwidth}
        \centering
        \includegraphics[width=\linewidth]{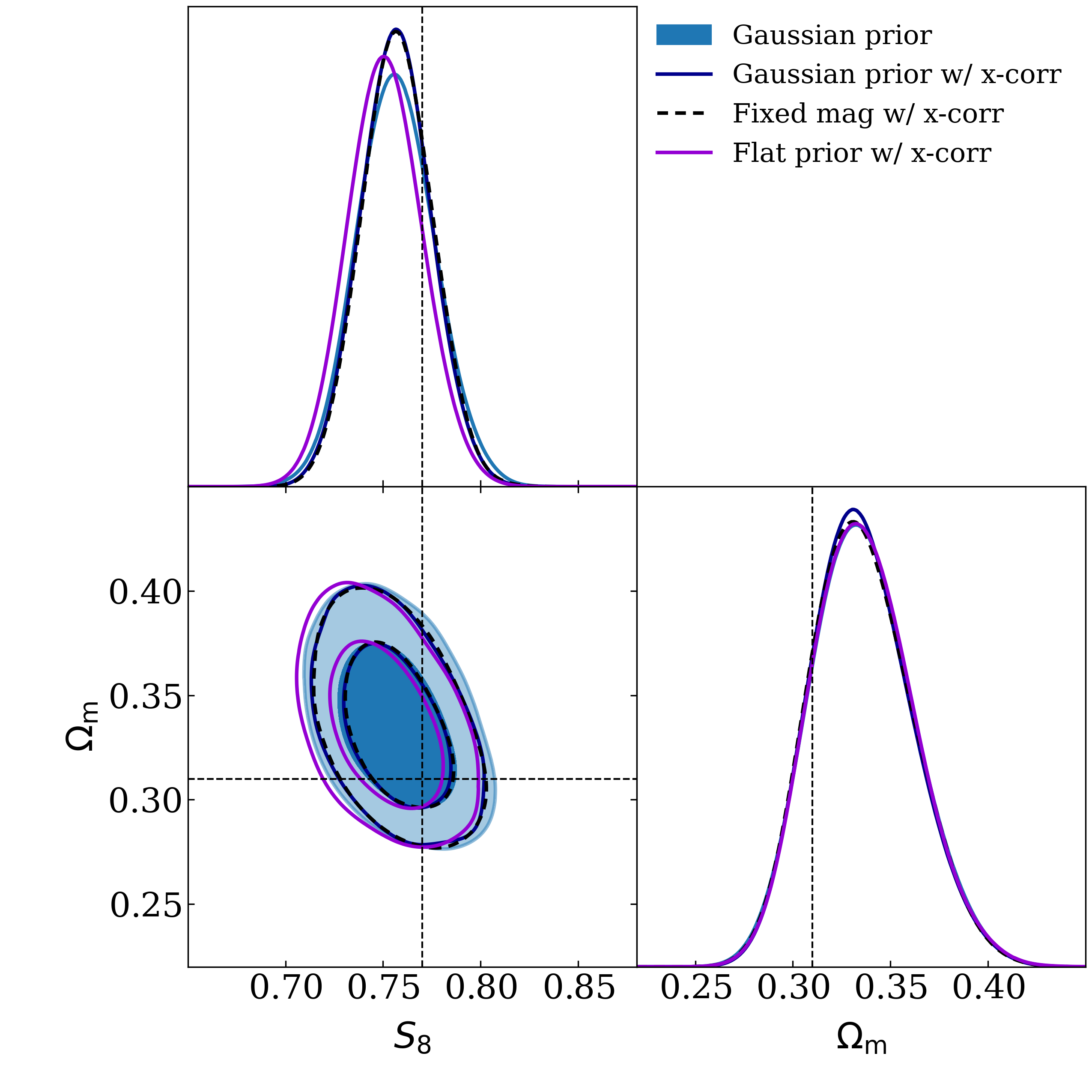}
        {(b) \textbf{Simulated 2$\times$2pt} w/ x-corr.}
        \label{fig:contours_xcorr}
    \end{minipage}
    \caption{Recovered 2D constraints on $S_8$ and $\Omega_m$ derived from a simulated, noiseless data vector, following the DES Y6 modeling strategy. The results explore five different priors on the magnification bias, $\Csample$, as outlined in Sec.~\ref{sec:exp_mag_cosmology}: (i) applying Gaussian priors centered on the true values with widths given by the \Balrog-derived uncertainties (``Gaussian prior''), the fiducial choice in the Y6 cosmological analyses; (ii) fixing $\Csample$ to its true values (``Fixed mag''); (iii) fixing $\Csample$ to 2, thereby removing magnification from the model (``No mag''); (iv) using Gaussian priors shifted by $3\sigma_{\Csample,\rm{tot}}$ (``Gaussian prior shifted''); and (v) wide uniform priors (``Flat prior''). The figure shows the outcomes for both a 2$\times$2pt analysis (Fig.~\ref{fig:contours}a), and a 2$\times$2pt analysis including the additional cross-clustering signal (``w/ x-corr'', Fig.~\ref{fig:contours}b) that further constrains the magnification effect. The dashed lines mark the true input values. Beyond prior volume effects, the results show that marginalizing over magnification with Gaussian priors recovers unbiased constraints on $S_8$ and $\Omega_m$, while omitting magnification or shifting the Gaussian priors introduces biases; adopting flat priors slightly increases uncertainties without significantly shifting the cosmological parameters, and including cross-clustering slightly improves cosmological constraints.}
    \label{fig:contours}
\end{figure*}

\begin{figure*}
    \centering
    \begin{minipage}{0.51\textwidth}
        \centering
        \includegraphics[width=\linewidth]{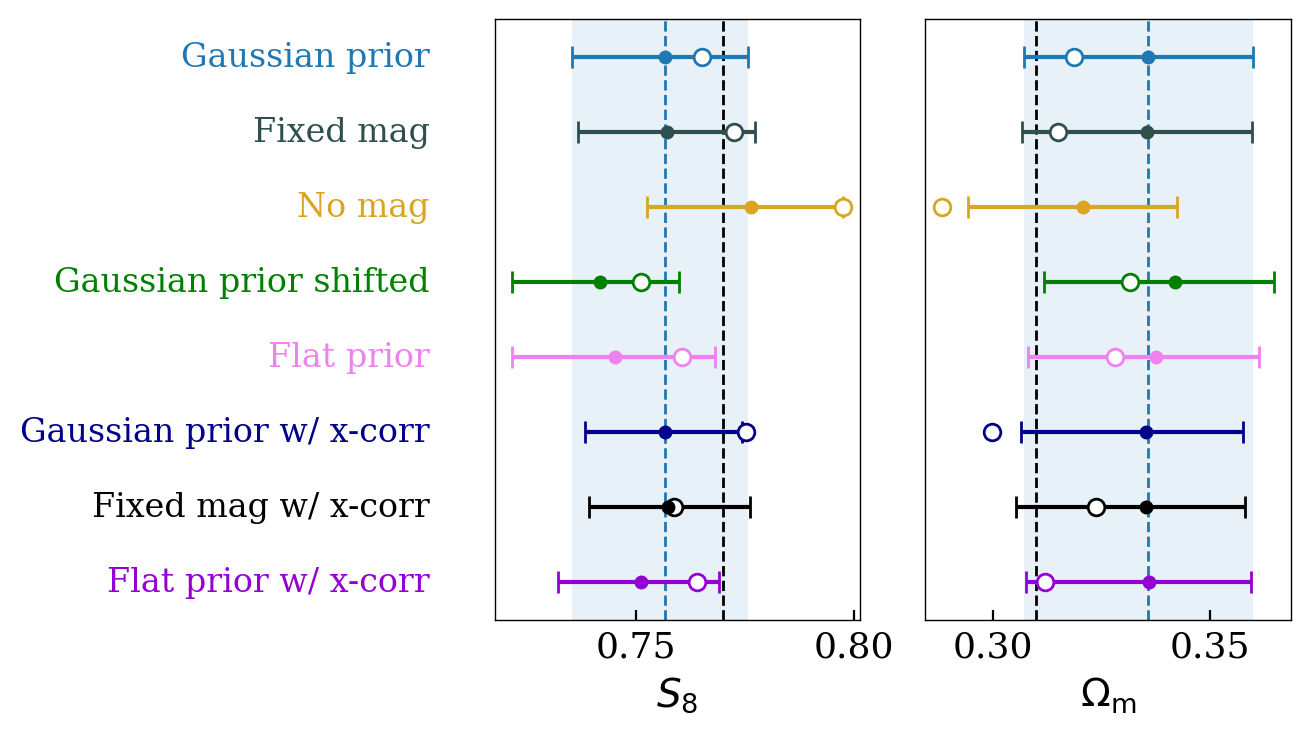}
        {(a) \textbf{Simulated 2$\times$2pt}.}
        \label{fig:contours_2x2pt}
    \end{minipage}\hfill
    \begin{minipage}{0.47\textwidth}
        \centering
        \includegraphics[width=\linewidth]{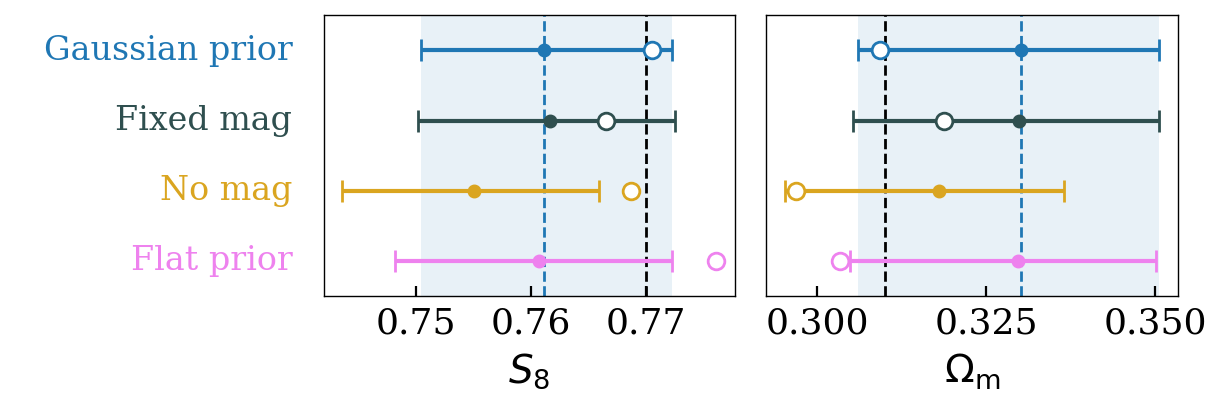}
        {(b) \textbf{Simulated 3$\times$2pt}.}
        \label{fig:contours_3x2pt}
    \end{minipage}
    \caption{Recovered 1D constraints on $S_8$ and $\Omega_m$ from a simulated, noiseless data vector, following the DES Y6 modeling strategy. The results consider the five different priors on the magnification bias, $\Csample$, outlined in Sec.~\ref{sec:exp_mag_cosmology}, and are presented for three analysis configurations: the fiducial 2$\times$2pt analysis, the 2$\times$2pt analysis including the additional cross-clustering signal (“w/ x-corr”), which further constrains magnification (Fig.~\ref{fig:contours_panel}a), and the full 3$\times$2pt analysis (Fig.~\ref{fig:contours_panel}b). The solid horizontal lines show the 68\% marginalized posterior intervals, with the filled circles marking the corresponding posterior means, and the open circles indicating the MAP values. The dashed black vertical lines mark the true input values. The dashed blue vertical lines indicate the posterior means of the fiducial analysis with Gaussian priors, and the shaded regions denote the corresponding $1\sigma$ uncertainty. Note that, because of noise, the MAP estimates may deviate from the true input values (see Appendix~\ref{app:map}).} 
    \label{fig:contours_panel}
\end{figure*}

\begin{figure*}
    \centering
    \includegraphics[width=0.93\textwidth]{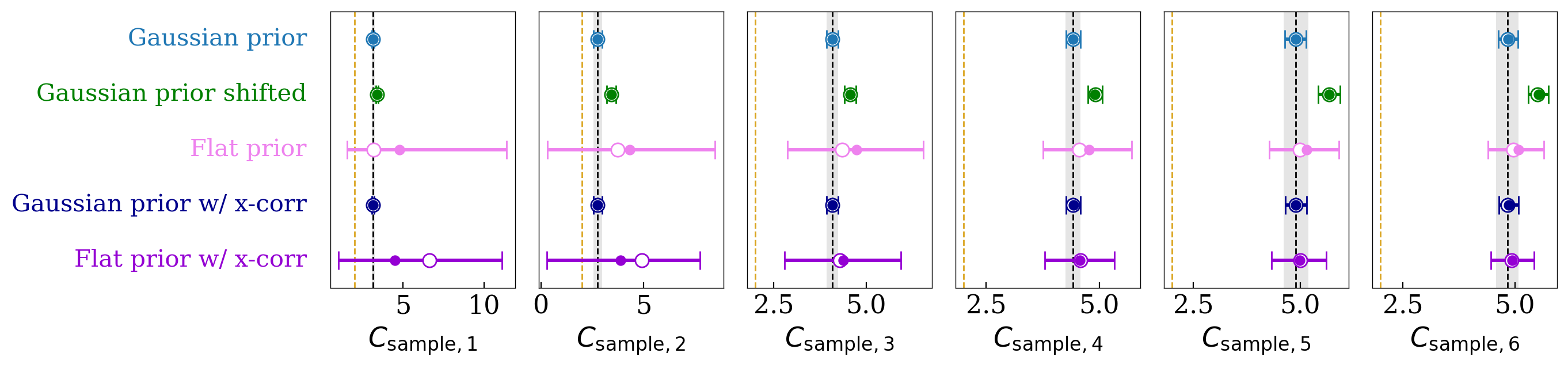} \\
    {(a) \textbf{Simulated 2$\times$2pt}.} \\
    \vspace{0.6cm}
    \includegraphics[width=0.92\textwidth]
    {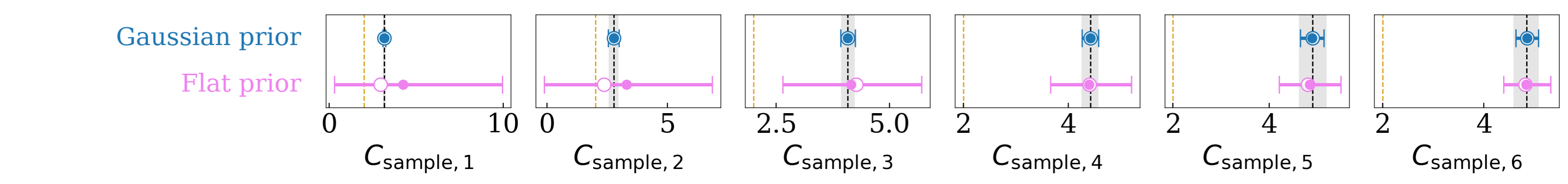} \\
    {(b) \textbf{Simulated 3$\times$2pt}.} \\
    \vspace{0.6cm}
    \caption{Constraints on the magnification coefficients $\Csample$ for each lens redshift bin. These constraints are derived from a simulated, noiseless 2$\times$2pt (Fig..~\ref{fig:contours_panel_mag_coeff}a) and 3$\times$2pt (Fig..~\ref{fig:contours_panel_mag_coeff}b) data vector, for the various analysis variations presented in Sec.~\ref{sec:exp_mag_cosmology} where $\Csample$ is allowed to vary. The solid horizontal bars represent the 68\% marginalized posterior intervals, with filled circles marking the posterior means and open circles showing the MAP estimates. The dashed black vertical lines denote the true input \Balrog values, and the gray shaded region corresponds to the total uncertainties in the \Balrog estimates (i.e. the width of the Gaussian prior used in the fiducial analysis). The dashed yellow vertical lines mark $\Csample$=2, corresponding to the case without magnification bias. The input magnification coefficients are accurately recovered even with flat, uninformative priors, while the inclusion of clustering cross-correlations provides a modest improvement in constraining power at the highest redshifts.}
    \label{fig:contours_panel_mag_coeff}
\end{figure*}

Fig.~\ref{fig:contours} shows the 2D posterior contours, and Fig.~\ref{fig:contours_panel} the corresponding 1D marginalized constraints, on $S_8 = \sigma_8 \sqrt{\Omega_m/0.3}$ and $\Omega_m$. Results are presented for different prior choices on the magnification coefficients, considering both the fiducial 2$\times$2pt analysis and the extended case including cross-clustering, as well as the full 3$\times$2pt analysis. The resulting constraints on the magnification coefficients, $\Csample$, for flat and Gaussian priors are shown in Fig.~\ref{fig:contours_panel_mag_coeff}\footnote{The mean values and their 68\% posterior bounds are computed using the \textsc{CosmoSIS} post-processing pipeline. The maximum a posteriori (MAP) values are obtained by running local optimizations initialized from 20 high-posterior samples from the chains. For more details on the MAP estimates, as well as a discussion of the impact of noise on the recovered parameters, refer to \citet*{y6-methods} and Appendix~\ref{app:map}.}.


When fixing the magnification coefficients in the 2$\times$2pt analysis, the means of the $S_8$ and $\Omega_m$ posteriors are biased relative to the input values by $-0.64\sigma$ and $0.89\sigma$, respectively (and by $-0.77\sigma$ and $0.81\sigma$, respectively, in the 3$\times$2pt analysis). However, the maximum a posteriori (MAP) parameter values remain close to the true cosmological parameters, suggesting that these biases do not arise from systematic errors but rather from \textit{prior volume} or \textit{projection effects}\footnote{Note that we found smaller projection effects when using a data vector generated at the best-fit of the cosmological analysis of real data (\citet{y3-3x2ptkp}) than with the simulated data vector used in this paper, which was generated before unblinding.}. Note that some deviation of the MAP estimates from the true values is expected even for a noiseless data vector, as discussed in Appendix~\ref{app:map}.

Even in the idealized scenario of applying the baseline analysis to a synthetic, noiseless data vector generated from the same model, the marginalized posteriors may still appear biased due to prior volume effects. These effects occur when the shape of the prior and the available parameter space influence the posterior, along with parameter degeneracies, where correlations between parameters shift the constraints---particularly when certain parameters are prior-constrained. The impact of projection effects is closely linked to the constraining power of the data: when strong constraints are present, the influence of priors diminishes, reducing these effects. However, when parameters are weakly constrained, the choice of priors can have a significant impact on the posterior distributions. As a result, marginalized constraints should be interpreted with caution, especially when comparing results across different analysis choices or datasets.

Beyond prior volume effects, we observe an additional bias of $0.81\sigma$ in $S_8$ and $-0.56\sigma$ in $\Omega_m$ when magnification is not included in the analysis, compared to the case where the correct magnification coefficients are assumed (and of $-0.61\sigma$ and $-0.64\sigma$, respectively, when cosmic shear is included). Marginalizing over the magnification coefficients with Gaussian priors shifted by 3$\sigma_{\Csample,\rm{tot}}$, introduces biases of $-0.76\sigma$ in $S_8$ and $0.21\sigma$ in $\Omega_m$ relative to the scenario where the coefficients are fixed at their true values. This highlights the importance of properly incorporating magnification effects in the theoretical model. Adopting a flat prior on $\Csample$ slightly degrades the constraining power, increasing the $1\sigma$ uncertainty in $S_8$ by 15\%, and amplifies the prior volume bias in the marginalized $S_8$ by $-0.50\sigma$ (for the 3$\times$2pt case, the uncertainty increases by only 8\% and the bias is reduced to $-0.07\sigma$). When marginalizing over the magnification coefficients with Gaussian priors, the recovered constraints remain nearly identical to the case where the coefficients are fixed at their true values ($0.015\sigma$ and $-0.006\sigma$ in $S_8$ and $\Omega_m$, respectively, and $-0.05\sigma$ and $-0.01\sigma$ when including cosmic shear). 

Including cross-clustering correlations does not mitigate projection effects when informative priors are used. However, when adopting a flat prior on $\Csample$, it slightly reduces the additional bias in $S_8$ relative to the case where the correct magnification coefficients are assumed, bringing it to $-0.32\sigma$, without degrading the constraining power compared to fixing $\Csample$. This improvement arises because cross-terms are more sensitive to magnification, as reflected in the recovered $\Csample$ constraints shown in Fig.~\ref{fig:contours_panel_mag_coeff}. In particular, when comparing the flat prior cases, adding cross-correlations tightens the 68\% posterior bounds on $\Csample$, especially at high redshift.

Among the approaches considered, the fiducial DES Y6 cosmology analysis adopts Gaussian priors on the \Balrog magnification coefficients, in contrast to the fixed values used in Y3, as this provides a conservative treatment that incorporates their uncertainties. The flat prior case is the least sensitive to the magnification coefficient measurements, but introduces additional prior volume effects, while fixing the coefficients is the most computationally convenient for inference yet ignores their uncertainty.


Overall, these results demonstrate that magnification is a significant systematic effect in DES Y6, which must be carefully modeled to avoid biases in cosmological constraints. This conclusion aligns with recent magnification studies for other survey specifications, such as those in \citet{duncan_cosmological_2022}, \citet{mahony_forecasting_2022}, and \citet{euclid_collaboration_euclid_2024}.  

Additional validation was performed in \citet*{y3-2x2ptmagnification} for DES Y3 using Buzzard simulations. Applying the same modeling to the redMaGiC lens galaxy sample in these simulations demonstrated unbiased recovery of both the magnification coefficients and the cosmological parameters across different prior choices, providing a robust foundation for its application in DES Y6 analyses.

\section{Impact on DES Y6 cosmology}
\label{sec:mag_cosmology}

\begin{figure*}
    \centering
    \includegraphics[width=0.6\textwidth]{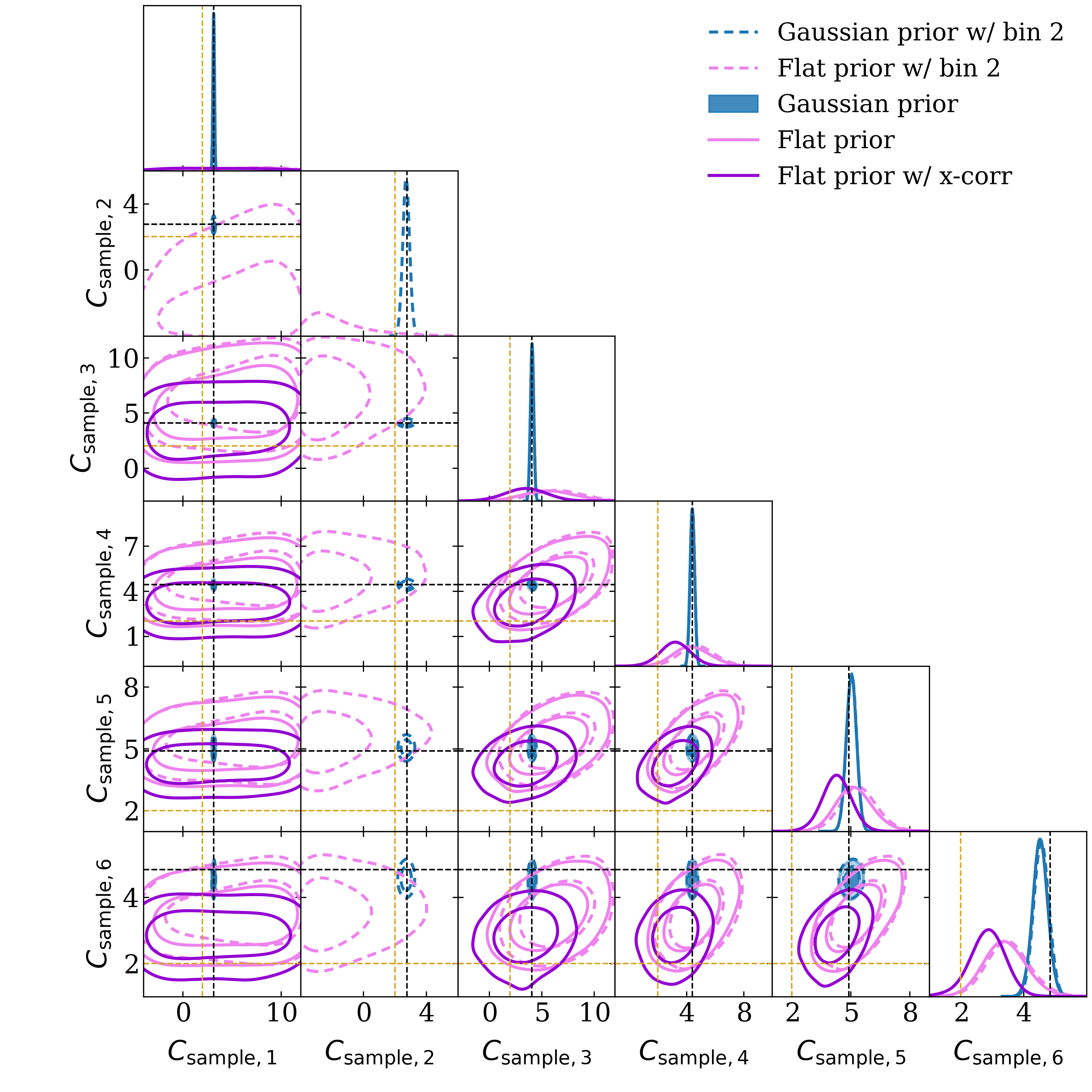}
    \\ \vspace{0.2cm}
    \includegraphics[width=0.99\textwidth]{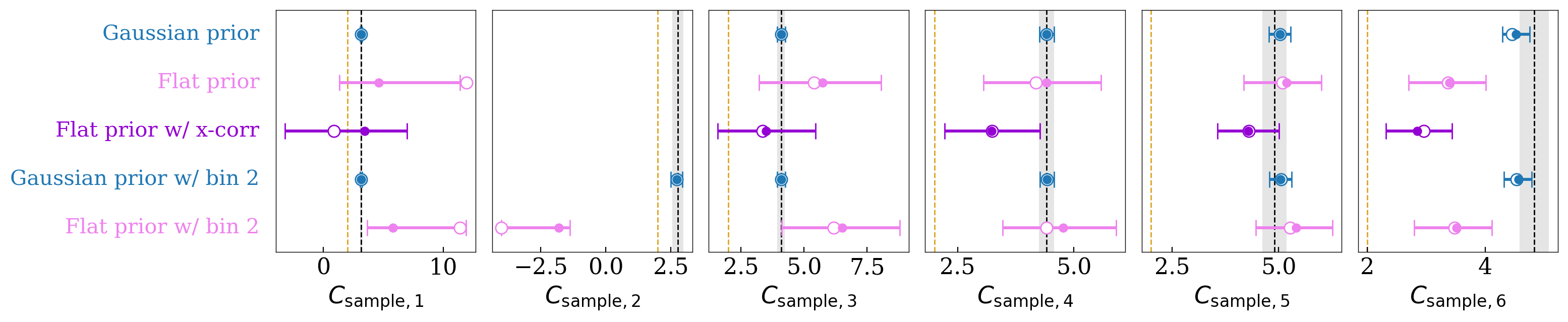}
    {(a) \textbf{Data 2$\times$2pt}.} \\
    \vspace{0.6cm}
    \includegraphics[width=0.99\textwidth]{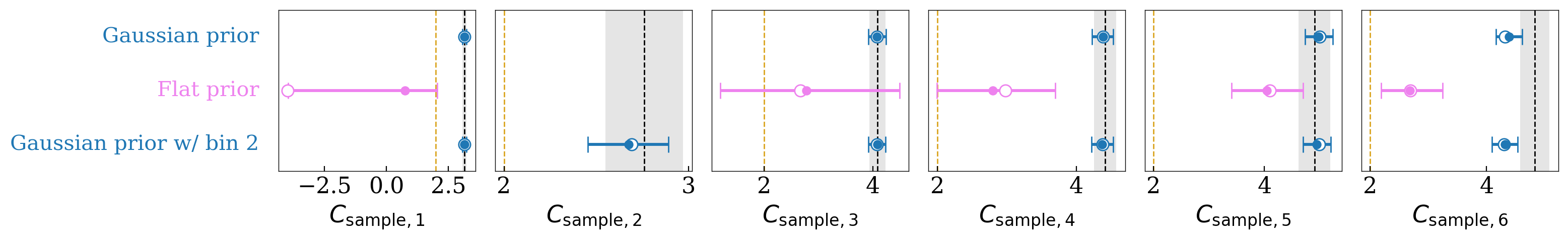}
    {(b) \textbf{Data 3$\times$2pt}.}
    \caption{Unblinded constraints on the magnification coefficients $\Csample$ for each lens redshift bin. These constraints are derived from Y6 data, for the various analysis variations presented in Sec.~\ref{sec:mag_cosmology}. The dashed yellow lines denote $\Csample$=2, corresponding to the absence of magnification bias. In the lower panels, the solid horizontal lines show the 68\% marginalized posterior intervals, with the filled circles marking the corresponding posterior means, and the open circles indicating the MAP values. Overall, for 2$\times$2pt data (see Fig.~\ref{fig:contours_mag_coeff_data}a), the constraints are consistent with the \Balrog estimates (dashed black lines, with shaded regions indicating the total uncertainties), except for the flat priors cases on bin 6, where the posteriors shift toward lower values, and on bin 2 when included in the analysis, where the posteriors shift to unphysical negative values. For 3$\times$2pt data (see Fig.~\ref{fig:contours_mag_coeff_data}b), allowing the magnification coefficients to vary freely results in posteriors that are generally lower than the \Balrog estimates.}
    \label{fig:contours_mag_coeff_data}
\end{figure*}

\begin{figure*}
    \centering
    \begin{minipage}{0.49\textwidth}
        \centering
        \includegraphics[width=\linewidth]{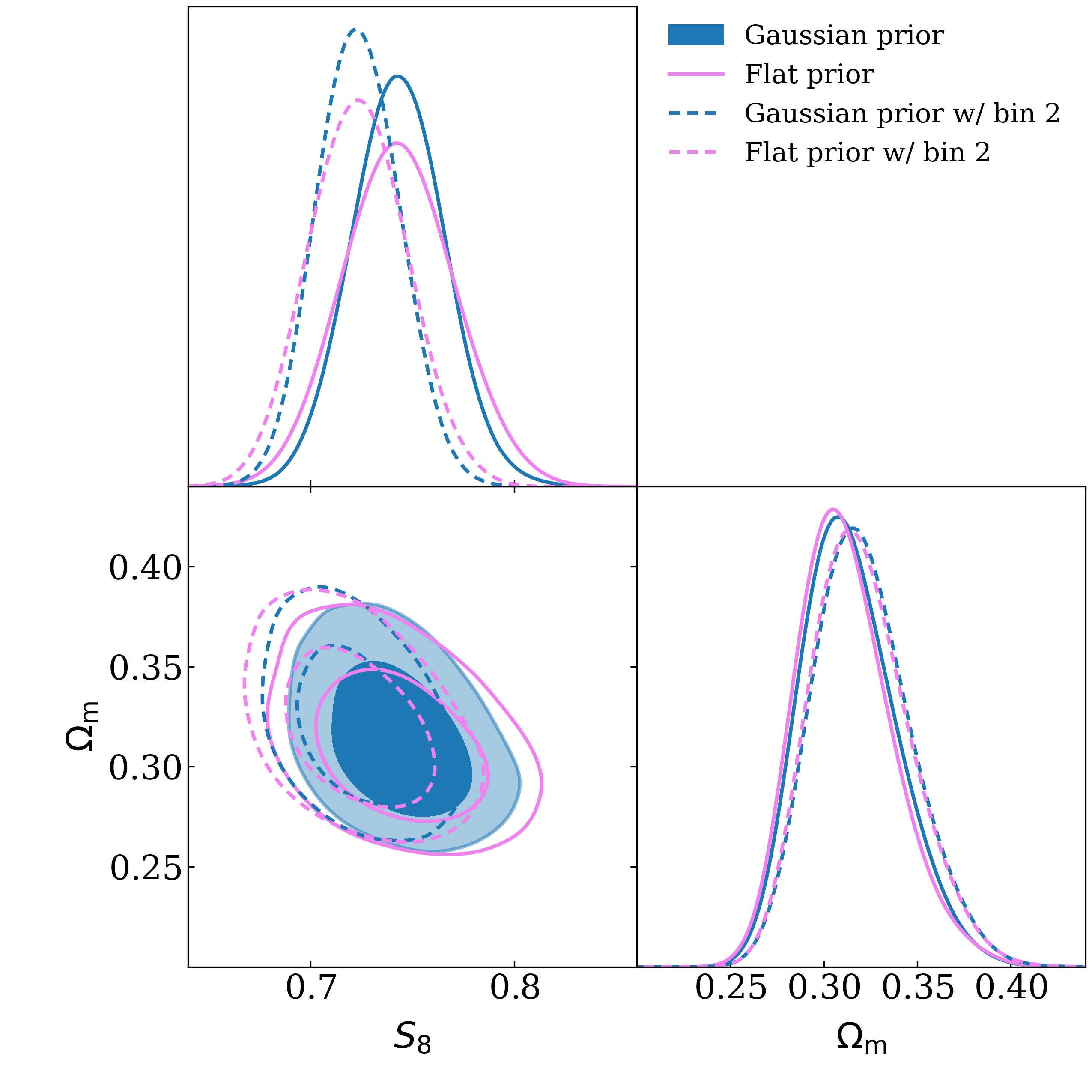}
        {(a) \textbf{Data 2$\times$2pt} including versus excluding lens bin 2.}
        \label{fig:contours_data_bin2}
    \end{minipage}\hfill
    \begin{minipage}{0.49\textwidth}
        \centering
        \includegraphics[width=\linewidth]{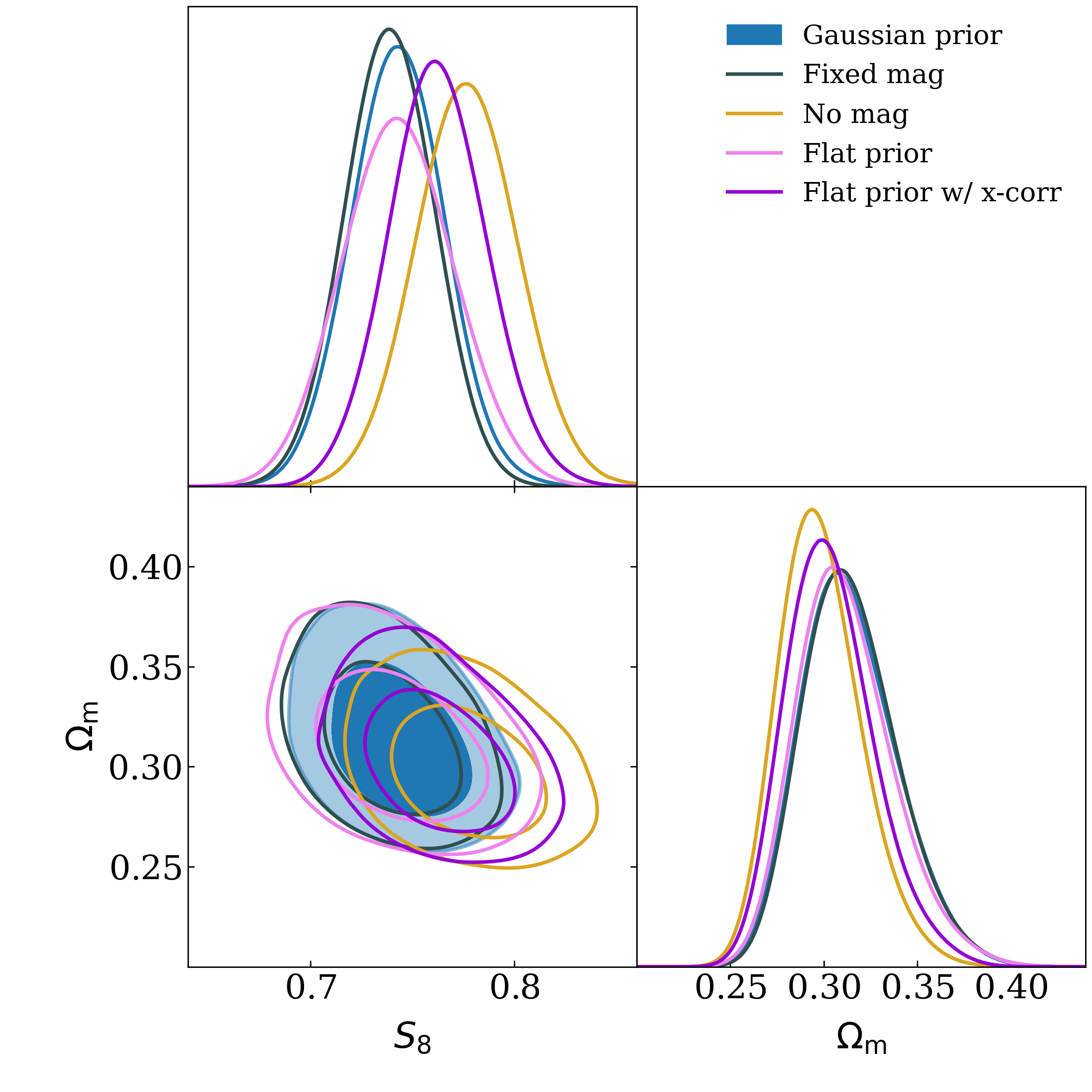}
        {(b) \textbf{Data 2$\times$2pt} different priors on $\Csample$.}
        \label{fig:contours_data_all}
    \end{minipage}
    \caption{2D constraints on $S_8$ and $\Omega_m$ from unblinded DES Y6 2$\times$2pt data, shown for different priors on the magnification bias, $\Csample$: (i) Gaussian priors centered on the \Balrog coefficients with widths given by the \Balrog-derived uncertainties (``Gaussian prior''), which is the fiducial choice in the Y6 cosmological analyses; (ii) fixed magnification coefficients set to $\Csample^{\rm{\Balrog}}$ (``Fixed mag''); (iii) $\Csample$ fixed to 2, thereby removing magnification from the model (``No mag''); and (iv) wide uniform priors (``Flat prior''), both excluding and including the additional cross-clustering correlations (``w/ x-corr''). Results from the alternative analysis with the inclusion of lens bin 2 are also shown (``w/ bin 2''). Note that, unless stated, the analysis excludes lens bin 2. Using flat priors slightly reduces the constraining power but does not introduce any significant shift in cosmology compared to the Gaussian priors case, regardless of whether lens bin 2 is included or excluded (see Fig.~\ref{fig:contours_data}a). When the magnification coefficients are fixed to the \Balrog estimates, the resulting contours are consistent with those obtained using Gaussian priors, while the largest deviation in cosmology occurs when magnification is fixed to zero (see Fig.~\ref{fig:contours_data}b).} 
    \label{fig:contours_data}
\end{figure*}

\begin{figure*}
    \centering
    \begin{minipage}{0.49\textwidth}
        \centering
        \includegraphics[width=\linewidth]{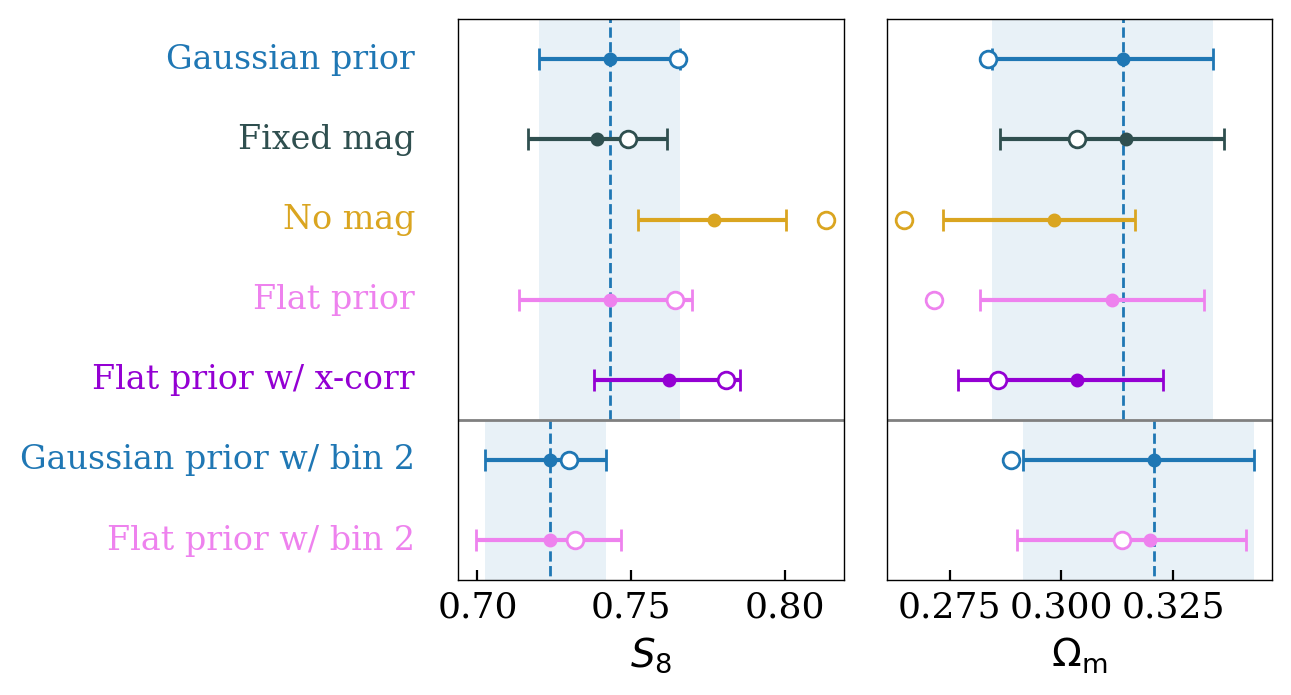}
        {(a) \textbf{Data 2$\times$2pt}.}
        \label{fig:contours_data_bin2}
    \end{minipage}\hfill
    \begin{minipage}{0.49\textwidth}
        \centering
        \includegraphics[width=\linewidth]{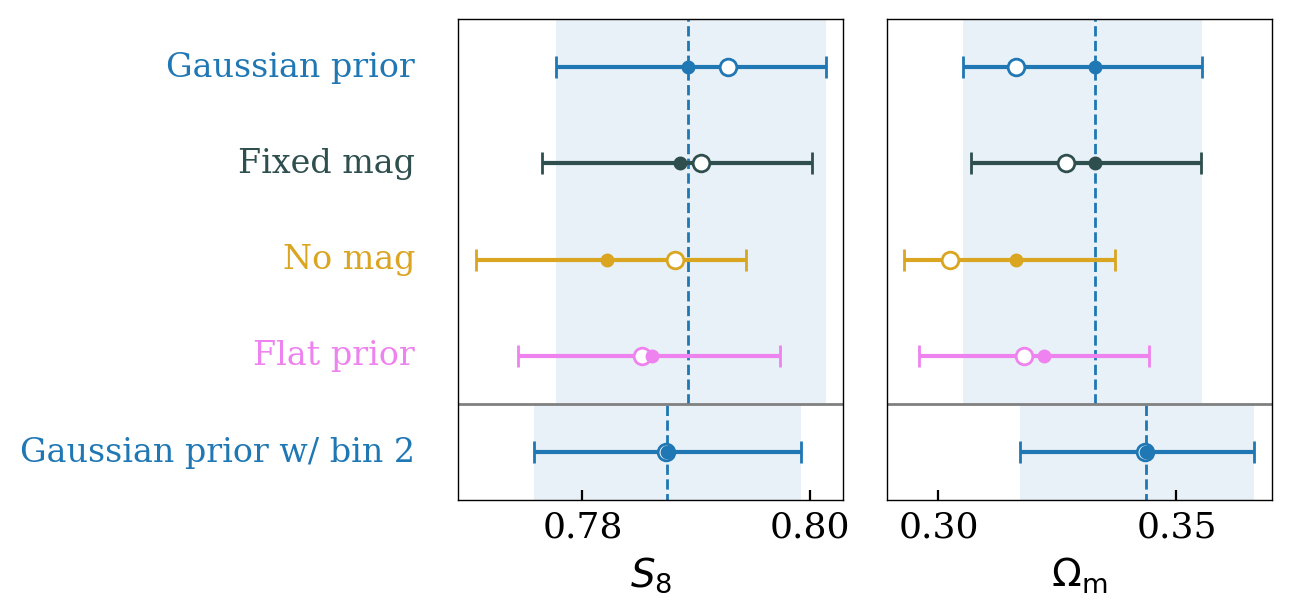}
        {(b) \textbf{Data 3$\times$2pt}.}
        \label{fig:contours_data_all}
    \end{minipage}
    \caption{1D constraints on $S_8$ and $\Omega_m$ from unblinded DES Y6 2$\times$2pt (Fig.~\ref{fig:contours_panel_data}a) and 3$\times$2pt (Fig.~\ref{fig:contours_panel_data}b) data. The results show the differences arising from different priors on the magnification bias, $\Csample$, as described in Sec.~\ref{sec:mag_cosmology}. The figure also shows results from the alternative analysis that includes lens bin 2 (``w/ bin 2'') in the lower panel. The dashed vertical lines mark the posterior means from the fiducial analysis with Gaussian priors, while the shaded areas indicate the associated $1\sigma$ uncertainties. The 68\% marginalized posterior intervals are illustrated with solid horizontal bars. Filled circles correspond to posterior means, while open circles denote the MAP values.} 
    \label{fig:contours_panel_data}
\end{figure*}

\begin{figure*}
    \centering
    \begin{minipage}{0.49\textwidth}
        \centering
        \includegraphics[width=\linewidth]{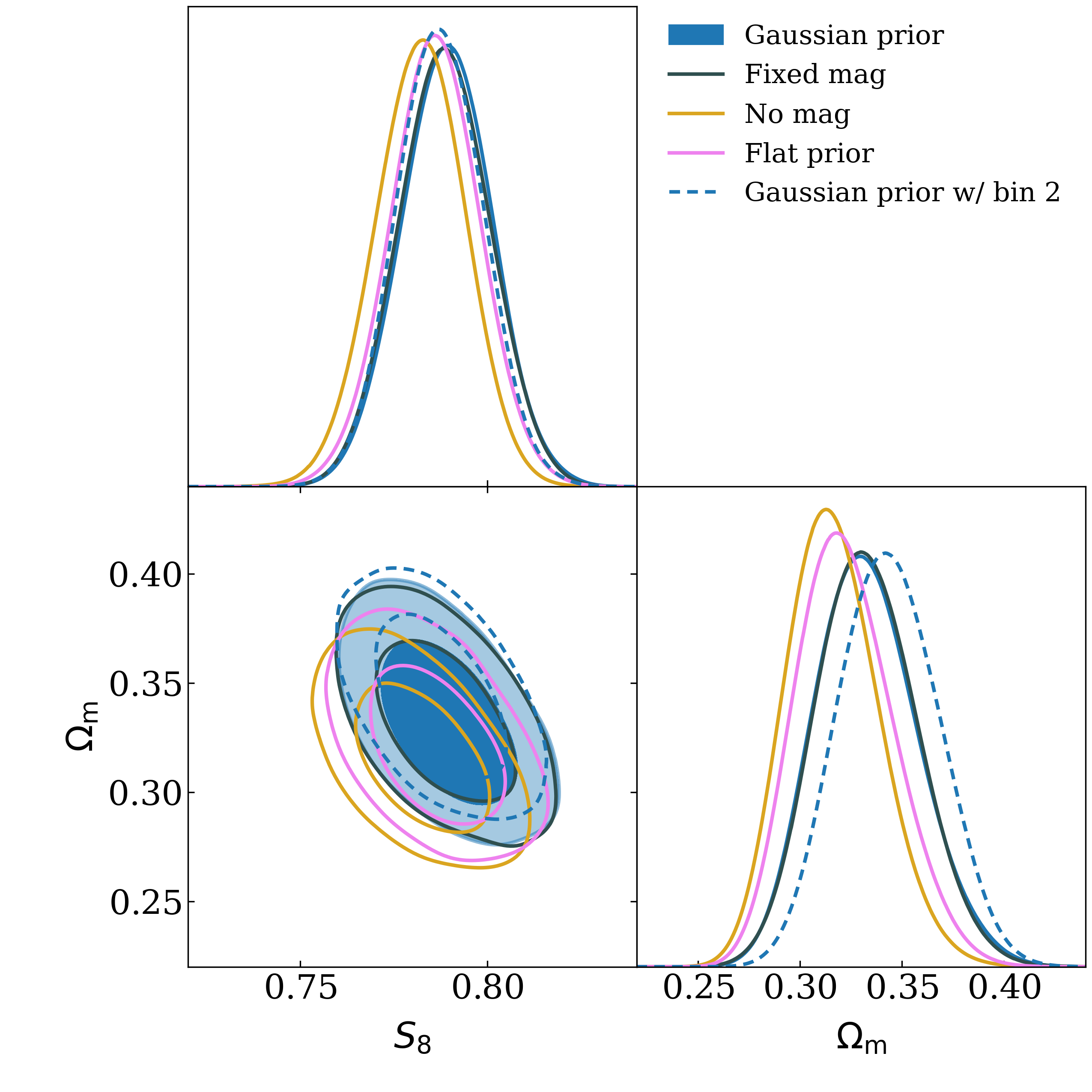}
        {(a) \textbf{Data 3$\times$2pt}.}
        \label{fig:contours_data_3}
    \end{minipage}\hfill
    \begin{minipage}{0.49\textwidth}
        \centering
        \includegraphics[width=\linewidth]{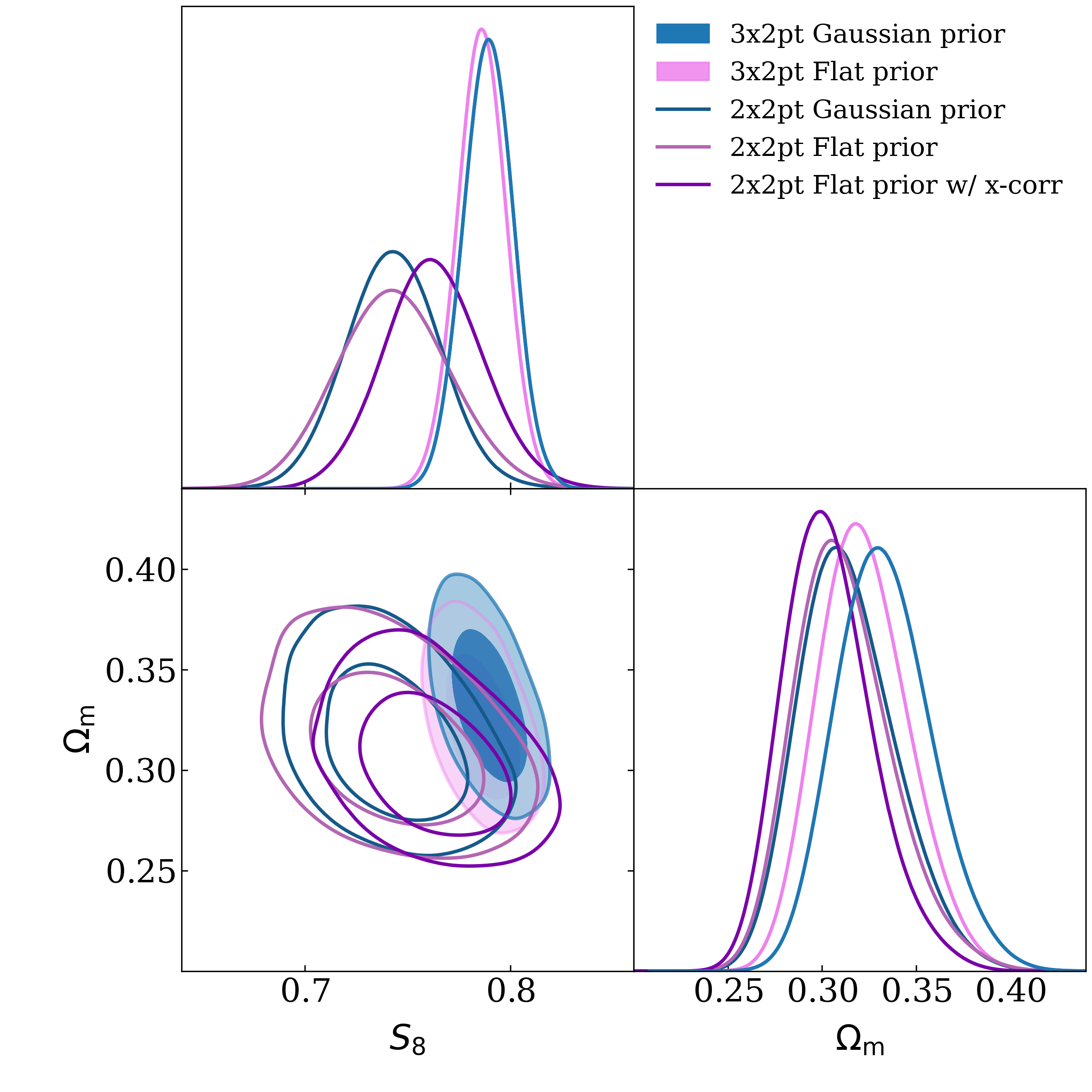}
        {(b) \textbf{Data 3$\times$2pt} and \textbf{2$\times$2pt} compared.}
        \label{fig:contours_data_32}
    \end{minipage}
    \caption{2D constraints on $S_8$ and $\Omega_m$ from unblinded DES Y6 3$\times$2pt data (Fig.~\ref{fig:contours_data_3x2pt}a) and a comparison with 2$\times$2pt (Fig.~\ref{fig:contours_data_3x2pt}b) shown for different priors on the magnification bias, $\Csample$: (i) Gaussian priors centered on the \Balrog coefficients with widths set by their uncertainties (``Gaussian prior''); (ii) magnification coefficients fixed to $\Csample^{\rm{\Balrog}}$ (``Fixed mag''); (iii) $\Csample$ fixed to 2, corresponding to no magnification (``No mag''); and (iv) broad uniform priors (``Flat prior''). The alternative analysis including lens bin 2 is also shown (``w/ bin 2''). 3$\times$2pt results are largely unchanged across magnification priors and remain consistent with the fiducial Gaussian prior analysis.} 
    \label{fig:contours_data_3x2pt}
\end{figure*}


We investigate the impact of magnification model choices on unblinded DES Y6 data and assess the robustness of our results under different assumptions. Our fiducial cosmological analysis models magnification by adopting Gaussian priors on the magnification coefficients $\Csample$, centered on the \Balrog estimates and with widths equal to the total uncertainty on those estimates. The fiducial analysis also excludes lens bin 2 (see \citet{y6-2x2pt} and \citet{y6-3x2pt}), a choice motivated by internal consistency tests. As we discuss below, the data indicate that this exclusion is unlikely to be related to magnification.

To test the sensitivity of the results, we repeat the analysis using the same alternative priors on $\Csample$ explored with the simulated noiseless data vector in Sec.~\ref{sec:exp_mag_cosmology}: using a wide flat prior on the magnification coefficients ($\Csample \in [-4,12]$), fixing them to the \Balrog values, fixing magnification to zero ($\Csample = 2$), and using flat priors while also including clustering cross-correlations. We also run chains including lens bin 2 to examine whether its exclusion affects the magnification posteriors or cosmological constraints. All analyses are performed for 2$\times$2pt and 3$\times$2pt statistics, adopting a linear galaxy bias model in $\Lambda$CDM.


As an indicator of our model's goodness-of-fit, we use the metric $\Delta_{\rm PPD}$ derived from the posterior predictive distribution (PPD) for each cosmology chain. The PPD represents the distribution of possible data realizations, conditioned on the observed data and a given model. $\Delta_{\rm PPD}$ is defined as the fraction of the PPD volume with a worse fit than the observed data, thereby quantifying how likely the particular realization of the data is under the assumed cosmology. Values close to unity indicate that the data is consistent with the model (or that the model is weakly constraining), while values near zero indicate significant tension. For more details, see \citet*{y6-ppd}. Table~\ref{tab:ppd_data} reports the resulting $\Delta_{\rm PPD}$ values for our analysis variations.

The posterior constraints on the magnification coefficients are stable to the inclusion of lens bin 2. As shown in Fig.~\ref{fig:contours_mag_coeff_data}, the posteriors shift very little when bin 2 is included (both for the Gaussian prior and flat prior cases), indicating that the issues motivating its exclusion do not arise from magnification. The dip in $\Csample(z)$ for bin 2 (Fig.~\ref{fig:mag_coeff}) is also far too small to explain the observed data behavior, and magnification effects are expected to be minimal at the low redshift of this bin (see Fig.~\ref{fig:predictions} for the bin 2 combinations in a simulated data vector). When using a flat prior, the $\Csample$ posterior for bin 2 extends to negative values, which are unphysical and inconsistent with expectations. Although the posterior remains broad, the Gaussian prior used in the fiducial analysis lies in the tail of this distribution. Given the good agreement between the prior estimates in this redshift bin, it is likely that the magnification parameter is instead capturing another systematic in the data that biases the two-point measurements. This behavior provides additional justification for excluding bin 2 from the fiducial analysis. $\Delta_{\rm PPD}$ values for the Gaussian prior cases are generally low (e.g., 0.006 for the 3$\times$2pt analysis).\footnote{For unblinding, we required $\Delta_{\rm PPD}$ to be grater than 0.001. Internal consistency tests failed this threshold---in particular $\Delta_{\rm PPD}(\gamma_t|\xi_{\pm} + w)$---ultimately motivating the exclusion of bin 2. For more details see \citet{y6-3x2pt}.} When allowing the magnification coefficients to vary under flat priors, the 2$\times$2pt goodness-of-fit improves only modestly, with $\Delta_{\rm PPD}$ increasing from 0.032 (Gaussian prior) to 0.074. For further discussion of the exclusion of lens bin 2, see \citet{y6-2x2pt} and \citet{y6-3x2pt}.

Cosmological constraints from 2$\times$2pt data are broadly insensitive to how magnification is treated. Fig.~\ref{fig:contours_data} shows the 2D constraints on $S_8$ and $\Omega_m$ obtained under different priors on $\Csample$, while Fig.~\ref{fig:contours_panel_data}a presents the 1D marginalized posteriors. Across all cases, the shifts relative to the fiducial results are small, with the exception of fixing magnification to zero, which shifts the contours on $S_8$ by $1.37\sigma$ and on $\Omega_m$ by $-0.84\sigma$. Using flat priors slightly broadens the posterior distributions compared to the fiducial Gaussian prior case (by 23\% for $S_8$), but does not lead to significant parameter shifts, either with or without bin 2. Fixing the magnification coefficients to the \Balrog values yields results that are nearly identical to the fiducial, Gaussian prior, constraints. Including clustering cross-correlations and adopting flat priors also produces a modest shift in the inferred cosmology compared to the fiducial analysis ($0.79\sigma$ in $S_8$ and $-0.53\sigma$ in $\Omega_m$), though the results remain consistent within uncertainties. All of these findings are consistent with trends seen in simulated data before unblinding. The $\Delta_{\rm PPD}$ values generally increase when the magnification coefficients are allowed to vary under flat priors (0.313), compared to case with Gaussian priors (0.173) or fixed magnification (0.108). However, this improvement is reduced once clustering cross-correlations are included (0.119). Our fiducial baseline, which excludes lens bin 2 from the analysis, results in a consistent improvement in $\Delta_{\rm PPD}$, regardless of the priors on the magnification coefficients.

The inferred $\Csample$ values (shown in Fig.~\ref{fig:contours_mag_coeff_data}a) exhibit consistent trends irrespective of whether lens bin 2 is included, and are generally consistent with the \Balrog estimates. Under flat priors, the data prefer a lower magnification coefficient in bin 6 relative to the \Balrog value, consistently with the Gaussian prior case also hitting the prior boundaries, while the coefficient for bin 2 shifts negative. Including clustering cross-correlations causes all $\Csample$ posteriors to shift slightly lower, although the majority---four out of five when excluding bin 2---remain consistent with the \Balrog predictions. This behavior may indicate that the \Balrog-derived magnification coefficients are imperfect or that an additional systematic effect is present---in particular, cross-correlations are likely more sensitive to the tails of the lens redshift distributions than auto-correlations alone. Moreover, these shifts more plausibly reflect unmodeled systematics because the corresponding cosmological shifts align with projection effects seen in simulated data, suggesting that large magnification-induced biases are unlikely. Despite these variations, all the changes have a negligible impact on the cosmological constraints.

\begin{table}
\setlength{\tabcolsep}{4pt}
\begin{tabular}{llc}
\hline
\hline
\rule{0pt}{1.2em}
Data & $\Csample$ prior & $\Delta_{\rm PPD}$ \rule[-0.8em]{0pt}{0pt} \\ 
\hline
2$\times$2pt & Gaussian prior & 0.173 \\ 
2$\times$2pt & Fixed mag & 0.108 \\
2$\times$2pt & No mag & 0.205 \\
2$\times$2pt & Flat prior & 0.313 \\
2$\times$2pt w/ x-corr & Flat prior & 0.119 \\
2$\times$2pt w/ bin 2 & Gaussian prior & 0.032 \\ 
2$\times$2pt w/ bin 2 & Flat prior & 0.074 \\
\hline 
3$\times$2pt & Gaussian prior & 0.059 \\ 
3$\times$2pt & Fixed mag & 0.022 \\
3$\times$2pt & No mag & 0.063\\
3$\times$2pt & Flat prior & 0.086 \\
3$\times$2pt w/ bin 2 & Gaussian prior & 0.006 \\ 
\hline 
\end{tabular}

\caption{Values of the metric $\Delta_{\rm PPD}$ for the PPD for each cosmology chain described in Sec.~\ref{sec:mag_cosmology}. $\Delta_{\rm PPD}$ measures the consistency between the observed data and the PPD, with values near 1 indicating good agreement and values near 0 indicating significant tension. Analyses that exclude lens bin 2 consistently yield higher $\Delta_{\rm PPD}$ values across all prior choices. Allowing the magnification coefficients to vary under flat priors generally leads to the best model-data agreement; however, this improvement disappears once clustering cross-correlations are included.}
\label{tab:ppd_data}
\end{table}

We repeat the analysis with the 3$\times$2pt data vector, although we explore the inclusion of lens bin 2 less extensively, based on the findings from the 2$\times$2pt analysis. The resulting 3$\times$2pt constraints on $S_8$ and $\Omega_m$ are shown in 2D (Fig.~\ref{fig:contours_data_3x2pt}a) and 1D (Fig.~\ref{fig:contours_panel_data}b). The 1D marginalized posteriors for the magnification coefficient are shown in Fig.~\ref{fig:contours_mag_coeff_data}b. For comparison, we show the 2$\times$2pt results alongside the 3$\times$2pt cosmology constraints (Fig.~\ref{fig:contours_data_3x2pt}b). Including cosmic shear reduces shifts in the cosmological parameters across different magnification coefficient priors, yielding results that are highly consistent with the fiducial analysis. As in the 2$\times$2pt case, fixing magnification to zero produces the largest deviations from the fiducial constraints ($-0.61\sigma$ in $S_8$ and $-0.71\sigma$ in $\Omega_m$). The goodness-of-fit is slightly improved when the magnification coefficients are varied under uninformative flat priors. In this scenario, the data prefers slightly lower values of $\Csample$ compared to the \Balrog estimates for all bins, which could arise from projection effects or other systematic effects in the data. These results should be interpreted with caution, as allowing the magnification coefficients to vary freely may effectively introduce more free parameters than the data can robustly constrain.

We now compare the results of this analysis to those found in the DES Y3 data in \citet*{y3-2x2ptmagnification}. In this analysis the 2$\times$2pt constraints on $C_{\rm{sample,3}}$ when using a flat prior on magnification parameters is fully consistent with the \Balrog estimate, unlike in Y3 where this bin showed a significant over-estimation from the 2$\times$2pt data in this bin. The Y3 data also found a similar low posterior on $C_{\rm{sample,6}}$, indicating that this could be caused by a common systematic in the two analyses.

In Appendix~\ref{app:sys_params_degeneracy} we examine the possible degeneracy between the magnification coefficients and other systematic parameters, such as those describing intrinsic alignments and galaxy bias. We find no evidence of significant degeneracy in the data.

\section{Conclusions}
\label{sec:conclusions}

In this work, we investigated the role of weak lensing magnification in the Dark Energy Survey Year 6 analysis, with a particular focus on the impact of lens magnification on cosmological parameter inference with galaxy clustering auto-correlations and galaxy-galaxy lensing. Magnification affects the observed number of selected galaxies in the \maglimplusplus lens sample, and must therefore be properly modeled to avoid biases in the inferred cosmology.

We derived lens magnification coefficients using different approaches and found good consistency among them (see Sec.~\ref{sec:mag_coeff_values}). For the fiducial Y6 analysis, we adopt estimates obtained from \Balrog synthetic source injection, with Gaussian priors centered on these values.

Validation tests conducted prior to unblinding on simulated data showed that our modeling choices, including the treatment of magnification, do not bias the recovered cosmological parameters. These tests also confirmed that magnification must be included in the modeling, as its omission leads to significant biases: $0.81\sigma$ in $S_8$ and $-0.56\sigma$ in $\Omega_m$ (and $-0.61\sigma$ and $-0.64\sigma$, respectively, when cosmic shear is included). We further explored the effects of different priors on the magnification coefficients, as well as the inclusion of clustering cross-correlations. Using uninformative priors slightly reduces the constraining power, increasing the $1\sigma$ uncertainty in $S_8$ by 15\% and introducing a bias of $-0.50\sigma$ (when including cosmic shear, the  uncertainty increases by 8\% and the bias is $-0.07\sigma$). Including clustering cross-correlations mitigates this effect, tightening constraints on magnification and reducing the bias in $S_8$ to $-0.32\sigma$. When marginalizing over the magnification coefficients with Gaussian priors (our fiducial choice), the recovered constraints remain nearly identical to the case where the coefficients are fixed at their true values. Projection effects themselves lead to biases of $-0.64\sigma$ in $S_8$ and $0.89\sigma$ in $\Omega_m$ (and of $-0.77\sigma$ and $0.81\sigma$, respectively, when including cosmic shear) when the magnification coefficients are fixed, and the biases quoted above are additional to these offsets (see Sec.~\ref{sec:exp_mag_cosmology}).


We repeated a similar exploration with the real, unblinded DES Y6 data (see Sec.~\ref{sec:mag_cosmology}). While allowing the magnification coefficients to vary under alternative priors produces some shifts in their inferred values, the resulting cosmological constraints remain stable. Allowing the coefficients to vary under uninformative flat priors leads to an improved goodness-of-fit, with some broadening of the posteriors (23\% in $S_8$, while minimal when including cosmic shear) and negligible cosmological shifts relative to the fiducial Gaussian-prior case (minor shifts when including cosmic shear: $-0.27\sigma$ in $S_8$ and $-0.41\sigma$ in $\Omega_m$). Fixing the magnification coefficients to the \Balrog values yields results nearly identical to the fiducial analysis, while fixing magnification to zero produces the largest deviation, shifting $S_8$ by $1.37\sigma$ and $\Omega_m$ by $-0.84\sigma$ (when cosmic shear is included: $-0.61\sigma$ in $S_8$ and $-0.71\sigma$ in $\Omega_m$). Including clustering cross-correlations and adopting flat priors leads to a modest shift of $0.79\sigma$ in $S_8$ and $-0.53\sigma$ in $\Omega_m$. The inferred magnification coefficients are broadly consistent with the \Balrog estimates, although the highest redshift bin shows deviations when flat priors are adopted, and even more so when clustering cross-correlations or cosmic shear are also included. These shifts are more likely driven by unmodeled systematics rather than incorrect inputs, as the cosmological changes align with prior volume effects observed in simulated data, suggesting that large biases from degeneracies with the magnification parameters are unlikely. 

Based on pre-unblinding goodness-of-fit tests, lens bin 2 was excluded from the fiducial analysis, and all data results presented above were therefore obtained without this bin. When included, the inferred magnification coefficient for lens bin 2 with flat priors shows a significant discrepancy from the \Balrog estimate, yielding unphysical negative values and further justifying its exclusion. Importantly, the inferred magnification coefficients for the remaining bins remain stable when bin 2 is included, confirming that its exclusion does not bias the magnification analysis. 

Overall, these results demonstrate that our fiducial modeling choices provide a reliable and conservative treatment of magnification effects in the DES Y6 analysis. Using a synthetic source injection framework to infer priors on the magnification coefficients is recommended, particularly when dealing with complex selection functions. Compared to simpler estimation methods---although for this specific sample they yield consistent results---\Balrog provides a more accurate characterization of the coefficients and yields more realistic uncertainty estimates. Moreover, while uninformative priors do not significantly affect the constraining power in the present analysis, employing informative priors is expected to become increasingly important for future surveys with greater depth and precision.

In conclusion, weak lensing magnification is an important effect that must be modeled for precision cosmology. Within the DES Y6 dataset and our modeling framework, its impact on cosmological inference is well controlled. Future analyses with larger datasets or grater constraining power will be more sensitive to magnification, making accurate modeling essential to avoid potential biases in cosmological parameters, while also enabling more precise measurements of magnification that can improve the modeling itself.


\section*{Data Availability}
\label{sec:data_availability}

All data from the Dark Energy Survey used in this work is publicly available through the cosmological data release at \url{https://des.ncsa.illinois.edu/releases}. The measurements of the magnification coefficients as well as the code used to produce them are available at \url{https://github.com/des-science/y6-lens-magnification}.

\section*{Author contributions}

All authors contributed to this paper and/or carried out infrastructure work that made this analysis possible. EL developed the code for measuring the DES Y6 magnification coefficients, building on the Y3 framework created by JE and NM. EL also performed all analyses and drafted the manuscript, with guidance from JE. DA contributed to the development of \Balrog for Y6, and coded the algorithm for \Balrog reweighting. DS and AF provided modeling guidance, produced the simulated data vectors, and assisted with the chain configuration and runs. NW carried out the \maglimplusplus data selection and helped with selection on \Balrog. AP, SA, and RM supervised EL throughout the project (with AP and SA serving as leads of the LSS SWG). The DNF catalogs were generated by JdV. JC ran most of the data chains. SS, WdA, AAl, and CS contributed through discussions, and JM assisted with goodness-of-fit analyses. JP, NM, and DB served as the internal review committee and provided valuable feedback on the writing and presentation of this paper. MAT, CC, MC, and MRB contributed to the development of this work as coordinators of the Y6 analysis. JB contributed to the modeling analysis planning and pipeline. MY, TS, and MRB contributed to the building of the \mdet sample. NW and MRM contributed to the definition of the \maglimplusplus sample and to the clustering measurements. JP produced the corresponding catalog versions, ang GG generated the the galaxy-galaxy lensing measurements. BY, AAm, AAl, GG, and WdA worked on determining the redshift distributions of the galaxy samples. KB and IS contributed to the development of the Y6 Gold sample. The remaining authors have made contributions to this paper that include, but are not limited to, the construction of DECam and other aspects of collecting the data; data processing and calibration; developing broadly used methods, codes, and simulations; running the pipelines and validation tests; and promoting the science analysis.
\section*{Acknowledgements}

EL acknowledges support from MCIN/AEI/10.13039/501100011033 and the FSE+ under the program Ayudas predoctorales of the Ministerio de Ciencia e Innovación PRE2022-101625. AP acknowledges financial support from the European Union's Marie Skłodowska-Curie grant agreement 101068581, and from the \textit{César Nombela} Research Talent Attraction grant from the Community of Madrid (Ref. 2023-T1/TEC-29011).

Funding for the DES Projects has been provided by the U.S. Department of Energy, the U.S. National Science Foundation, the Ministry of Science and Education of Spain, the Science and Technology Facilities Council of the United Kingdom, the Higher Education Funding Council for England, the National Center for Supercomputing Applications at the University of Illinois at Urbana-Champaign, the Kavli Institute of Cosmological Physics at the University of Chicago, the Center for Cosmology and Astro-Particle Physics at the Ohio State University, the Mitchell Institute for Fundamental Physics and Astronomy at Texas A\&M University, Financiadora de Estudos e Projetos, Funda{\c c}{\~a}o Carlos Chagas Filho de Amparo {\`a} Pesquisa do Estado do Rio de Janeiro, Conselho Nacional de Desenvolvimento Cient{\'i}fico e Tecnol{\'o}gico and the Minist{\'e}rio da Ci{\^e}ncia, Tecnologia e Inova{\c c}{\~a}o, the Deutsche Forschungsgemeinschaft and the Collaborating Institutions in the Dark Energy Survey. 

The Collaborating Institutions are Argonne National Laboratory, the University of California at Santa Cruz, the University of Cambridge, Centro de Investigaciones Energ{\'e}ticas, Medioambientales y Tecnol{\'o}gicas-Madrid, the University of Chicago, University College London, the DES-Brazil Consortium, the University of Edinburgh, the Eidgen{\"o}ssische Technische Hochschule (ETH) Z{\"u}rich, Fermi National Accelerator Laboratory, the University of Illinois at Urbana-Champaign, the Institut de Ci{\`e}ncies de l'Espai (IEEC/CSIC), the Institut de F{\'i}sica d'Altes Energies, Lawrence Berkeley National Laboratory, the Ludwig-Maximilians Universit{\"a}t M{\"u}nchen and the associated Excellence Cluster Universe, the University of Michigan, NSF NOIRLab, the University of Nottingham, The Ohio State University, the University of Pennsylvania, the University of Portsmouth, SLAC National Accelerator Laboratory, Stanford University, the University of Sussex, Texas A\&M University, and the OzDES Membership Consortium.

Based in part on observations at NSF Cerro Tololo Inter-American Observatory at NSF NOIRLab (NOIRLab Prop. ID 2012B-0001; PI: J. Frieman), which is managed by the Association of Universities for Research in Astronomy (AURA) under a cooperative agreement with the National Science Foundation.

The DES data management system is supported by the National Science Foundation under Grant Numbers AST-1138766 and AST-1536171. The DES participants from Spanish institutions are partially supported by MICINN under grants PID2021-123012, PID2021-128989 PID2022-141079, SEV-2016-0588, CEX2020-001058-M and CEX2020-001007-S, some of which include ERDF funds from the European Union. IFAE is partially funded by the CERCA program of the Generalitat de Catalunya.

We  acknowledge support from the Brazilian Instituto Nacional de Ci\^encia e Tecnologia (INCT) do e-Universo (CNPq grant 465376/2014-2).

Part of this research was carried out at the Jet Propulsion Laboratory, California Institute of Technology, under a contract with the National Aeronautics and Space Administration (80NM0018D0004).

This document was prepared by the DES Collaboration using the resources of the Fermi National Accelerator Laboratory (Fermilab), a U.S. Department of Energy, Office of Science, Office of High Energy Physics HEP User Facility. Fermilab is managed by Fermi Forward Discovery Group, LLC, acting under Contract No. 89243024CSC000002.

We acknowledge the use of Spanish Supercomputing Network (RES) resources provided by the Barcelona Supercomputing Center (BSC) in MareNostrum 5 under allocations  2025-1-0045 and 2025-2-0046.


\appendix
\section{Source magnification}
\label{app:source_mag}

In addition to affecting the lens galaxies, magnification also impacts the source sample by increasing the number of observed source galaxies behind regions of high mass density. However, its impact on the 3$\times$2pt data vector is generally smaller than that of lens magnification and can be neglected, as demonstrated in the DES Y3 analysis by \citet{y3-gglensing}. Furthermore, source magnification has a minimal impact on the estimation of source redshift uncertainties when applying the clustering redshift (WZ) method, as shown in \citet{y6-wz}. While magnification is included in our WZ modeling---since it significantly affects the WZ data vector---its overall influence on the combined SOMPZ+WZ redshift calibration remains small, as the SOMPZ component is insensitive to magnification.

In this appendix, we present the magnification coefficients $\Csample$ estimates for source galaxies. Since source magnification is not included in the baseline 3$\times$2pt analysis model from \citet*{y6-methods} but is only used in the WZ calibration, we report only the values obtained with the \Balrog method and do not perform a complete systematic error analysis. 


We generate a source sample in \Balrog that is equivalent to the \mdet shape catalog described in \citet*{y6-metadetect}, following the methodology outlined in \citet*{y6-balrog}. To assess the redshift dependence of the source magnification signal, we divide the sample into four tomographic bins using the \textsc{SOMPZ} photometric redshift binning method from \citet{y6-sourcepz}, which is also applied to the wide-field data objects.

The resulting $\Csample$ values are provided in Table \ref{tab:mag_coeff_source}. The source magnification coefficients increase with redshift, as expected from the growing impact of magnification bias, and are broadly comparable to Y3 results, although differences may arise from sample selection and the greater depth of Y6. The associated error bars are smaller than those for the lens coefficients, reflecting the larger size of the \Balrog source sample.

\begin{table}
    \begin{tabular}{cc}
    \hline
    \hline
    \rule{0pt}{1.5em}
    Bin & $C_\text{sample}^\text{\Balrog}$ \rule[-0.8em]{0pt}{0pt} \\ 
    \hline
     1 & $1.628 \pm 0.045$ \\ 
     2 & $1.504 \pm 0.052 $ \\ 
     3 & $1.757 \pm 0.047$ \\ 
     4 & $2.731 \pm 0.055$ \\ 
    \hline 
    \end{tabular}
    \caption{Magnification coefficient estimates, along with their associated statistical errors, for the four redshift bins of the \mdet source sample, computed using the fiducial \Balrog method outlined in Sec.~\ref{sec:mag_coeff}.}
    \label{tab:mag_coeff_source}
\end{table}

\section{Balrog injection and weighting scheme}
\label{app:balrog_injweights}

The Y6 \Balrog simulations introduce an improved injection scheme designed to enhance the statistical power of cosmological analyses while ensuring a realistic representation of survey data. Building on the framework used in DES Y3 (\citet{y3-balrog}), where sources are randomly drawn from the deep-field sample and injected into relevant CCD images, the Y6 scheme introduces targeted weighting for different cosmological samples. Instead of treating all sources equally, the injections are divided into five categories, each reflecting the primary scientific goals of the survey:

\begin{itemize}
    \item All sources (Y3-scheme): One-third of injections follow the unweighted Y3 method, including all galaxies and stars without selection.
    \item LSS sample: One-sixth of injections are weighted toward brighter galaxies ($m_i < 21.5$), applying a sigmoid weighting function and star-galaxy selection.  
    \item LSS high-quality redshifts sample: Another sixth uses the LSS sample weights but is further restricted to galaxies with high-precision redshifts from COSMOS (\citet{cosmos1, cosmos2}), PAUS (\citet{paus}), or C3R2 (\citet{c3r21}, \citet{c3r22}, \citet{c3r23}).
    \item WL sample: One-sixth of the injections prioritize fainter galaxies relative to the LSS sample, down-weighting those with $m_i > 23.5$, in line with weak lensing sample requirements. 
    \item WL high-quality redshifts sample: The final sixth combines the weak lensing selection with high-quality redshifts.
\end{itemize}

\begin{figure}
    \centering
    \includegraphics[width=0.45\textwidth]{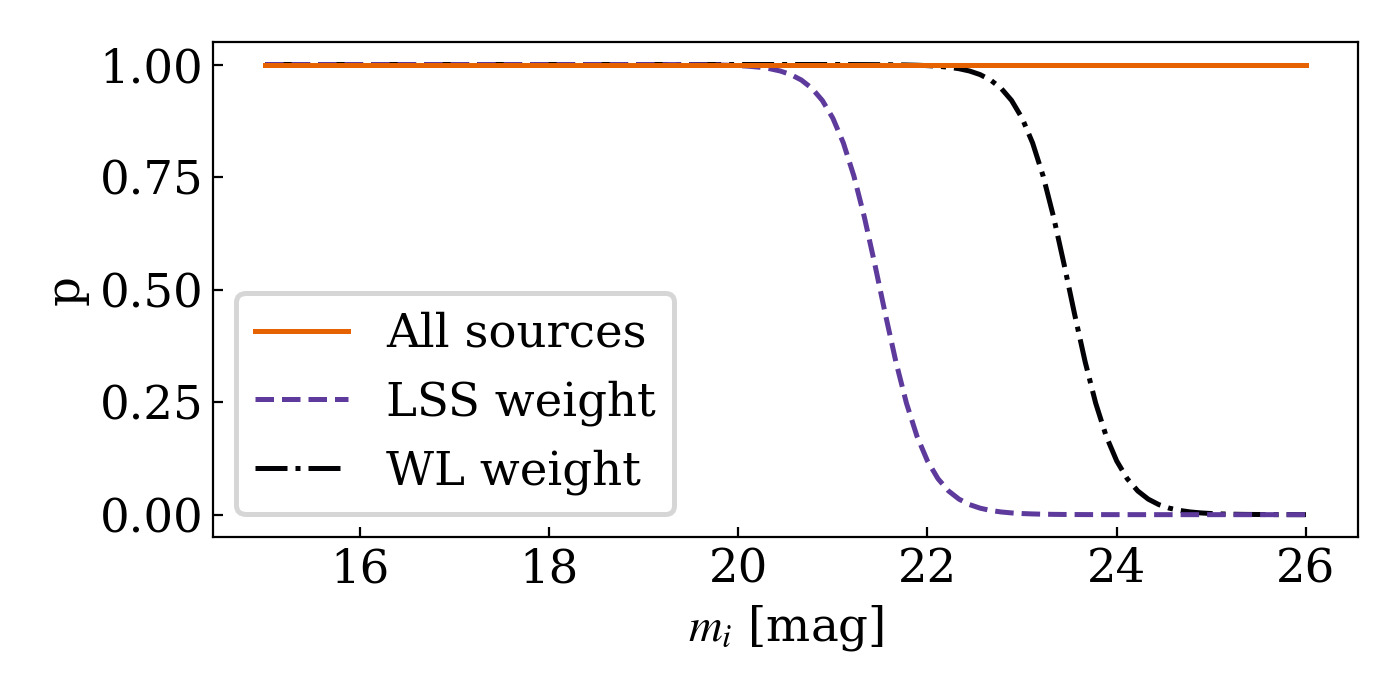}
    \caption{Magnitude-based weighting functions used for injecting the LSS and WL samples in \Balrog. Both follow a sigmoid profile, with transition points centered at $i = 21.5$ for the LSS sample and $m_i = 23.5$ for the WL sample.} 
    \label{fig:inj_weights}
\end{figure}

See Fig.~\ref{fig:inj_weights} for a visualization of the weighting scheme.

This refined injection scheme greatly increases the number of usable injections---those that are successfully detected and included in the relevant samples---compared to the Y3 approach, with improvements by factors of 12 (LSS) and 2.7 (WL). These enhancements provide a more complete and representative synthetic catalog, improving the robustness of DES Y6 cosmology measurements, reducing statistical uncertainties, and enhancing the precision of constraints on cosmological parameters.

It is important to note that these weights must be taken into account when calculating the magnification coefficients with \Balrog. Since the improved injection scheme increases the number of detected sources, this effect---combined with the extension of \Balrog to the full 5000 deg$^2$ survey footprint (146 million injections across the full survey area, compared to about 30 million over a $\sim$20\% subset in Y3) and the expansion of the magnification \Balrog run from 500 to 2000 tiles (covering $\sim$1000 deg$^2$, or about 20\% of the full footprint)---results in smaller error bars for the magnification coefficient estimates in Y6 \Balrog compared to those obtained in Y3.

For a detailed description of the injection scheme and its implementation, refer to \citet*{y6-balrog}.
\section{Balrog reweighting}
\label{app:balrog_reweighting}

One of the main systematic uncertainties in our magnification coefficient estimates comes from differences between the \Balrog catalog and the real data. If the distribution of galaxy properties---such as magnitudes, colors, and sizes--differs between the two, the magnification inferred from \Balrog may not fully capture the magnification effects present in the actual dataset. For instance, an underrepresentation of bright galaxies in \Balrog compared to the data could lead to a biased estimate of magnification.  

To address this, we apply a reweighting procedure that adjusts the contribution of each deep-field galaxy. By assigning appropriate weights, we ensure that the weighted \maglimplusplus sample in \Balrog closely matches the observed distribution of galaxy properties in the corresponding data-selected sample, thereby improving the accuracy of our magnification estimates. 

We define a set of galaxy properties, $\textbf{F}$, that we aim to explicitly match between \Balrog and the data. In our case $\textbf{F}$ consists of the key properties used in selecting the \maglimplusplus sample:
\begin{equation}
\textbf{F} \in \{z, m_i, m_r - m_z, m_z - m_{W1}, T\}, 
\end{equation}
corresponding to photometric redshift point estimate from DNF, $i$-band magnitude, colors used for star–galaxy separation, and galaxy size. An ideal, theoretical weighting function would be:
\begin{equation}
    w = \frac{{\rm P}^{\rm data}(\textbf{F})}{{\rm P}^{\Balrog}(\textbf{F})}.
\end{equation}
However, accurately estimating this function across the full $N$-dimensional space requires far more data than available. To circumvent this, we discretize the function by binning galaxies into coarse groups:
\begin{equation}
    w_i = \frac{N^{\rm data}_i}{N^{\Balrog}_i},
\label{eq:weights_group}
\end{equation}
where $w_i$ is the weight assigned to all galaxies in group $i$, and $N^a_i$ represents the number of galaxies from sample $a \in \{\rm data, {\Balrog}\}$ in group $i$.

To define these groups, we employ the k-means clustering algorithm from \textsc{Sci-kit Learn}, using the galaxy property distribution from the data sample. Before clustering, all galaxy properties are rescaled so that their 99.7\% bounds map to the interval $[0, 1]$, preventing any one feature from dominating due to differences in normalization.\footnote{We avoid using the absolute minimum and maximum values to minimize the influence of outliers.} The data sample is then divided into 100 clusters based on the five selected properties. Once the clusters are established, \Balrog galaxies are assigned to them accordingly.

Note that a direct application of Eq.~\ref{eq:weights_group} would inadvertently suppress the magnification signal, because both the regular and magnification samples in \Balrog would be forced to match the data distribution. To preserve the magnification signal, we assign weights to the unique deep-field galaxies in \Balrog rather than all wide-field realizations. 

For each deep-field galaxy $j$, we define its weight based on its wide-field realizations:
\begin{equation}
    w^{\rm deep}_j = \frac{1}{N_{{\rm wide}, j}}\sum_i^{N_{{\rm wide}, j}} \frac{N^{\rm data}_i}{N^{\Balrog}_i}.
\label{eq:weights_avg_deep}
\end{equation}
Then, every \Balrog galaxy is assigned the weight $w^{\rm deep}_j$ corresponding to the deep-field galaxy from which its wide-field realization was generated. This procedure is repeated consistently for both the regular and magnification samples.

However, as magnification alters the sample selection, some deep-field galaxies that were not selected in the regular sample may appear in the magnification sample. These galaxies do not have pre-assigned weights from Eq.~\ref{eq:weights_avg_deep}, as these are estimated using the regular \Balrog sample. For such cases, we assign the average weight of all deep-field galaxies in the same cluster:
\begin{equation}\label{eq:weights_avg}
    w^{\rm avg}_j = \frac{1}{N}\sum_i^{N} w^{\rm deep}_i.
\end{equation}

The entire weighting procedure is performed bin-by-bin, meaning a given deep-field galaxy may receive different weights depending on the specific bin under consideration. This added flexibility enhances our ability to match the data more accurately.

\begin{figure*}
    \centering
    \includegraphics[width=0.98\textwidth]{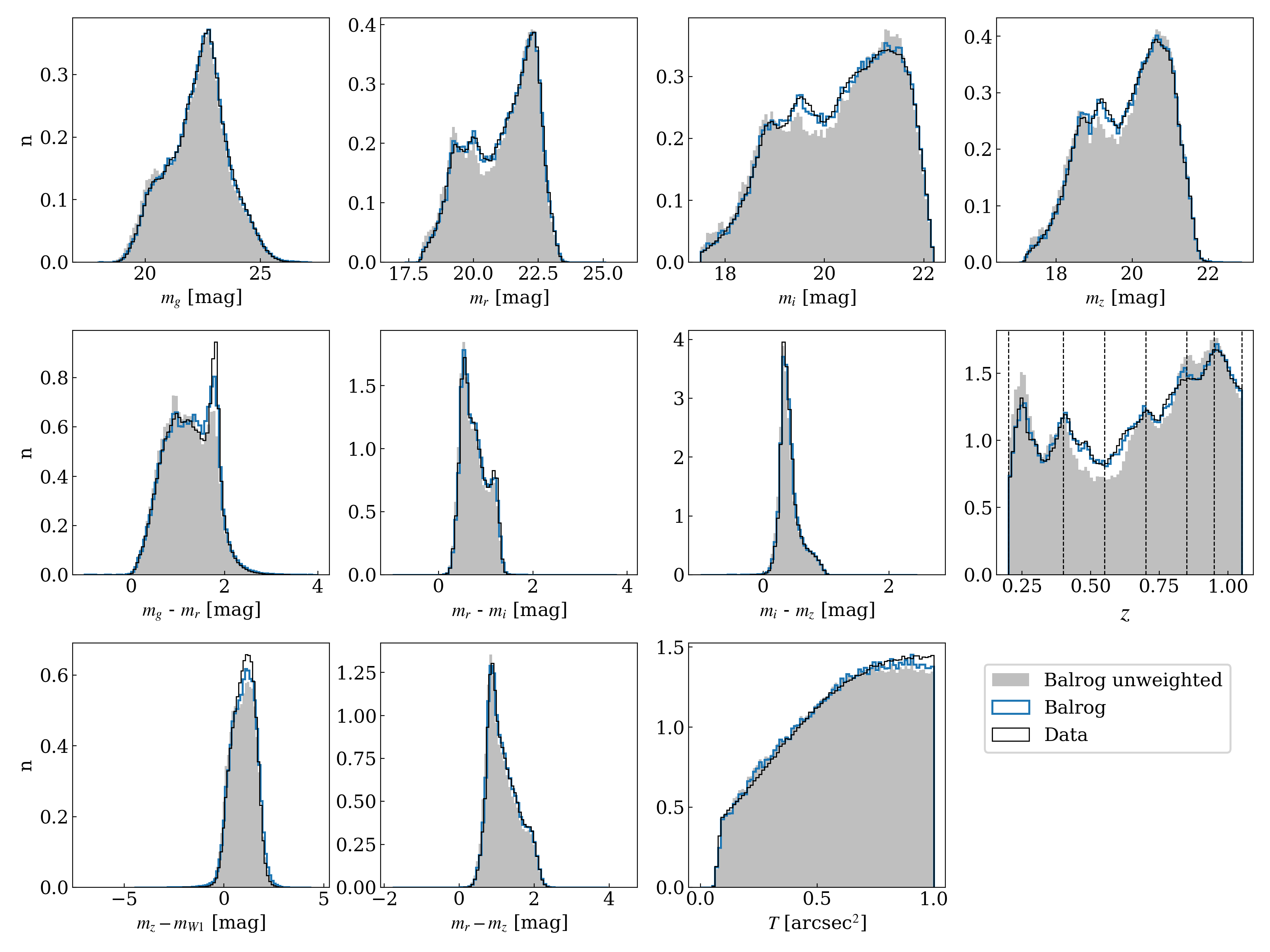}
        \caption{Histograms comparing various properties of the \maglimplusplus lens sample between real data (in black) and \Balrog (including all injection weighting schemes; weighted in blue, unweighted shown as filled gray). The plotted quantities include magnitudes and colors in the \textit{g}, \textit{r}, \textit{i}, and \textit{z} bands; the photometric redshift point estimate from the DNF algorithm used for selection and binning; colors in the \textit{r}, \textit{z}, and \textit{W1} bands, and size estimates \textit{T} used for star-galaxy separation. The agreement between the two samples improves when applying weights to \Balrog.} 
    \label{fig:hists_balrog_data}
\end{figure*}

Fig.~\ref{fig:hists_balrog_data} compares the real data with the \Balrog \maglimplusplus sample. We find good agreement in the color, magnitude, photometric redshift, and galaxy size distributions between the two samples, with the agreement improving when \Balrog is weighted using the procedure described here. Any remaining differences are accounted for by incorporating a systematic error in $\Csample^\text{\Balrog}$, equal to the difference between the flux-only measurements in \Balrog and in the real data.

\begin{figure}
    \centering
    \includegraphics[width=0.48\textwidth]{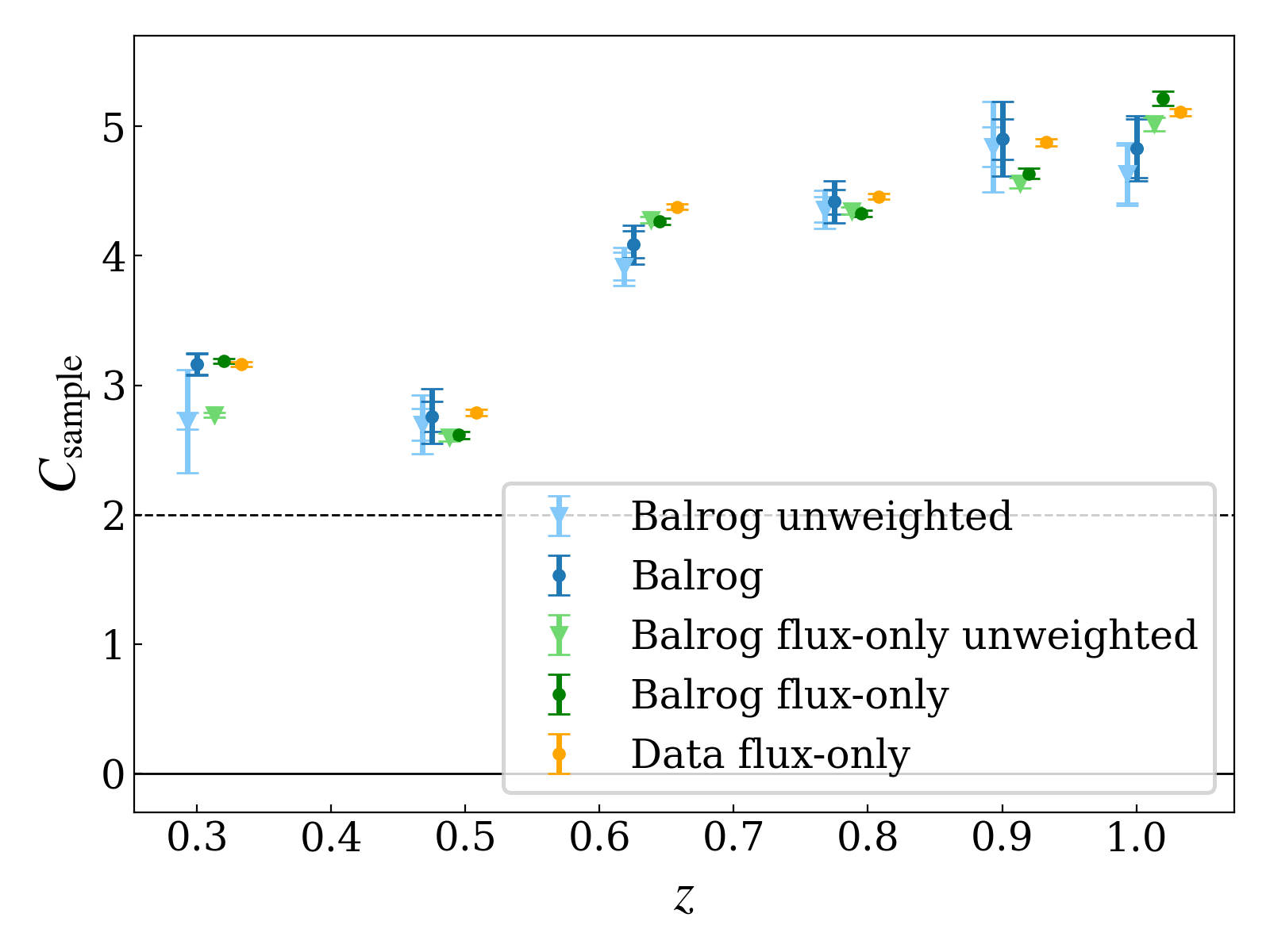}
        \caption{Magnification coefficient estimates for the six redshift bins of the \maglimplusplus lens sample, computed using the three methods described in Sec.~\ref{sec:mag_coeff} (\Balrog, \Balrog flux-only and data flux-only). The lighter points with triangles indicate the estimates obtained from the \Balrog samples prior to reweighting. Overall, applying weights has minimal impact on the estimates, with the exception of bin 1, where it improves the agreement of flux-only \Balrog coefficients with the data estimates.} 
    \label{fig:mag_coeff_reweighting}
\end{figure}

Fig.~\ref{fig:mag_coeff_reweighting} illustrates how reweighting the \Balrog samples affects the magnification coefficients. While most redshift bins show little change, bin 1 exhibits improved agreement between the flux-only \Balrog estimates and the data when weights are applied. This improvement is reflected in a reduction of the systematic uncertainty on $\Csample$ across most bins, from $[0.393, 0.193, 0.099, 0.109, 0.315, 0.092]$ to $[0.026,  0.175, 0.110, 0.130, 0.242, 0.104]$.

\section{Statistical uncertainty}
\label{app:stat_uncertainty}

In this appendix, we derive the statistical uncertainty for the magnification coefficients estimates in Eq.~\ref{eq:stat_error} and provide the explicit equation used when galaxy weights are included.

We denote the number of objects common to both the regular and magnified samples as $N(\delta\kappa + 0)$, while the numbers of objects unique to each sample are labeled $N(0 \text{ only})$ and $N(\delta\kappa \text{ only})$, respectively. We assume these three quantities are independent and follow Poisson distributions with associated uncertainties:
\begin{equation}
    \begin{aligned}
        \sigma^2_{N(\delta\kappa + 0)} &= N(\delta\kappa + 0), \\
        \sigma^2_{N(\delta\kappa \text{ only})} &= N(\delta \kappa \text{ only}), \\
        \sigma^2_{N(0 \text{ only})} &= N(0 \text{ only}).
    \label{eq:variances}
    \end{aligned}
\end{equation}
The total number of objects in each sample can be expressed as
\begin{equation}
    \begin{aligned}
        N(\delta\kappa) &= N(\delta\kappa \text{ only}) +  N(\delta\kappa + 0), \\
        N(0) &= N(0 \text{ only}) +  N(\delta\kappa + 0).
    \label{eq:Ns}
    \end{aligned}
\end{equation}

If we introduce the quantity
\begin{equation}
    X = N(\delta\kappa) - N(0) = N(\delta\kappa \text{ only}) - N(0 \text{ only}),
\end{equation}
we can rewrite $\Csample$ form Eq.~\ref{eq:csample} as
\begin{equation}
    \Csample = \frac{N(\delta\kappa) - N(0)}{N(0) \delta\kappa} = \frac{X}{N(0) \delta\kappa},
    \label{eq:csample_short}
\end{equation}
and its uncertainty as
\begin{equation}
    \begin{aligned}
        \sigma_{\Csample}^2 =& \left( \frac{1}{N(0) \delta\kappa} \right)^2 \sigma_X^2 + \left( - \frac{X}{N(0)^2 \delta\kappa} \right)^2 \sigma_{N(0)}^2 \\& + 2 \left( \frac{1}{N(0) \delta\kappa} \right) \left( - \frac{X}{N(0)^2 \delta\kappa} \right) \sigma_{XN(0)},
    \end{aligned}
    \label{eq:stat_error_short_derivation}
\end{equation}
and thus, dividing by $\Csample$,
\begin{equation}
    \frac{\sigma^2_{\Csample}}{\Csample^2} = \frac{\sigma^2_X}{X^2} + \frac{\sigma^2_{N(0)}}{N(0)^2} - \frac{2 \sigma_{X N(0)}}{X N(0)}.
    \label{eq:stat_error_short}
\end{equation}
We can then derive the error for $\Csample$ in Eq.~\ref{eq:stat_error} by substituting the following relations into the previous equation:
\begin{equation}
    \begin{aligned}
        \sigma_X^2 &= \sigma_{N(\delta\kappa \text{ only})}^2 + \sigma_{N(0 \text{ only})}^2, \\
        \sigma_{N(0)}^2 &= \sigma_{N(0 \text{ only})}^2 + \sigma_{N(\delta\kappa + 0)}^2,  \\
        \sigma_{XN(0)} &= -\sigma_{N(0 \text{ only})}^2, \\
    \end{aligned}
    \label{eq:variances2}
\end{equation}
and get the final statistical uncertainty,
\begin{equation}
    \resizebox{\columnwidth}{!}{$
    \frac{\sigma_{\Csample}}{\Csample} = \sqrt{ \frac{N(0 \text{ only}) + N(\delta\kappa \text{ only})}{[N(\delta\kappa)-N(0)]^2} + \frac{1}{N(0)} + \frac{2N(0 \text{ only})}{N(0)[N(\delta\kappa) - N(0)]}}.
    $}
    \label{eq:stat_error_long}
\end{equation}

When galaxy weights are included, such as those from the \Balrog injection weighting scheme or \Balrog weights used to better match data property distributions, they must be taken into account when measuring the statistical uncertainty.

A Poisson process with mean $M$ can be thought of as $M$ independent Poisson processes, each with mean $1$:
\begin{equation}
    \text{Pois}(\lambda = M) = \sum_{i = 0}^{M - 1} \text{Pois}(\lambda = 1).
    \label{eq:poisson}
\end{equation}
One way to think of this is that each galaxy measurement (for $N(0)$, $N(\delta \kappa)$ etc.) is defined by a Poisson process of mean 1. The variance of the left-hand side is simply the sum of the variances on the right-hand side:
\begin{equation}
    \text{Var}(\text{Pois}(\lambda = M)) = \sum_{i = 0}^{M - 1} \text{Var}(\text{Pois}(\lambda = 1)).
    \label{eq:variance_poisson}
\end{equation}
Rewriting this, a weighted count $Y$ can be seen as the sum of $M$ weighted Poisson distributions with mean 1:
\begin{equation}
    Y = \sum_{i = 0}^{M - 1} w_i \text{Pois}(\lambda = 1).
    \label{eq:poisson_weighted}
\end{equation}
The variance of the weighted sum is:
\begin{equation}
    \begin{aligned}
        \text{Var}(Y) = & \sum_{i = 0}^{M - 1} \text{Var}(w_i \text{Pois}(\lambda = 1))\\
        = & \sum_{i = 0}^{M - 1} w^2_i \text{Var}(\text{Pois}(\lambda = 1))\\
        = & \sum_{i = 0}^{M - 1} w^2_i,
    \end{aligned}
    \label{eq:variance_poisson_weighted}
\end{equation}
where $\text{Var}(\text{Pois}(\lambda = 1)) = 1$. If there are no weights, the variance simplifies to $\text{Var}(Y) \equiv M$, which is the original Poisson result.

We assume that the number of objects common to both samples, $N(\delta\kappa + 0)$, and the number of objects unique to each sample, $N(\delta\kappa \text{ only})$ and $N(0 \text{ only})$, are independent Poisson-distributed quantities. These are given by:

\begin{equation}
    \begin{aligned}
        & N(\delta\kappa \text{ only}) =  \sum_{i ~\in ~\delta\kappa\text{ only}} w_i, \\
        & N(0 \text{ only}) = \sum_{j ~\in ~0 \text{ only}} w_j, \\
        & N(\delta\kappa + 0) = \sum_{k ~\in ~\delta\kappa + 0} w_k .
    \end{aligned}
    \label{eq:Ns_weighted_2}
\end{equation}

The variances for each of these quantities are:
\begin{equation}
    \begin{aligned}
        & \sigma_{N(\delta\kappa \text{ only})}^2 = \sum_{i} w_i^2, \\
        & \sigma_{N(0 \text{ only})}^2 = \sum_{j} w_j^2, \\
        & \sigma_{N(\delta\kappa + 0)}^2 = \sum_{k} w_k^2, \\
    \end{aligned}
    \label{eq:variances_weighted}
\end{equation}

and then,
\begin{equation}
    \begin{aligned}
        & \sigma_X^2 =  \sigma_{N(\delta\kappa \text{ only})}^2 + \sigma_{N(0 \text{ only})}^2 = \sum_{i} w_i^2 + \sum_{j} w_j^2, \\
        & \sigma_{N(0)}^2 = \sigma_{N(0 \text{ only})}^2 + \sigma_{N(\delta\kappa + 0)}^2  = \sum_{j} w_j^2 + \sum_{k} w_k^2, \\
        & \sigma_{XN(0)} = -\sigma_{N(0 \text{ only})}^2 = - \sum_{j} w_j^2. \\
    \end{aligned}
    \label{eq:variances_weighted_2}
\end{equation}

Thus, the statistical error for the magnification coefficient, including the galaxy weights, is:
\begin{equation}
    \begin{aligned}
        \frac{\sigma_{\Csample}^2}{\Csample^2} =& \frac{\sum_{i} w_i^2 + \sum_{j} w_j^2}{(N(\delta\kappa) - N(0))^2} + \frac{\sum_{j} w_j^2 + \sum_{k} w_k^2}{N(0)^2} \\& + \frac{2 \sum_{j} w_j^2}{N(0)(N(\delta\kappa) - N(0))}.
    \end{aligned}
    \label{eq:stat_error_weighted}
\end{equation}
This is the expression we use when any form of galaxy weights are included.

We also compare the theoretical error estimates with those obtained via a jackknife resampling over the ~2000 tiles of the magnified \Balrog sample footprint and find good agreement.
\section{Impact of magnification on photometric redshift}
\label{deltaz}

This appendix examines the case where magnification effects on photometric redshift estimates are neglected, by applying the flux-only method without recalculating redshifts to reflect flux changes, i.e., without using Eq.~\eqref{eq:deltaz}.

\begin{figure}
    \centering
    \includegraphics[width=0.48\textwidth]{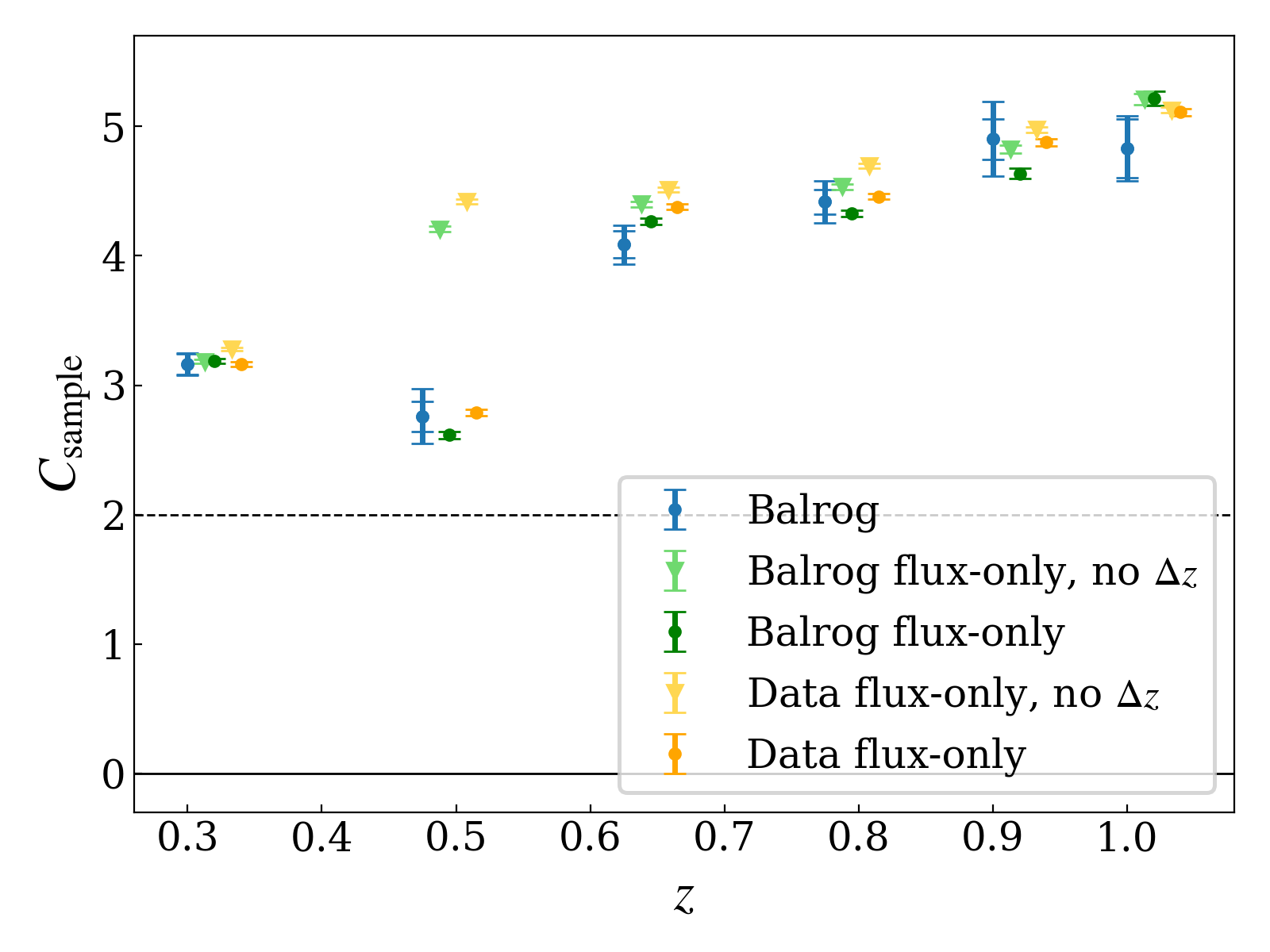}
    \caption{Magnification coefficient estimates for the six redshift bins of the \maglimplusplus lens sample, computed using the three methods described in Sec.~\ref{sec:mag_coeff} (\Balrog, \Balrog flux-only and data flux-only). The lighter points with triangles indicate the estimates obtained without accounting for the impact of magnification on photometric redshift estimates, i.e., when the redshifts are not recalculated according to the changes in flux for the flux-only estimates. Without this correction, the flux-only coefficients in bin 2 are biased and deviate from the fiducial \Balrog estimates.} 
    \label{fig:mag_coeff_deltaz}
\end{figure}

Comparing these estimates with the baseline ones in Fig.~\ref{fig:mag_coeff_deltaz} reveals that the low fiducial $\Csample$ value in bin 2 is primarily driven by the photo-z assignment. As in Y3 (\citet*{y3-2x2ptmagnification}), the discrepancy between the flux-only and fiducial \Balrog estimates in bin 2 is more pronounced when photo-zs are not recalculated for the flux-only methods. If the photometric redshift assignments do not account for magnification-induced flux changes, the $\Csample$ measurements become biased. This highlights the importance of properly updating redshift estimates in response to flux changes due to magnification, thereby improving accuracy and reducing discrepancies in the measurements.

Similarly, assuming the same redshift estimates for both the regular and magnification \Balrog runs (i.e., using the photo-zs from the regular run for the magnification objects, which we know to be inaccurate) shifts the $\Csample$ estimates, with the fiducial \Balrog estimate in bin 2 moving to higher values around 4.5.

\begin{figure}
    \centering
    \includegraphics[width=0.48\textwidth]{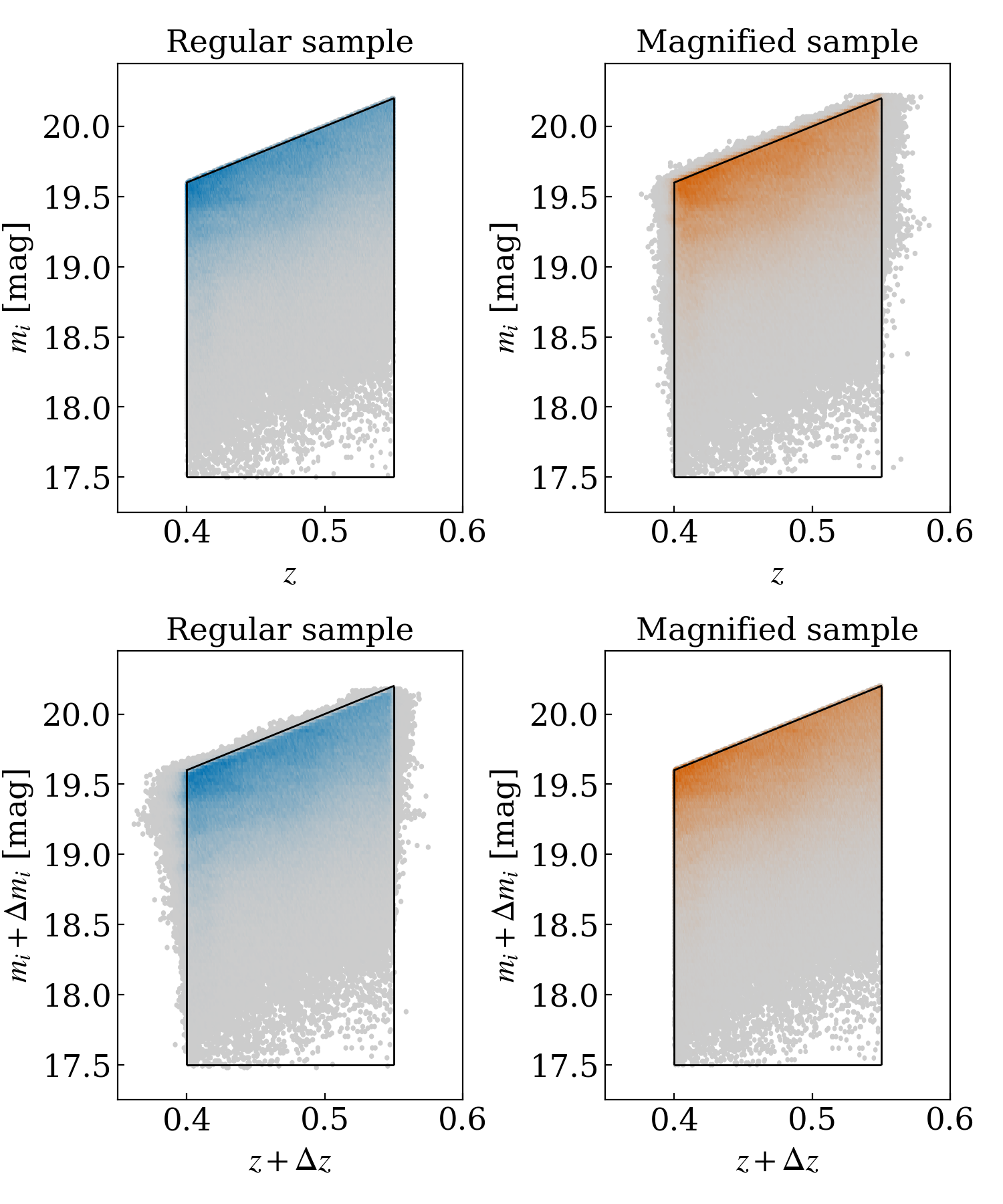}
    \caption{Scatter plots of galaxy magnitudes in the $i$ band versus redshift for the \maglimplusplus second redshift bin, selected from data. Black lines in all panels indicate the \maglimplusplus selection cuts for this bin. The top row shows unmagnified quantities, while the bottom row displays magnified quantities. The left column corresponds to the unmagnified sample, while the right column shows the magnified sample, where magnification is applied using the flux-only method (i.e., adding a constant offset to magnitudes and redshifts). Color represents galaxy density: gray indicates low density, while blue and orange represent high-density regions for the unmagnified and magnified samples, respectively. White regions indicate areas with no objects. The gray regions outside the selection boundaries (black lines) in the bottom-left and top-right panels highlight objects that exit or enter the sample due to magnification effects on magnitudes and redshifts.} 
    \label{fig:bin2_i_dnfz}
\end{figure}

We observe from Fig.~\ref{fig:bin2_i_dnfz} and Table~\ref{tab:bin2_i_dnfz} that the main factor contributing to the difference in the $\Csample$ measurement for the second redshift bin when applying magnification to the redshift estimates is the movement of objects near the $z > 0.4$ cut. Magnification tends to shift more objects from bin 2 to bin 1 than the reverse, leading to a redistribution of galaxies across redshift bins. This shift lowers the $\Csample$ measurement for bin 2 (see the "Net change" values in Table~\ref{tab:bin2_i_dnfz} when ignoring versus applying magnification to the redshift estimates).

\begin{table}
\centering
\begin{tabular}{cccc}
    \hline
    \hline
    Cut type & Entering (\%) & Exiting (\%) & Net change (\%) \\
    \hline
    \hline
    \multicolumn{4}{c}{flux-only magnification applied to magnitudes only} \\
    \hline
    All cuts & 3.41 & 0.00 & 3.41 \\
    $z > 0.2$ & 0.00 & 0.00 & 0.00 \\
    $z \leq$ 0.4 & 0.00 & 0.00 & 0.00 \\
    $m_i < 18 + 4z$ & 3.41 & 0.00 & 3.41 \\
    \hline
    \multicolumn{4}{c}{flux-only magnification applied to magnitudes and redshifts} \\
    \hline
    All cuts & 4.67 & 1.33 & 3.33 \\
    $z > 0.2$ & 0.27 & 0.44 & $-0.17$ \\
    $z \leq$ 0.4 & 1.51 & 0.50 & 1.01 \\
    $m_i < 18 + 4z$ & 2.91 & 0.40 & 2.51 \\
    \hline
\end{tabular}
\caption{Object counts entering or exiting redshift bin 1 of the \maglimplusplus data sample ($0.2 < z \leq 0.4$), due to the bin boundaries and the \maglimplusplus upper selection cut $m_i < 18 + 4z$. The percentages show changes under flux-only magnification applied to magnitudes alone (with redshifts held constant, i.e., unchanged from the unmagnified sample), and under magnification applied to both magnitudes and redshifts.}
\label{tab:bin1_i_dnfz}
\end{table}

\begin{table}
\centering
\begin{tabular}{ccccc}
    \hline
    \hline
    Cut type & Entering (\%) & Exiting (\%) & Net change (\%) \\
    \hline
    \hline
    \multicolumn{4}{c}{flux-only magnification applied to magnitudes only} \\
    \hline
    All cuts & 4.42 & 0.0 & 4.42 \\
    $z >$ 0.4 & 0.0 & 0.0 & 0.0 \\
    $z \leq$ 0.55 & 0.0 & 0.0 & 0.0 \\
    $m_i < 18 + 4z$ & 4.42 & 0.0 & 4.42 \\
    \hline
    \multicolumn{4}{c}{flux-only magnification applied to magnitudes and redshifts} \\
    \hline
    All cuts & 5.57 & $-2.78$ & 2.79 \\
    $z >$ 0.4 & 0.76 & $-2.12$ & $-1.36$ \\
    $z \leq$ 0.55 & 0.51 & $-0.41$ & 0.1 \\
    $m_i < 18 + 4z$ & 4.38 & $-0.28$ & 4.1 \\
    \hline
\end{tabular}
\caption{Object counts entering or exiting redshift bin 2 of the \maglimplusplus data sample ($0.4 < z \leq 0.55$), based on the cuts illustrated in Fig.~\ref{fig:bin2_i_dnfz}, namely the bin boundaries and the \maglimplusplus upper selection cut $m_i < 18 + 4z$. The percentages indicate changes when flux-only magnification is applied to magnitudes only (with redshifts fixed to the unmagnified values), and when magnification is applied to both magnitudes and redshifts.}
\label{tab:bin2_i_dnfz}
\end{table}

\begin{figure}
    \centering
    \includegraphics[width=0.45\textwidth]{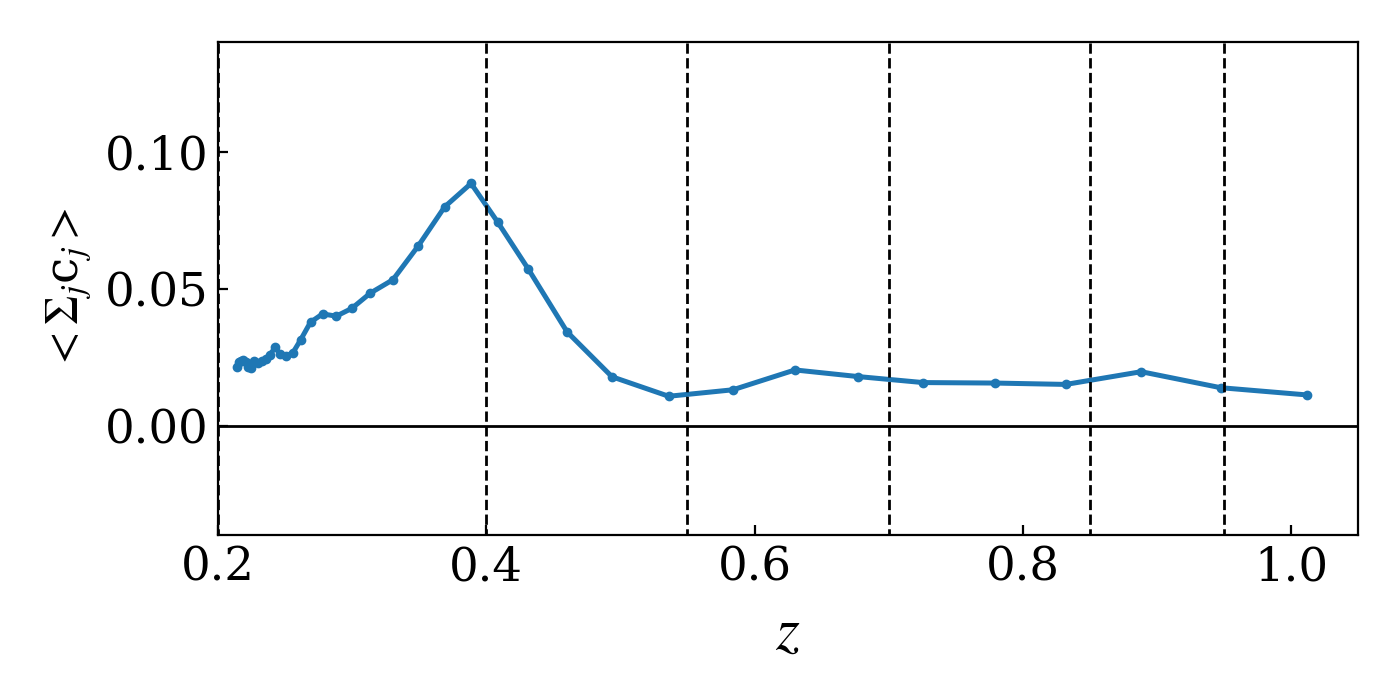}
    \caption{Mean of the DNF fit parameters, $\sum_jc_j$, where $j$ iterates over the \textit{griz} bands, as a function of redshift for \maglimplusplus selected from data. The vertical dashed lines represent the redshift bins boundaries at $z = [0.20, 0.40, 0.55, 0.70, 0.85, 0.95, 1.05]$. The curve peaks near $z=0.4$, indicating increased sensitivity of redshift estimates to magnitude shifts at the boundary between bins 1 and 2.} 
    \label{fig:bin2_mean-dnfc_z}
\end{figure}

A similar conclusion can be drawn by examining the DNF fit parameters. The mean of the DNF parameters, $\sum_j c_j$, where $j$ iterates over the \textit{griz} bands, as a function of redshift for the \maglimplusplus data sample is shown in Fig.~\ref{fig:bin2_mean-dnfc_z}. The DNF fit parameters are larger around $z = 0.4$, indicating that redshift estimates are more sensitive to magnitude shifts in this region. The redshift shift due to flux-only magnification is given by $\Delta z = \sum_j c_j \ \Delta m_j$, with $\Delta m_j = \Delta m$ for all bands $j$ (see Eq.~\eqref{eq:deltaz}). The plot explains the observed behavior in the second redshift bin: when the redshift shift is applied, the selection condition $m_i + \Delta m_i < 18 + 4 (z + \Delta z)$ selects fewer objects in bin 2, as compared to $m_i + \Delta m_i < 18 + 4z $; moreover, due to the applied $\Delta z$, some objects shift into bin 1. In bin 1, the reduction in selected objects is compensated by those entering from the boundary with bin 2 (see Table~\ref{tab:bin1_i_dnfz}). Few objects move from bin 3 to bin 2 due to the DNF parameters being small in bin 3. For the other bins, the impact is minimal as $\sum_j c_j$ is smaller. This explains the reduced $\Csample$ measurement for bin 2 when magnification effects on redshift estimates are taken into account.

The increased sensitivity of redshift estimates to magnitude shifts around $z=0.4$ arises from greater uncertainties in photometric redshift assignments in this region. In particular, this could be due to the 4000\r{A} break transitioning between the $g$ and $r$ bands, weak spectral features that make color-based redshift determination harder, and color degeneracies from the limited number of filters. Consequently, small magnitude shifts due to magnification can significantly impact redshift estimates, affecting sample selection.

This underscores the importance of properly accounting for magnification effects on redshift estimates, as neglecting them can lead to biases, especially in regions with higher photo-z uncertainty.

\section{Magnification with size selection}
\label{app:size_selection}

In this appendix, we explore the impact of object size selection on the magnification coefficient estimates, $\Csample$, for the \maglimplusplus sample. We consider two different ways of selecting subsamples based on object size and analyze how these selections affect the magnification coefficient estimates.

\begin{figure}
    \centering
    \includegraphics[width=0.48\textwidth]{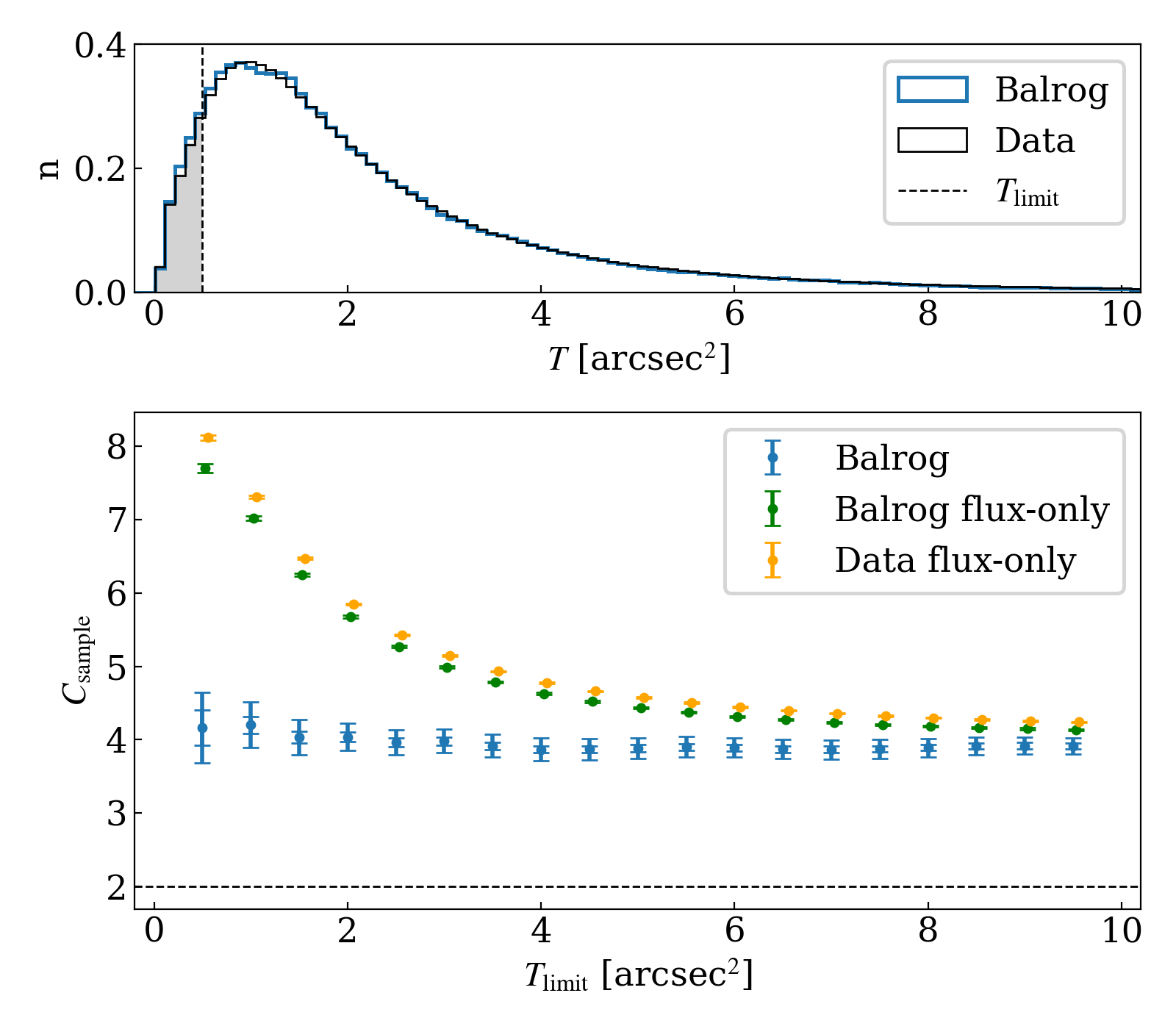}
    \caption{The upper panel shows the distribution of object sizes $T$ for the \Balrog and real data \maglimplusplus samples, highlighting a subsample defined by a size cut at $T_{\rm{limit}}$ (shaded in gray). The lower panel presents the magnification coefficient estimates, $\Csample$, computed using the three methods from Sec.~\ref{sec:mag_coeff}, as a function of the size threshold of each selected subsample. Unlike the fiducial analysis, these subsamples are not binned in redshift. The flux-only estimates are biased for size-selected subsamples, particularly for those dominated by small objects, and only align with the fiducial \Balrog results when most of the size distribution is included.} 
    \label{fig:mag_coeff_T-cuts}
\end{figure}

The first method involves applying a size cut at a threshold value, $T_{\rm{limit}}$, to divide the sample into subsamples of objects smaller than this threshold, with the number of objects increasing as $T_{\rm{limit}}$ increases. Importantly, in this analysis, the subsamples are not binned in redshift, which isolates the effects of size selection on the magnification coefficient estimates. The estimates of the magnification coefficient for these subsamples are shown in Fig.~\ref{fig:mag_coeff_T-cuts}. We find that the flux-only estimates, when a selection on size is applied, are biased. These estimates do not agree with the fiducial \Balrog results, except when the selected subsample includes a large portion of the size distribution, and the bias is more pronounced when the subsample consists of smaller sizes.

\begin{figure}
    \centering
    \includegraphics[width=0.48\textwidth]{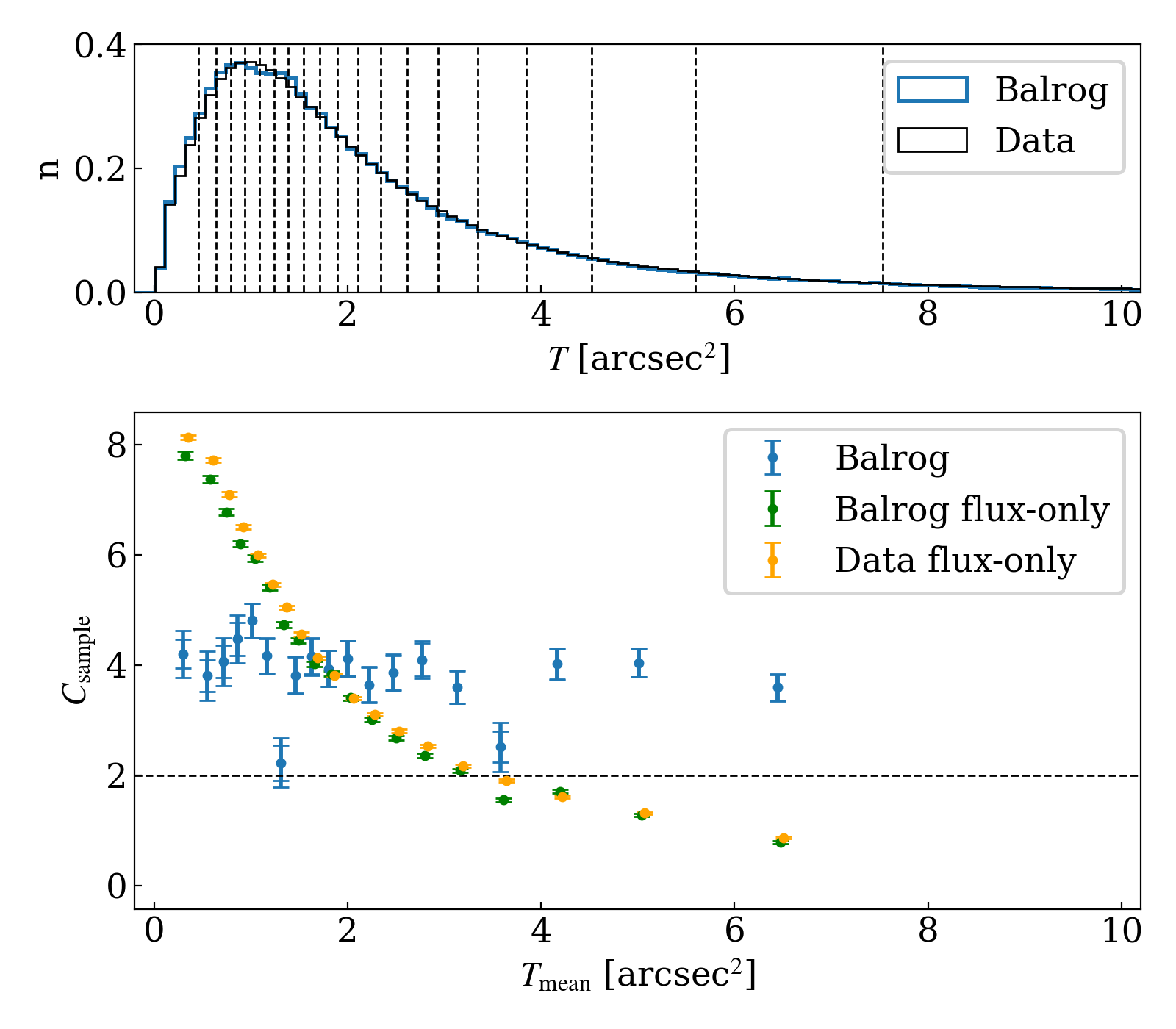}
    \caption{The upper panel shows the distribution of object sizes $T$ for the \Balrog and real data \maglimplusplus samples, with vertical lines marking bins of roughly equal number density. The lower panel presents the magnification coefficient estimates, $\Csample$, for each bin, computed using the three methods from Sec.~\ref{sec:mag_coeff}, plotted as a function of the mean size $T_{\rm{mean}}$ in the bin. Unlike the fiducial analysis, these \maglimplusplus subsamples are not binned in redshift. The flux-only estimates are biased and largely inconsistent with the fiducial \Balrog values, with agreement observed only for subsamples with $T_{\rm{mean}} \sim 1.5$, corresponding to the peak of the size distribution.}
    \label{fig:mag_coeff_T-bins}
\end{figure}

The second method creates subsamples by binning object sizes into intervals of roughly equal number density, rather than using a fixed size cut. This technique aims to provide more balanced subsamples by distributing objects across bins based on their size. Similarly to the first method, these subsamples are also not binned in redshift, allowing for a focused analysis of size selection. Fig.~\ref{fig:mag_coeff_T-bins} shows the magnification coefficients for these subsets as a function of the mean size in each bin. Again, we observe that the flux-only estimates are biased and do not match the fiducial \Balrog estimates, except for the bins at $T \sim 1.5$,  where the size distribution peaks.

These analyses demonstrate that size selection can significantly impact the magnification coefficient estimates, highlighting the need for caution when using approximate methods such as flux-only, which do not account for this effect. The fiducial \Balrog estimates, which incorporate the full range of magnification effects, remain more consistent and accurate, especially when the selection criteria extend beyond a simple flux cut.
\section{Maximum a posteriori estimates}
\label{app:map}

The maximum a posteriori (MAP) parameters are estimated by first selecting the 20 highest-posterior samples from the chains and then refining them through local optimization searches using the Powell algorithm; the result with the highest posterior is chosen as the MAP estimate. Because of the high dimensionality of the parameter space, simply taking the highest-posterior sample gives a poor MAP estimate, so a two-step optimization procedure is used instead. 

Tests on simulated data with multiple noise realizations show that while this method reliably identifies parameter sets that yield good fits to the data, the recovered MAP parameters themselves can be noisy and do not necessarily reflect the true underlying values. This behavior persists across different optimization algorithms and arises not from optimization failures but from the fact that many parameter combinations can, by chance, fit a particular noise realization better than the true model. For more details, see \citet*{y6-ppd}.

\begin{figure}
    \centering
    \includegraphics[width=0.38\textwidth]{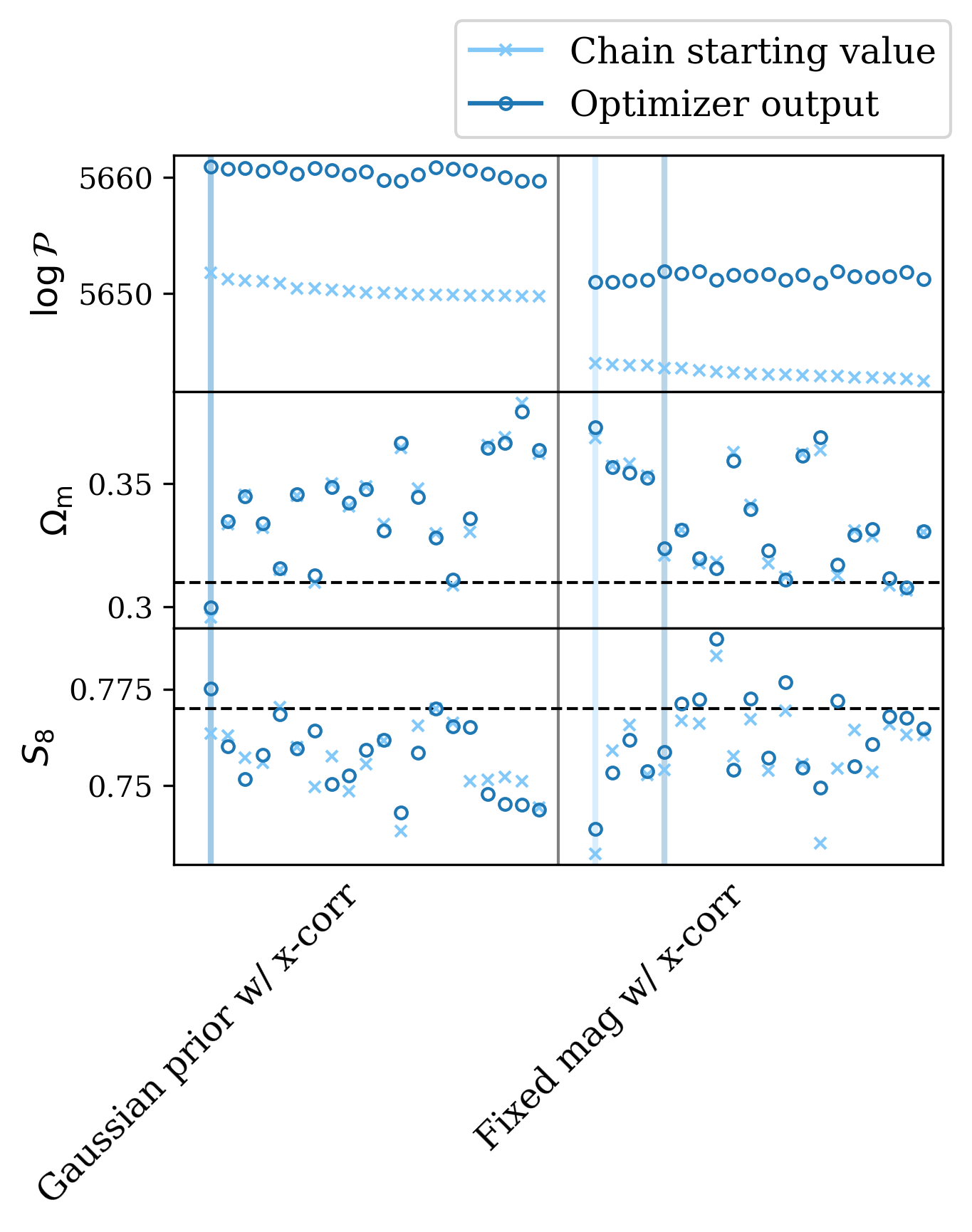}
    \caption{Scatter of the 20 highest-posterior samples for the “Gaussian prior w/ x-corr” and “Fixed prior w/ x-corr” chains. The top panel shows the log-posterior, and the lower panels show $\Omega_m$ and $S_8$. Light blue crosses indicate the original highest-posterior samples, while blue open circles show the results after the optimization search. Vertical lines mark the highest-posterior sample from each group. The dashed horizontal lines mark the true input values. Although the optimized points have higher log-posterior values, the scatter in $\Omega_m$ and $S_8$ remains similar, illustrating the variability arising from noise in the MAP estimation, even when the underlying data and priors are very similar.} 
    \label{fig:map_wcross_fix_gauss}
\end{figure}

As a result, MAP estimates can differ even for identical underlying data. For example, in Fig.~\ref{fig:contours_panel}, the MAP values differ between the "Gaussian prior w/ x-corr" and "Fixed prior w/ x-corr" cases, despite both data vectors being generated from the same fiducial cosmology and the priors on the magnification coefficients being very similar. This discrepancy arises from the inherent noise in the MAP estimation procedure. Fig.~\ref{fig:map_wcross_fix_gauss} further illustrates this effect by showing the scatter among the 20 individual optimization runs for these two cases, highlighting the resulting variability in the recovered cosmological parameters.

\section{Systematic parameters degeneracy}
\label{app:sys_params_degeneracy}

This appendix provides additional details on parameter degeneracies that arise when varying the magnification parameters $\Csample$ under different priors. Our aim is to identify the features in the data that lead to cases where the $\Csample$ posteriors deviate from the \Balrog estimates---for instance, when flat priors are used in bin 6.

\begin{figure*}
    \centering
    \includegraphics[width=0.99\textwidth]{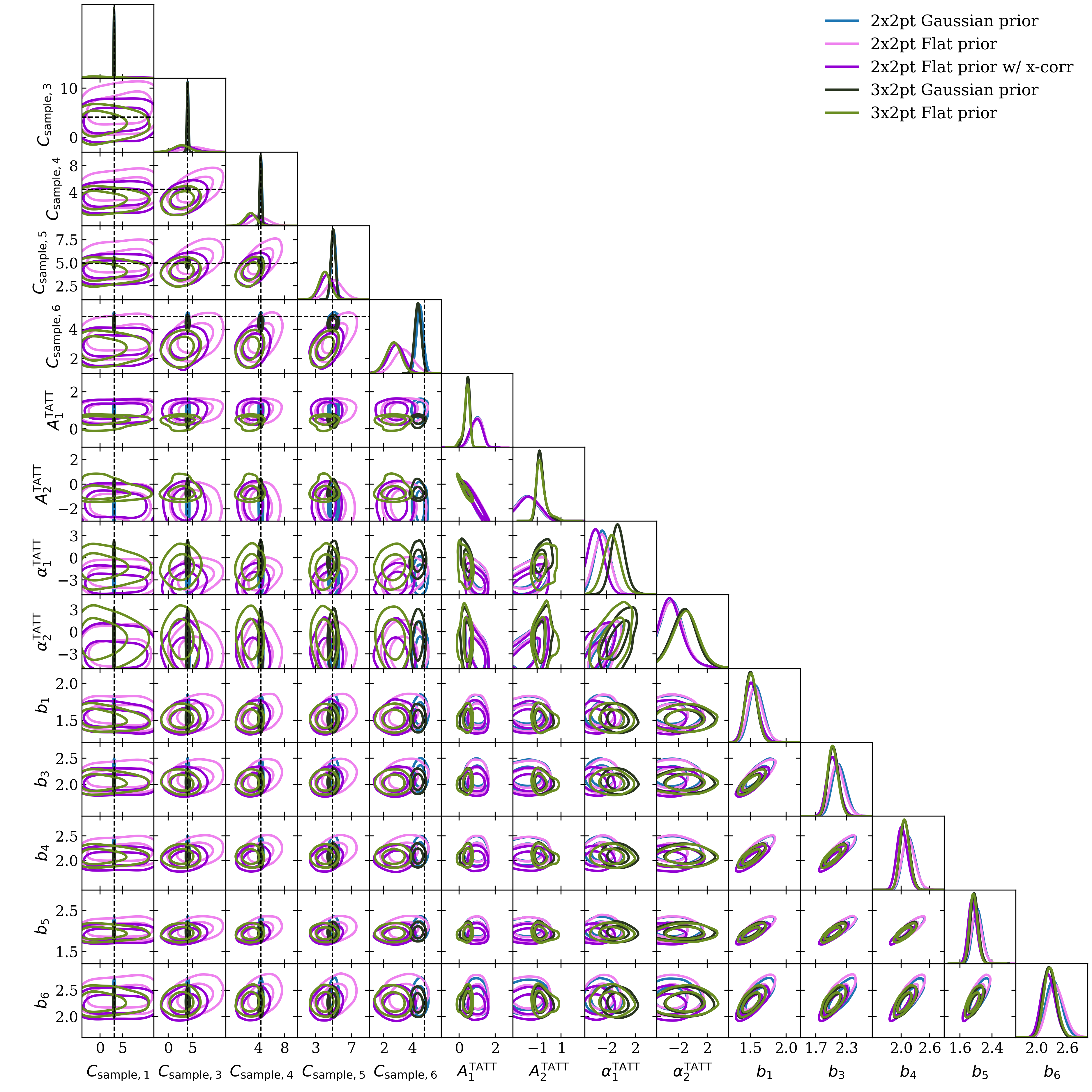}
    \caption{\textbf{Data 3$\times$2pt} and \textbf{2$\times$2pt}. Unblinded constraints on the magnification coefficients $\Csample$ for each lens redshift bin, the intrinsic alignment parameters of the TATT model ($A_1^{\rm TATT}$, $A_2^{\rm TATT}$, $\alpha_1^{\rm TATT}$ and $\alpha_2^{\rm TATT}$), and the galaxy bias parameters $b$ for each lens redshift bin. These constraints are derived from Y6 2$\times$2pt  and 3$\times$2pt data, for the various analysis variations presented in Sec.~\ref{sec:mag_cosmology}. Estimates of $\Csample$ from the \Balrog image simulations are shown as dashed black lines. There are no strong degeneracies among these parameters.} 
    \label{fig:contours_mag_coeff_data_3vs2x2pt_sys-params}
\end{figure*}

Fig.~\ref{fig:contours_mag_coeff_data_3vs2x2pt_sys-params} shows the 2D parameter constraints illustrating the correlations between the magnification coefficients $\Csample$ and other systematic parameters, including the intrinsic alignment and galaxy bias parameters, for both the 2$\times$2pt and 3$\times$2pt analyses.

We do not find significant correlation between the magnification $\Csample$ parameters and the other free systematic parameters in the analysis, indicating that the cause of the low $\Csample$ posteriors cannot be described by the existing systematics model.

\bibliographystyle{apsrev4-2}
\bibliography{bib_des, bib_des_y3kp, bib_des_y6kp, bib_non_des}

\end{document}